\documentclass[ review]{elsarticle}
\usepackage{setspace}
\usepackage{enumerate}
\usepackage{graphicx}
\usepackage{amsmath, bm}
\usepackage{float}
\usepackage{mdframed}
\usepackage{multirow}
\usepackage{textcomp}
\usepackage{subcaption}
\usepackage{url}
\usepackage{makecell}
\usepackage{amssymb}
\usepackage{bm}
\usepackage{textcmds}
 \usepackage{hyperref}
\biboptions{numbers,sort&compress}

\usepackage{tikz}
\usetikzlibrary{shapes.geometric, arrows}
\usepackage{caption}  

\def\be{\begin{equation}}\def\ee{\end{equation}}
\def\ba{\begin{array}}\def\ea{\end{array}}
\def\bfg{\begin{figure}}\def\efg{\end{figure}}
\makeatletter
\def\fps@figure{htbp}
\makeatother

\usepackage{stackengine}
\stackMath
\newcommand\tenq[2][1]{%
 \def\useanchorwidth{T}%
  \ifnum#1>1%
    \stackunder[0pt]{\tenq[\numexpr#1-1\relax]{#2}}{\scriptscriptstyle\sim}%
  \else%
    \stackunder[1pt]{#2}{\scriptscriptstyle\sim}%
  \fi%
}

\RequirePackage[margin=20mm]{geometry}

\journal{}
\begin{document}

\begin{frontmatter}

\title{Hilbert Transform Technique for Analyzing Mode I Crack Growth in an Pre- Stressed Monoclinic Crystalline Strip Under Punch Pressure}

\author[label1]{Diksha}
\author[label1]{Soniya Chaudhary*}
\author[label1]{Pawan Kumar Sharma}
\cortext[cor1]{Corresponding author: soniyachaudhary18@gmail.com}
\address[label1]{Department of Mathematics and Scientific Computing, National Institute of Technology Hamirpur, Himachal Pradesh, 177005, India}
\begin{abstract}
The crux of the present study is to analyze the Mode I crack propagation behavior in a pre-stressed monoclinic crystalline strip of finite thickness and infinite extent. The investigation focuses on the effects of collinear Griffith cracks and dynamic punch loading induced by plane wave propagation. The cracks are assumed to be in motion, and a Galilean transformation is employed to formulate the problem within a moving coordinate system. The boundary value problem is transformed into a system of coupled Cauchy-type singular integral equations, which are solved analytically using the Hilbert transform method. This approach yields elegant closed-form solutions for both the stress intensity factor and the crack opening displacement.
The study considers two monoclinic crystalline materials—Lithium Niobate (LiNbO$_3$) and Lithium Tantalate (LiTaO$_3$)—and compares their behavior with that of an isotropic material to assess the role of material anisotropy.
Numerical simulations and graphical analysis are performed for the crystalline materials with monoclinic symmetry to evaluate the influence of crack velocity, punch loading, material anisotropy, initial stress, and crack geometry on the fracture parameters. As a special case, the system is analyzed under the action of point loading from the punch pressure, and a comparative assessment is conducted between point loading and constant normal punch pressure. The results unveil critical insights into the dynamic fracture behavior of anisotropic materials under localized loading. This understanding enhances failure prediction in high-precision fields such as geomechanics, MEMS, surface acoustic devices, and biosensors.
\end{abstract}

\begin{keyword}
Griffith cracks \sep monoclinic crystalline strip \sep stress intensity factor \sep crack opening displacement \sep Hilbert transform \sep dynamic punch loading \sep fracture mechanics
\end{keyword}

\end{frontmatter}
\section{Introduction}
\subsection{Punch pressure in elastic and anisotropic solids}
The study of punch pressure is a classical yet continually evolving topic in the field of solid mechanics, particularly within the framework of fracture and contact mechanics. It involves analyzing the behavior of elastic or anisotropic bodies subjected to localized compressive forces applied by a rigid or semi-rigid punch. This type of loading condition is encountered in various engineering applications such as metal forming, forging, blanking, stamping, and ballistic impacts, where understanding the stress and strain fields around the contact region is critical.
When a punch presses against the surface of a solid, especially under dynamic or high-speed conditions, it induces complex stress distributions that can lead to the initiation and propagation of cracks. The nature of the material, the geometry of the punch, and the loading rate all influence the mechanical response of the system. In particular, dynamic punching introduces additional complexity due to the generation and propagation of elastic waves, which interact with internal heterogeneities or flaws, producing scattered wave fields and localized stress concentrations. These interactions are especially significant when dealing with materials exhibiting anisotropic or layered characteristics, where wave behavior deviates from that in homogeneous isotropic solids.
Practical implications of such analyses are far-reaching. In the aerospace and defense sectors, for instance, understanding the effect of impact and punching on composite and anisotropic panels is crucial for designing armor systems and protective structures \cite{zukas1993some,huang2011dynamic}. In microelectronic and microelectromechanical systems (MEMS) device fabrication, precise control of punch-induced forces plays a critical role in wafer dicing and thin-film processing, where fracture due to stress concentration can compromise device performance \cite{xu2012development,xu2013surface,xiao2020ultrasonic}. Moreover, in geomechanical applications, punch tests are used to estimate the strength parameters of rock and soil, especially in anisotropic formations \cite{lips2012experimental,simoes2016punching}. These examples underscore the significance of accurate stress-field prediction under punch pressure in real-world materials and structures.
\subsection{Dynamic effects and wave propagation in punch-induced loading}
When a punch impacts a solid—especially under dynamic or transient loading conditions—it generates a spectrum of elastic waves that propagate through the material. These waves interact with internal features such as cracks, voids, or inclusions, giving rise to complex scattering phenomena and localized stress concentrations. Such interactions are critical for understanding the onset and propagation of fracture zones, particularly in anisotropic or heterogeneous media, where wave velocities and directions depend on material orientation.
The classical work by Galin \textit{et al.} \cite{galin1961contact} laid the foundation for analyzing punch problems with different geometries, demonstrating how localized contact stresses evolve in both static and dynamic regimes. In dynamic impact scenarios, the motion of the punch introduces time-dependent boundary conditions that couple with wave propagation effects, significantly complicating the stress analysis. Unlike quasi-static cases, dynamic punches generate transient fields that evolve rapidly in space and time, requiring careful treatment of both contact mechanics and wave mechanics.
These complexities are not merely of academic interest—they have direct practical implications in fields such as metal forming, ballistic impact analysis, and geomechanics. For instance, in geophysical exploration, the principles governing dynamic punch problems closely resemble those of seismic wave propagation in anisotropic geological formations, where accurate modeling is essential for subsurface imaging and resource identification \cite{zheng2013seismic}. Moreover, contemporary research highlights the need to incorporate wave-induced effects to improve the prediction of failure mechanisms in engineered materials with anisotropic or layered microstructures \cite{song2020stress,feng2025investigation,chaudhary2025integral}. Such insights are indispensable in the design of durable materials and structures subjected to high-speed or impact loading.
\subsection{Monoclinic materials: structure and applications}
The study of elastic wave propagation in anisotropic solids, especially in the presence of cracks or material discontinuities, plays a pivotal role in various scientific and engineering disciplines such as geophysical exploration, earthquake seismology, and geo-mathematical prospecting \cite{maupin2007theory,tsvankin2012seismic}. Among the different classes of anisotropic materials, monoclinic crystals have attracted particular interest in the fields of solid mechanics and fracture mechanics due to their complex symmetry and direction-dependent elastic properties. Crystalline solids are broadly classified into seven systems: isometric, tetragonal, hexagonal, orthorhombic, trigonal, triclinic, and monoclinic. Notably, the monoclinic crystal system accounts for nearly one-third of all naturally occurring minerals, highlighting its structural relevance and practical importance in mineralogy and material science \cite{destrade2004explicit}.
Monoclinic crystals are defined by their distinct crystallographic structure, consisting of three unequal axes—two of which intersect at an oblique angle, while the third remains perpendicular to the other two. This unique geometry results in a single plane of elastic symmetry and requires 13 independent elastic constants to fully describe their mechanical behavior \cite{nye1985physical}. Foundational studies by Ekstein \cite{ekstein1945high}, Newman \textit{et al.} \cite{newman1957vibrations}, and Kaul \textit{et al.} \cite{kaul1962frequency} have significantly contributed to the mathematical modeling and understanding of the dynamic response of monoclinic materials under various mechanical loading conditions.
In practical applications, monoclinic materials such as Lithium Tantalate (LiTaO\textsubscript{3}) and Lithium Niobate (LiNbO\textsubscript{3}) are extensively utilized in advanced device fabrication. These materials serve as essential elements in surface acoustic wave (SAW) systems, optical switches, laser second-harmonic generation, non-contact ultrasonic sensors, and vibration suppression technologies \cite{hashimoto2000surface}. Due to their intrinsic piezoelectric and pyroelectric properties, they are also widely employed in the development of sensors, telecommunication devices, and aerospace systems \cite{tiersten2013linear}. The dual capability of these materials to interact with both mechanical and electrical fields renders them especially suitable for multifunctional smart materials and the miniaturized architecture of MEMS.

\subsection{Voids, cracks, and crack propagation in monoclinic materials}
Crack propagation in monoclinic materials presents intricate challenges due to their anisotropic elasticity and direction-dependent fracture behavior. The unique crystallographic structure leads to asymmetric stress fields near crack tips, making their analysis crucial for structural integrity in high-performance applications.
Early foundational work by Yuan \cite{yuan1998determination} addressed stress coefficient terms in cracked monoclinic solids with plane symmetry, followed by Nair and Sotiropoulos \cite{nair1999interfacial}, who examined interfacial wave behavior in monoclinic systems with an interlayer. Exner and Dabrowski \cite{exner2010monoclinic} investigated three-dimensional (3D) flanking structures around elliptical cracks, while Banks-Sills and Ikeda \cite{banks2011stress} studied interface cracks between orthotropic and monoclinic materials, focusing on stress intensity factor (SIF) mismatches.
Chaudhuri’s contributions \cite{chaudhuri2012three,chaudhuri2024three} provided detailed 3D analyses of singular stress fields and interfacial instability in monoclinic and bicrystalline systems under antiplane shear loading. 
Computational approaches have further enriched understanding; Zhao \textit{et al.} \cite{zhao2016study} applied the phase field method to model crack propagation in single-crystal zirconia, a framework extendable to monoclinic systems. Varna \textit{et al.} \cite{varna2025cod} proposed a crack opening displacement (COD) and crack sliding displacement (CSD)-based model to characterize in-plane stiffness degradation in symmetric laminates containing cracks and local delaminations. Singh \textit{et al.}
These developments collectively advance the theoretical and practical understanding of voids, collinear cracks, and crack dynamics in monoclinic materials—crucial for applications in geomechanics, aerospace structures, and MEMS.

\subsection{Collinear cracks in layered and half-space media: evolution and modelling approaches}
The study of collinear cracks in layered and half-space configurations plays a pivotal role in evaluating the integrity of advanced engineering structures and materials such as functionally graded materials (FGMs), magnetoelectroelastic composites, and piezoelectric laminates. These cracks, when subjected to dynamic or static loading, give rise to complex stress interactions influenced by material gradation, anisotropy, and interfacial behavior.
Lowengrub and Srivastava established initial theoretical foundations \cite{lowengrub1968two}, who investigated two coplanar Griffith cracks in an infinitely long elastic strip. This was extended by Parihar and Kushwaha \cite{parihar1975stress}, who calculated SIF for two symmetrically located Griffith cracks in an elastic strip under symmetrical body forces. Later, Itou \cite{itou1980diffraction} studied the diffraction of antiplane shear waves by two Griffith cracks in an infinite medium. Generalizations to multi-crack systems were made by Das and Ghosh \cite{das1992four}, analyzing four coplanar Griffith cracks in an infinite elastic domain. In a layered context, Singh \textit{et al.} \cite{singh2009closed} derived closed-form solutions for two collinear anti-plane cracks in a magnetoelectroelastic layer, while Hiriyur et al. \cite{hiriyur2012quasi} introduced a quasi-algebraic multigrid method to solve fracture problems using extended finite element methods, offering a computational advantage for large-scale crack simulations.
 Bagheri and Ayatollahi \cite{bagheri2012multiple} addressed multiple moving cracks in functionally graded strips.
Li et al. \cite{li2013dynamic} focused on the dynamic initiation and propagation of multiple cracks in brittle materials, offering insights into transient fracture behaviors.
Zhu and Liu \cite{9019244} used the layering method to perform fracture analysis of multiple cracks in functionally graded piezoelectric materials. Jangid \cite{jangid2021two} extended this to examine the effects of electric and magnetic poling on crack behavior in magneto-electro-elastic (MEE) materials.
Recent studies have brought in advanced formulations. Mandal \textit{et al.} \cite{mandal2021dispersion} explored wave dispersion by four coplanar cracks, and Babadjian and Bonhomme \cite{babadjian2023discrete} contributed to the numerical modeling domain by developing a discrete approximation of the Griffith functional using adaptive finite elements, offering new computational strategies for fracture simulation. Sharma \textit{et al.} \cite{sharma2024influence} incorporated strain gradient elasticity to study two unequal mode-III cracks in FGMs. Bagheri \cite{bagheri2023several} employed the dislocation method to address steady harmonic loading in cracked piezoelectric FGMs. Kumar \textit{et al.} \cite{kumar2024singular} presented a singular integral method for collinear cracked MEE materials under moving  electric-magnetic polarization saturation loading conditions.  
These investigations provide a progressively refined understanding of crack interaction, wave scattering, and intensity factors in complex media, offering practical implications for smart composites and layered structural components.
\subsection{Reseach gap and objective of the study}
Several distinguished researchers have extensively studied fracture mechanics to evaluate SIF, COD, and wave interactions in materials with single or multiple cracks \cite{das1992four,bagheri2012multiple,bagheri2023several,kumar2021analytical,babadjian2023discrete,chaudhary2025crack}. However, a critical gap remains in incorporating the combined influence of pre-existing (initial) stresses on the material and their effect on COD under dynamic loading.
Previous studies such as \cite{das1992four,bagheri2012multiple,bagheri2023several} do not consider the role of dynamic loading or moving loads in the analysis, limiting their applicability to more realistic engineering scenarios. On the other hand, works like \cite{kumar2021analytical,babadjian2023discrete,chaudhary2025crack} neglect the influence of initial stress fields on crack behavior, particularly in relation to COD and its evolution over time. To bridge these gaps, the present study develops a comprehensive analytical model to examine Mode I Griffith crack propagation in a monoclinic crystalline strip. The model considers the presence of collinear cracks subjected to both horizontal and vertical initial (pre-) stresses, along with the action of a moving punch load. The loading is applied in the direction of crack propagation, simulating the dynamic interaction between the propagating wave and the moving load.
This investigation introduces a novel approach by combining pre-stress effects and dynamic loading in anisotropic media. The Galilean transformation is applied to convert the stationary frame to a moving reference frame, simplifying the formulation under a traveling load. The Hilbert transform technique is then used to derive closed-form expressions for the SIF and crack opening displacement COD.
The main objective of this study is to analyze the behavior of collinear Mode I Griffith cracks in a monoclinic crystalline strip under the combined influence of dynamic punch loading and initial stresses. Special consideration is also given to the case of concentrated (point) loading. This formulation provides valuable insights into fracture behavior in anisotropic, pre-stressed media subjected to dynamic forces.
\subsection{Motivation of the Study}
In advanced systems like MEMS actuators and SAW biosensors, punch-type loading commonly arises from mechanical actuation or contact forces, acting as a localized stress concentrator that promotes crack initiation and growth. Simultaneously, residual or initial stresses—due to fabrication, thermal effects, or mechanical constraints—modify the stress field near crack tips, influencing fracture behavior. The combined impact of punch loading and pre-stress significantly alters both the SIF and COD, especially under wave propagation triggered by dynamic loading.
For instance, in SAW-based biosensors used for detecting biological elements like circulating tumor cells (CTCs), any structural defect due to unnoticed crack initiation can impair signal fidelity or lead to catastrophic failure. Similarly, in piezoelectric crystalline wafers used in MEMS actuators, the combined action of surface wave propagation, punch pressure, and internal stress may lead to premature device degradation if not properly modeled.  Thus, this study is driven by the need to establish an analytical framework that captures the coupled effects of initial stress, punch pressure, and wave propagation on fracture parameters in monoclinic crystalline materials, ensuring improved reliability and performance in real-world applications.
\subsection{Novelty of the Approach}
The novel contributions of the present study are listed below:
\begin{itemize}
    \item The study models the interaction between pre-existing stresses and dynamic punch loading on Mode I collinear Griffith crack propagation in monoclinic crystalline materials.
    \item It incorporates both horizontal and vertical initial stresses along with moving punch loading, providing a more realistic representation of crack behavior under practical conditions.
    \item The Hilbert transform method is utilized to obtain explicit expressions for the SIF and COD.
    \item The influence of various parameters such as initial stress, punch load, crack geometry, and material properties on SIF and COD is thoroughly investigated.
    \item The framework offers insights into crack growth prediction, material integrity, and device durability under dynamic and pre-stressed conditions.
\end{itemize}
\subsection{Key applications of the present study}
The understanding and quantification of SIF and COD under combined dynamic loading and initial stress conditions are critical in ensuring material reliability and performance. The findings of this study have valuable implications in several domains:
\begin{itemize}
    \item {SAW sensors and biosensors:} Enhancing sensitivity and reliability in detecting biological agents like circulating tumor cells.
    \item {MEMS devices:} Improving fracture resistance in microactuators made of piezoelectric crystalline materials.
   \item {Aerospace engineering:} Aiding the prediction of crack growth in anisotropic structural components under dynamic loading conditions.
    \item {Microelectronics:} Assisting in the design of durable substrates subject to mechanical and thermal stresses.
    \item {Structural health monitoring:} Enabling early detection and modeling of crack evolution in stressed, anisotropic materials.
\end{itemize}
\subsection{Layout of the manuscript} The organization of the paper is as follows: Section \ref{Analytical approach} presents the analytical method used to derive closed-form expressions for the SIF and COD for the propagation of a Mode I collinear Griffith crack. Subsection \ref{Formulation of the problem} describes the geometry of the considered model. Subsection \ref{Mathematical modeling for the problem} outlines the governing equations and constitutive relations for the present problem in a moving reference frame, utilizing the Galilean transformation. Subsection \ref{Boundary conditions} presents the boundary conditions for the problem governing stable crack propagation. In Subsection \ref{Solution of the problem}, the solution procedure is discussed, assuming displacement components in integral form. Subsection \ref{Stress intensity factor and crack opening displacement} provides the explicit expressions for the SIF and COD at both crack tips.
Subsection \ref{Special cases} explores a special case incorporating point loading instead of constant normal pressure, and the results are verified against well-established prior studies. Section \ref{Numerical simulation and graphical analysis} presents numerical simulations and graphical analysis, where the effects of various material parameters on the SIF and COD are analyzed. Finally, Section \ref{Conclusion} concludes the study and discusses the findings, which will be useful for material failure prediction.
\section{Analytical approach}
\label{Analytical approach}
\subsection{Formulation of the problem}
\label{Formulation of the problem}
\begin{figure}[h]
    \centering
    \includegraphics[width=0.9\linewidth]{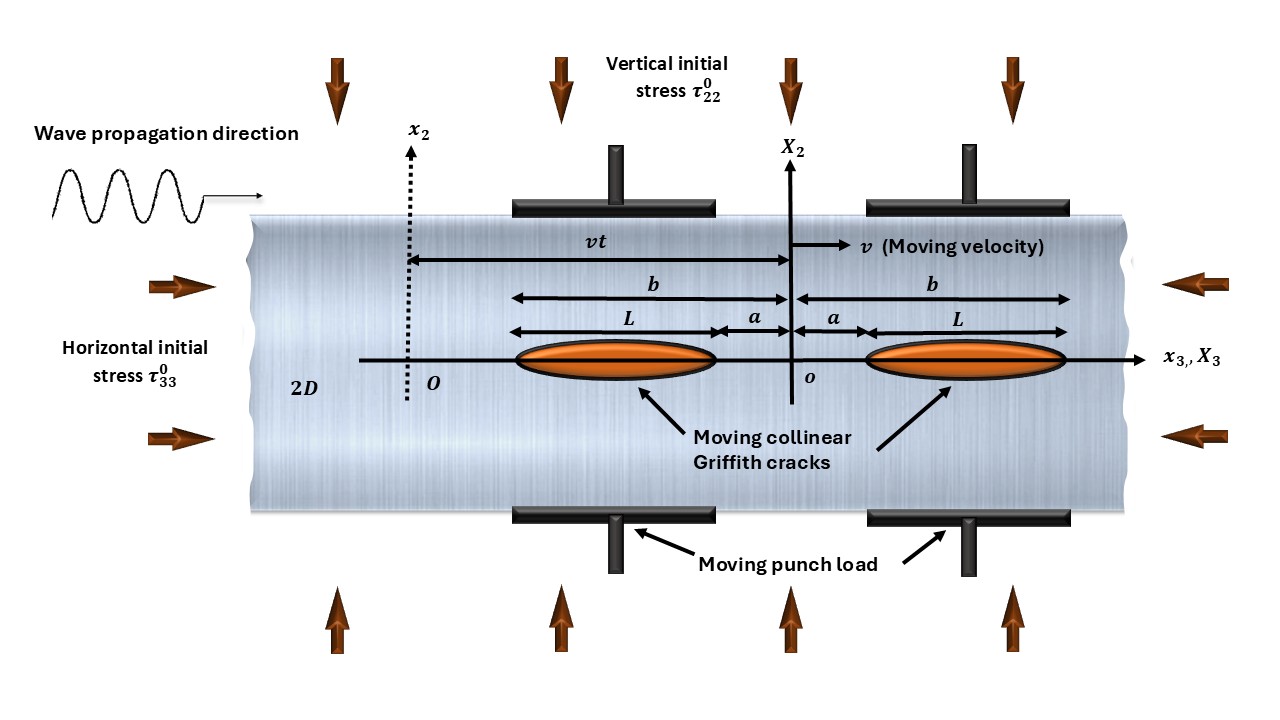}
    \caption{Schematic of a collinear cracks propagating in a monoclinic crystalline strip.}
    \label{geometry}
\end{figure}

Consider an initially stressed, infinite, crystalline monoclinic strip containing two collinear Griffith cracks of identical finite length \(L\), symmetrically located along its midplane. The strip is subjected to a propagating plane wave, which interacts with the Griffith cracks and induces tensile stress perpendicular to the crack surfaces. This interaction results in symmetric crack opening, thereby giving rise to Mode I crack propagation. Simultaneously, the strip is influenced by a smooth parallel punch pressure applied along its boundary at the same location as the propagating cracks, further affecting the stress state near the crack tips.
A Cartesian coordinate system \(O x_1 x_2 x_3\) is introduced such that the strip extends infinitely in the \(x_3\)-direction, while its finite thickness is confined within \(-D \leq x_2 \leq D\). The mechanical configuration considers initial stresses: a horizontal initial stress \(\tau_{33}^0\) acting in the \(x_3\)-direction and a vertical initial stress \(\tau_{22}^0\) acting in the \(x_2\)-direction. Symmetrically embedded Griffith cracks are positioned along the midplane (\(x_2 = 0\)), and a moving parallel punch pressure is applied on the boundary surface (\(x_2 = \pm D\)), distributed over the same interval as the crack zone in the \(x_3\)-direction.
It is assumed that the cracks propagate uniformly along the positive \(x_3\)-axis with a constant speed \(v\), parallel to the direction of both the wave propagation and the applied punch pressure. A schematic diagram illustrating the problem setup is presented in Fig.~\ref{geometry}.

In the subsequent subsection, the mathematical modeling of the problem is carried out by transforming the fixed reference frame into a translating (moving) reference frame. The governing equation is then derived in this moving reference frame, capturing the dynamics of the system under the applied conditions.

\subsection{Mathematical modeling for the problem}
\label{Mathematical modeling for the problem}
The constitutive relation for a crystalline strip with monoclinic symmetry, oriented with a rotated $x_2$-cut where the $x_1$-axis corresponds to the diagonal axis, is given by \cite{chattopadhyay1987diffraction}:
\begin{equation}
\begin{bmatrix}
\tau_{x_1 x_1} \\
\tau_{x_2 x_2} \\
\tau_{x_3 x_3} \\
\tau_{x_2 x_3} \\
\tau_{x_1 x_3} \\
\tau_{x_1 x_2}
\end{bmatrix}
=
\begin{bmatrix}
\mu_{11} & \mu_{12} & \mu_{13} & \mu_{14} & 0 & 0 \\
\mu_{12} & \mu_{22} & \mu_{23} &\mu_{24} & 0 & 0 \\
\mu_{13} & \mu_{23} & \mu_{33} & \mu_{34} & 0 & 0 \\ 
\mu_{14} & \mu_{24} & \mu_{34} & \mu_{44} & 0 & 0\\
0 & 0 & 0 & 0 &\mu_{55} & \mu_{56}  \\
0 & 0 & 0 &  0 &\mu_{56} & \mu_{66} 

\end{bmatrix}
\begin{bmatrix}
\epsilon_{x_1} \\
\epsilon_{x_2} \\
\epsilon_{x_3} \\
\gamma_{x_2 x_3} \\
\gamma_{x_1 x_3} \\
\gamma_{x_1 x_2}
\end{bmatrix},
\label{constitutive}
\end{equation}
\noindent
where, \( \tau_{x_1 x_1}, \tau_{x_2 x_2}, \tau_{x_3 x_3} \) represent the normal stresses, while \( \tau_{x_2 x_3}, \tau_{x_1 x_3}, \tau_{x_1 x_2} \) denote the shear stresses. Similarly, \( \epsilon_{x_1}, \epsilon_{x_2}, \epsilon_{x_3} \) are the normal strains, while \( \gamma_{x_2 x_3}, \gamma_{x_1 x_3}, \gamma_{x_1 x_2} \) represent the shear strains. 
These strain components are further related to the displacement field through the following expression:  
\be
\begin{aligned}
\epsilon_{x_1} &= \frac{\partial u_1}{\partial x_1}, \quad
\epsilon_{x_2} = \frac{\partial u_2}{\partial x_2}, \quad
\epsilon_{x_3} = \frac{\partial u_3}{\partial x_3}, \quad  
\gamma_{x_2 x_3} = \frac{\partial u_2}{\partial x_3} + \frac{\partial u_3}{\partial x_2}, \quad
\gamma_{x_1 x_3} = \frac{\partial u_1}{\partial x_3} + \frac{\partial u_3}{\partial x_1}, \quad
\gamma_{x_1 x_2} = \frac{\partial u_1}{\partial x_2} + \frac{\partial u_2}{\partial x_1},
\end{aligned}
\label{strain_relation}
\ee
where \( u_i \) represents the displacement component in the \( x_i \)-direction. The terms \( \mu_{ij} \) are the elastic constants that characterize the material's stiffness properties in the monoclinic system. 

The governing equation of motion for elastic disturbances in a pre-stressed crystalline monoclinic strip (in the absence of body forces), as derived by Biot~\cite{biot1940influence}, is given by:
\begin{equation}
    \tau_{ij,j} + (u_{i,k} \tau_{kj}^0)_{,j} = \rho \ddot{u}_i,
    \label{governing_equation}
\end{equation}
where \( \rho \) is the material density,
    \( \ddot{u}_i \) represents the second-order time derivative of the displacement components, accounting for the acceleration, and $\tau_{ij}^0$ denotes the initial stress components.
     The notation \( (\,\cdot\,)_{,j} \) represents the partial derivative with respect to the coordinate \( x_j \).
In this formulation, the term \( \tau_{ij,j} \) corresponds to the divergence of the stress field arising from elastic disturbances, while the term \( (u_{i,k} \tau_{kj}^0)_{,j} \) accounts for the influence of the initial stress field on the elastic disturbances.

The condition for plane strain deformation in a crystalline monoclinic strip containing a moving collinear Griffith cracks, influenced by plane wave propagation parallel to the \(x_2 x_3\)-plane, is given by:  
\be
u_1=0, \quad u_2 =u_2(x_2,x_3,t), \quad u_3=u_3(x_2,x_3,t), \quad\frac{\partial}{\partial x_1} ()\equiv 0.
\label{plane_deformation_condition}
\ee
This condition implies that displacement components vary only in the \(x_2\)- and \(x_3\)-directions, with no dependence on the \(x_1\)-coordinate.
Using Eqs. (\ref{constitutive}), (\ref{strain_relation}), and (\ref{plane_deformation_condition}), the governing equation of motion (\ref{governing_equation}) for the present problem, considering the initial stress only in the $x_2$ and $x_3$ directions, is given by:
\begin{equation}
   ( \mu_{22} + \tau_{22}^0)\frac{\partial^2 u_2}{\partial x_2^2} 
    + (\mu_{44}+\tau_{33}^0) \frac{\partial^2 u_2}{\partial x_3^2} 
    + \mu_{24} \frac{\partial^2 u_3}{\partial x_2^2} 
    + \mu_{34} \frac{\partial^2 u_3}{\partial x_3^2} 
    + 2\mu_{24} \frac{\partial^2 u_2}{\partial x_2 \partial x_3} 
    + (\mu_{23} + \mu_{44}) \frac{\partial^2 u_3}{\partial x_2 \partial x_3} 
    = \rho \frac{\partial^2 u_2}{\partial t^2},
    \label{gv1}
\end{equation}
\begin{equation}
    \mu_{24} \frac{\partial^2 u_2}{\partial x_2^2} 
    + \mu_{34} \frac{\partial^2 u_2}{\partial x_3^2} 
    + (\mu_{44}+\tau_{22}^0) \frac{\partial^2 u_3}{\partial x_2^2} 
    + (\mu_{33}+\tau_{33}^0) \frac{\partial^2 u_3}{\partial x_3^2} 
    + (\mu_{23} + \mu_{44}) \frac{\partial^2 u_2}{\partial x_2 \partial x_3} 
    + 2\mu_{34} \frac{\partial^2 u_3}{\partial x_2 \partial x_3} 
    = \rho \frac{\partial^2 u_3}{\partial t^2}.
    \label{gv2}
\end{equation}
Since the collinear Griffith cracks are assumed to propagate along the midline of the strip with a uniform speed \(v\), a steady-state stress distribution is established. To simplify the analysis, introduce a translating coordinate system \((X_1, X_2, X_3)\) that moves in the direction of the propagating cracks. According to the Galilean transformation law, the relationship between the stationary Cartesian coordinate system \((x_1, x_2, x_3)\) and the moving coordinate system is given by:  
\begin{equation}
    X_1 = x_1, \quad X_2 = x_2, \quad X_3 = X = x_3 - vt.
    \label{galilean_transformation}
\end{equation}  
By applying the Galilean transformation described in Eq. (\ref{galilean_transformation}), the governing equations of motion given in Eqs. (\ref{gv1}) and (\ref{gv2}) are transformed into the following form:
\begin{equation}
   ( \mu_{22} + \tau_{22}^0)\frac{\partial^2 u_2}{\partial X_2^2} 
    + (\mu_{44}+\tau_{33}^0-v^2 \rho) \frac{\partial^2 u_2}{\partial X_3^2} 
    + \mu_{24} \frac{\partial^2 u_3}{\partial X_2^2} 
    + \mu_{34} \frac{\partial^2 u_3}{\partial X_3^2} 
    + 2\mu_{24} \frac{\partial^2 u_2}{\partial X_2 \partial X_3} 
    + (\mu_{23} + \mu_{44}) \frac{\partial^2 u_3}{\partial X_2 \partial X_3} 
    = 0,
    \label{GV1}
\end{equation}
\begin{equation}
    \mu_{24} \frac{\partial^2 u_2}{\partial X_2^2} 
    + \mu_{34} \frac{\partial^2 u_2}{\partial X_3^2} 
    + (\mu_{44}+\tau_{22}^0) \frac{\partial^2 u_3}{\partial X_2^2} 
    + (\mu_{33}+\tau_{33}^0 -v^2 \rho) \frac{\partial^2 u_3}{\partial X_3^2} 
    + (\mu_{23} + \mu_{44}) \frac{\partial^2 u_2}{\partial X_2 \partial X_3} 
    + 2\mu_{34} \frac{\partial^2 u_3}{\partial X_2 \partial X_3} 
    =0.
    \label{GV2}
\end{equation}

Equations (\ref{GV1}) and (\ref{GV2}) represent the final system of governing equations for the present problem in the translating coordinate system. In the next subsection, the boundary conditions of the analysis are described, which are subsequently used to derive the stress intensity factor and crack opening displacement for the Mode I Griffith crack analysis.
\subsection{Boundary conditions}
\label{Boundary conditions}
In the present problem, two collinear Griffith cracks propagating along the midline of a monoclinic crystalline strip are considered. The cracks are assumed to move with a uniform speed \( v \), and the corresponding stress field is considered in the moving coordinate system.  
At a fixed time \( t \), the moving coordinate system defines the crack positions within the interval \( a \leq |X_3| \leq b \) at \( X_2 = 0 \) as illustrated in Fig. \ref{geometry}. These cracks are subjected to an internal normal traction distribution, denoted by \( \sigma_1(X_3) \), acting along the crack surfaces. Additionally, a normal punch pressure load \( \sigma_2(X_2) \) is applied at the boundary of the strip, i.e., \( a \leq |X_3| \leq b \) at \( X_2 =\pm D \).  
Due to the symmetric nature of the problem with respect to the \( X_3 \)-axis, it is sufficient to restrict the analysis to the upper half of the strip, defined by \( 0 \leq X_2 \leq D \). This symmetry simplifies the boundary conditions without loss of generality.  

The boundary conditions for the proposed model, formulated in the moving coordinate system, are expressed as follows:

At the boundary of the strip ($X_2 = D$):

\begin{enumerate}[(i)]

\item The vertical displacement $u_2(X_2,X_3)$ is constrained to zero outside the cracked region, implying no vertical motion of the boundary in the uncracked segments.
\begin{equation}
u_2(X_2,X_3) = 0, \quad |X_3|<a,\, |X_3|>b. \tag{10a} \label{bc1}
\end{equation}
\item The longitudinal displacement $u_3(X_2,X_3)$ along the $X_3$-direction is constrained to zero outside the cracked region, ensuring no tangential slip occurs in the uncracked segments.
\begin{equation}
u_3(X_2,X_3) = 0, \quad |X_3|<a,\, |X_3|>b. \tag{10b} \label{bc2}
\end{equation}

 \item On the cracked region subjected to the punch pressure, the tangential shear stress $\tau_{23}(X_2,X_3)$ is assumed to vanish, implying a shear stress-free boundary condition.
    \begin{equation}
    \tau_{23}(X_2,X_3) = 0, \quad a \le |X_3| \le b. \tag{10c} \label{bc3}
    \end{equation}
    
   \item At the upper boundary, within the cracked region, the material strip is subjected to an applied punch pressure, represented by \( \sigma_2(X_2) \), which induces a normal stress at the crack faces.
\begin{equation}
\tau_{22}(X_2,X_3) = -\sigma_2(X_2), \quad \text{for } a \le |X_3| \le b. \tag{10d} \label{bc4}
\end{equation}
    \end{enumerate}

At the midline of the strip ($X_2 = 0$):

\begin{enumerate}[(i)]
\item Outside the crack region, at the midplane, the vertical displacement is zero, indicating that there is no vertical motion in the regions beyond the crack faces.
\begin{equation}
u_2(X_2,X_3) = 0,  \quad |X_3|<a,\, |X_3|>b. \tag{11a}\label{bc5}
\end{equation}
\item Similarly, outside the cracked region, the longitudinal displacement is constrained to zero, indicating no longitudinal motion beyond the crack faces.
\begin{equation}
u_3(X_2,X_3) = 0,  \quad |X_3|<a,\, |X_3|>b. \tag{11b} \label{bc6}
\end{equation}

\item The strip is free of shear traction along the crack faces, satisfying the stress-free condition in the tangential direction within the cracked region.
\begin{equation}
\tau_{23}(X_2,X_3) = 0, \quad a \le |X_3| \le b. \tag{11c} \label{bc7}
\end{equation}
\item In the region occupied by the cracks, the midplane experiences an internal normal stress distribution represented by $p(X_3)$, which captures the effect of internal mechanical loading within the crack zone.
\begin{equation}
\tau_{22}(X_2,X_3) = -p(X_3), \quad a \le |X_3| \le b. \tag{11d} \label{bc8}
\end{equation}
\end{enumerate}

This subsection presents the boundary conditions imposed on the physical system. In the following subsection, these boundary conditions are further utilized to derive the explicit expressions for the stress intensity factor and the crack opening displacement.

\subsection{Solution of the problem}
\label{Solution of the problem}
The solution of the Eqs. (\ref{GV1}) are (\ref{GV2}) are assumed in the following form given by:
\be
u_3(X_2,X_3)= \int_0^ \infty M(X_2, \zeta) \sin (\zeta X_3) d \zeta,
\label{u_3}
\ee
\be
u_2(X_2,X_3)= \int_0^ \infty N(X_2, \zeta) \cos (\zeta X_3) d \zeta,
\label{u_2}
\ee
where \( M(X_2, \zeta) \) and \( N(X_2, \zeta) \) are determined using boundary conditions. Substituting the assumed displacement components from Eqs.~(\ref{u_3}) and (\ref{u_2}) into the governing equations (\ref{GV1}) and (\ref{GV2}) yields the following system of differential equations for \( M(X_2, \zeta) \) and \( N(X_2, \zeta) \):
\begin{equation}
    (\mu_{22}+\tau_{22}^0)\frac{d^2 N}{d X_2^2}+(\mu_{23}+\mu_{44})\zeta \frac{d M}{d X_2}- (\mu_{44}+\tau_{33}^0- \rho v^2)\zeta^2 N=0,
    \label{differential1}
\end{equation}
\begin{equation}
    (\mu_{44}+\tau_{22}^0)\frac{d^2 M}{d X_2^2}-(\mu_{23}+\mu_{44}) \zeta\frac{d N}{d X_2}-(\mu_{33}+\tau_{33}^0- \rho v^2) \zeta^2 M=0.
    \label{differentia2}
\end{equation}
The general solutions for \( M(\zeta, X_2) \) and \( N(\zeta, X_2) \), satisfying Eqs. (\ref{differential1}) and (\ref{differentia2}), are expressed as:
\begin{align}
M(\zeta, X_2)&=M_1(\zeta) \cosh(\alpha_1\zeta X_2)+M_2(\zeta) \cosh(\alpha_2\zeta X_2)+M_3(\zeta) \sinh(\alpha_1\zeta X_2)+M_4(\zeta) \sinh(\alpha_2\zeta X_2),
\label{M}  \\[2ex]
N(\zeta, X_2)&=N_1(\zeta) \sinh(\alpha_1\zeta X_2)+N_2(\zeta) \sinh(\alpha_2\zeta X_2)+N_3(\zeta) \cosh(\alpha_1\zeta X_2)+N_4(\zeta) \cosh(\alpha_2\zeta X_2), 
\label{N}
\end{align}
where \( \alpha_i \) represent the positive roots obtained from solving the following biquadratic equation:
\be
\left.
\begin{aligned}
    \displaystyle \frac{\mu_{22}+\tau_{22}^0}{\mu_{44}+\tau_{22}^0} \:\alpha_i^4+ \left \{ \left( \frac{\mu_{23}+\mu_{44}}{\mu_{44}+\tau_{22}^0}\right)^2 -\left( \frac{\mu_{22}+\tau_{22}^0}{\mu_{44}+\tau_{22}^0}\right) \left( \frac{\mu_{33}+\tau_{33}^0}{\mu_{44}+\tau_{22}^0}-\frac{v^2}{v_c^2}\right)-\left( \frac{\mu_{44}+\tau_{33}^0}{\mu_{44}+\tau_{22}^0}-\frac{v^2}{v_c^2}\right)\right\} \alpha_i^2
    \\[2ex]
    \displaystyle+\left( \frac{\mu_{33}+\tau_{33}^0}{\mu_{44}+\tau_{22}^0}-\frac{v^2}{v_c^2}\right) \left( \frac{\mu_{44}+\tau_{33}^0}{\mu_{44}+\tau_{22}^0}-\frac{v^2}{v_c^2}\right)=0,
    \end{aligned}
\right.
\label{biquadaratic_equation}
\ee
where $\displaystyle v_c=\sqrt{\frac{\mu_{44}+\tau_{22}^0}{\rho}}$; \( M_i \) and \( N_i \) are arbitrary functions, and the relationship between them is given by: 
\begin{equation}
N_i(\zeta) = \frac{\xi_1}{\alpha_1} M_i(\zeta), \quad i = 1,3, \quad \text{and} \quad  
N_i(\zeta) = \frac{\xi_2}{\alpha_2} M_i(\zeta), \quad i = 2,4.
\label{relation_in_MN}
\end{equation}
where  
\begin{equation}
\xi_i = \frac{\alpha_i^2 - \left( \frac{\mu_{33}+\tau_{33}^0}{\mu_{44}+\tau_{22}^0} - \frac{v^2}{v_c^2} \right)}
{\frac{\mu_{23}+\mu_{44}}{\mu_{44}+\tau_{22}^0}}.
\label{zi_expression}
\end{equation}
By substituting Eqs. (\ref{u_3}), (\ref{u_2}), (\ref{M}), (\ref{N}), and (\ref{relation_in_MN}) into Eq. (\ref{constitutive}), the stress components can be rewritten in terms of convective coordinates as:
\be
\begin{aligned}
    \tau_{23}=&\int_0^\infty \left\{  (\mu_{34}+\xi_1 \mu_{24})\:M_1 \cosh(\alpha_1\zeta X_2)+(\mu_{34}+
    \xi_2 \mu_{24})\:M_2 \cosh(\alpha_2\zeta X_2)+(\mu_{34}+\xi_1 \mu_{24})\:M_3 \sinh(\alpha_1\zeta X_2) \right.\\
    & \left. +(\mu_{34}+
    \xi_2 \mu_{24})\:M_4 \sinh(\alpha_2\zeta X_2)\right\} \:\zeta \cos(\zeta X_3)\:d \zeta +\int_0^\infty \left\{ \Omega_1 M_1 \sinh(\alpha_1\zeta X_2)+ 
    \Omega_2 M_2 \sinh(\alpha_2\zeta X_2)\right.\\
    & \left. 
    +\Omega_1 M_3 \cosh(\alpha_1\zeta X_2)+\Omega_2 M_4 \cosh(\alpha_2\zeta X_2) \right\} \:\zeta \sin(\zeta X_3)\: d \zeta,
    \label{eq21}
    \end{aligned}
\ee
\be
\begin{aligned}
    \tau_{22}=&\int_0^\infty \left\{ (\mu_{23}+\xi_1 \mu_{22}) \:M_1 \cosh(\alpha_1\zeta X_2) + (\mu_{23}+\xi_2 \mu_{22}) \:M_2 \cosh(\alpha_2\zeta X_2)+ (\mu_{23}+\xi_1 \mu_{22}) \:M_3 \sinh (\alpha_1\zeta X_2) \right. \\
    & \left.+(\mu_{23}+\xi_2 \mu_{22}) \:M_4 \sinh(\alpha_2\zeta X_2)\right\} \: \zeta \cos(\zeta X_3) \: d \zeta+\frac{\mu_{24}}{\mu_{44}}\: \int_0^\infty \left \{ \Omega_1 M_1 \sinh(\alpha_1 \zeta X_2) +\Omega_2 M_2 \sinh (\alpha_2 \zeta X_2) \right. \\
    & \left. + \Omega_1 M_3 \cosh(\alpha_1 \zeta X_2) +\Omega_2 M_4 \cosh (\alpha_2 \zeta X_2) \right\} \: \zeta \sin(\zeta X_3) \: d\zeta,
    \label{eq22}
\end{aligned} 
\ee
where $\Omega_i=\mu_{44}\left(\alpha_i-\frac{\xi_i}{\alpha_i}\right)$; $i=1,2$.
Imposing the boundary condition (\ref{bc7}) in Eq. (\ref{eq21}) yields:
\be
M_4=-\frac{\Omega_1}{\Omega_2} M_3.
\label{relation_M4M3}
\ee
Using the displacement boundary conditions imposed at \(X_3=D\) (Eq.~\eqref{bc1}) and \(X_3=0\) (Eq.~\eqref{bc5}), substitution into Eqs.~\eqref{u_2} and \eqref{N} leads to the following relation:
\be
 \int_0^\infty M_3 \cos(\zeta X_3) d\zeta=0, \quad |X_3|<a,\, |X_3|>b,
 \label{eq24}
 \ee
 \be
 \begin{aligned}
 \int_0^\infty &\left\{  \frac{\xi_1}{\alpha_1} M_1(\zeta)\sinh(\alpha_1 \zeta D)+\frac{\xi_2}{\alpha_2} M_2(\zeta)\sinh(\alpha_2 \zeta D) + \frac{\xi_1}{\alpha_1} M_3(\zeta)\cosh(\alpha_1 \zeta D)\right. \\
 & \left.+\frac{\xi_2}{\alpha_2} M_4(\zeta)\cosh(\alpha_2 \zeta D)\right\}\cos(\zeta X_3) d \zeta=0, \quad  |X_3|<a,\, |X_3|>b.
 \label{eq25}
  \end{aligned}
 \ee
To find the arbitrary functions appearing in Eqs. (\ref{eq24}) and (\ref{eq25}), following relation are assumed:
\be
M_3(\zeta)= \frac{1}{\zeta} \int_a^b \Phi_1(s^2) \sin(\zeta s) ds,
\label{eq26}
\ee
\be
\begin{aligned}
&\frac{\xi_1}{\alpha_1} M_1(\zeta)\sinh(\alpha_1 \zeta D)+\frac{\xi_2}{\alpha_2} M_2(\zeta)\sinh(\alpha_2 \zeta D) + \frac{\xi_1}{\alpha_1} M_3(\zeta)\cosh(\alpha_1 \zeta D) \\
 &+\frac{\xi_2}{\alpha_2} M_4(\zeta)\cosh(\alpha_2 \zeta D)= \frac{1}{\zeta} \int_a^b \Phi_2(s^2) \sin(\zeta s) ds,
 \label{eq27}
 \end{aligned}
 \ee
 where \(\Phi_1(s^2)\) and \(\Phi_2(s^2)\) denote the unknown functions. To satisfy the conditions given in Eqs.~\eqref{eq24} and \eqref{eq25}, these functions must obey the following integral constraints:
\begin{equation}
\int_a^b \Phi_1(s^2) \, ds = 0,\qquad \int_a^b \Phi_2(s^2) \, ds = 0.
\label{eq28}
\end{equation}
Equations (\ref{eq26}) and (\ref{eq27}) identically satisfy the integral property:
\[
\int_0^\infty \frac{\sin(\zeta s)\,\cos(\zeta X_3)}{\zeta}\,d\zeta =
\begin{cases}
\frac{\pi}{2}, & s > X_3, \\[1ex]
0, & s < X_3.
\end{cases}
\]
Upon solving Eq.~(\ref{eq27}) together with the equation obtained by substituting the boundary condition (\ref{bc3}) into Eq.~(\ref{eq21}), the following relations are derived:
\be
M_1(\zeta)=-\frac{[1+ \Delta_1(\zeta)]}{\zeta} \int_a^b \Phi_1(s^2) \sin(\zeta s) ds + \frac{\Delta_2(\zeta)}{\zeta} \int_a^b \Phi_2(s^2) \sin(\zeta s) ds,
\label{eq29}
\ee
\be
M_2(\zeta)= \frac{\Omega_1}{\Omega_2}\frac{[1+\Delta_3(\zeta)]}{\zeta}  \int_a^b \Phi_1(s^2) \sin(\zeta s) ds - \frac{\Delta_4(\zeta)}{\zeta} \int_a^b \Phi_2(s^2) \sin(\zeta s) ds ,
\label{eq30}
\ee
 where,
\be
\Delta_1(\zeta)= \frac{e^{-\alpha_1 \zeta D}}{\sinh({\alpha_1 \zeta D)}}, \quad \Delta_2(\zeta)
= \frac{\Psi_1}{\Psi_2 \sinh({\alpha_1 \zeta D)}}, \quad
\Delta_3(\zeta)= \frac{e^{-\alpha_2 \zeta D}}{\sinh({\alpha_2 \zeta D)}}, \quad \Delta_4(\zeta)=\frac{\Omega_1}{\Omega_2}\frac{\Psi_1}{\Psi_2 \sinh({\alpha_2 \zeta D)}},
\ee
\be
\Psi_1=\frac{\Omega_2}{\Psi^*}, \quad \Psi_2= \frac{1}{\Psi^*}\left[\frac{\xi_1 \Omega_2}{\alpha_1}-\frac{\xi_2 \Omega_1}{\alpha_2} \right], \quad \Psi^*= \Omega_1 \gamma_2-\Omega_2 \gamma_1, \quad \gamma_i= \mu_{23}+ \xi_i \:\mu_{22}; \quad i=1,2.
\ee
Imposing the boundary conditions (\ref{bc4}) and (\ref{bc8}) into Eq.~(\ref{eq22}) yields the following relation:
\be
\int_0^\infty [\gamma_1 M_1+\gamma_2 M_2] \:\zeta  \cos(\zeta X_3) \:d \zeta+\frac{\mu_{24}}{\mu_{44}} \int_0^\infty [\Omega_1 M_3+\Omega_2 M_4] \: \zeta  \sin(\zeta X_3) \: d \zeta=-\sigma_1(X_3),  \quad a \le |X_3| \le b, 
\label{eq33}
\ee
\be
\begin{aligned}
\int_0^\infty [\gamma_1 M_1 \cosh(\alpha_1 \zeta D)+\gamma_2 M_2 \cosh(\alpha_2 \zeta D)+\gamma_1 M_3 \sinh(\alpha_1 \zeta D) 
+\gamma_2 M_4 \sinh(\alpha_2 \zeta D)] \: \zeta \: \cos(\zeta X_3) \: d \zeta \\ 
+\frac{\mu_{24}}{\mu_{44}} \int_0^\infty [\Omega_1 M_1 \sinh(\alpha_1 \zeta D)+\Omega_2 M_2 \sinh(\alpha_2 \zeta D) +\Omega_1 M_3 \cosh(\alpha_1 \zeta D)  +\Omega_2 M_4 \cosh(\alpha_2 \zeta D)] \:\zeta \sin(\zeta  X_3) d \zeta\\
=-\sigma_2(X_3);  \quad a \le |X_3| \le b.
\end{aligned}
\label{eq34}
\ee
Substituting the values of $M_i \:(i=1,2,3,4)$ from Eqs.~(\ref{relation_M4M3}), (\ref{eq26}), (\ref{eq29}), and (\ref{eq30}) into Eqs.~(\ref{eq33}) and (\ref{eq34}) yields the following set of double integral equations:
\be
\int_a^b \frac{s  \: \Phi_1(s^2)}{s^2-X_3^2}ds+\frac{1}{2} \int_a^b \Upsilon_{11}(X_3,s) \Phi_1(s^2) ds+ \frac{1}{2} \int_a^b \Upsilon_{12}(X_3,s) \Phi_2(s^2) ds= -\Psi_1 \sigma_1(X_3), \quad a \le |X_3| \le b,
\label{eq35}
\ee
\be
\int_a^b \frac{s \: \Phi_2(s^2)}{s^2-X_3^2}ds+\frac{1}{2} \int_a^b \Upsilon_{21}(X_3,s) \Phi_1(s^2) ds+ \frac{1}{2} \int_a^b \Upsilon_{22}(X_3,s) \Phi_2(s^2) ds= \Psi_2 \sigma_2(X_3), \quad a \le |X_3| \le b,
\label{eq36}
\ee
where,
\be
\Upsilon_{ij}(X_3,s)= \int_0^\infty \varphi_{ij}(\zeta) \: [\sin\zeta(s+X_3)+\sin \zeta (s-X_3)] \:d \zeta; \quad  i,j = 1,2 
\ee
\begin{equation*}
    \varphi_{11}(\zeta)=\varphi_{22}(\zeta)= \frac{1}{\Psi^*}[\Omega_1 \gamma_2 \Delta_3(\zeta)-\Omega_2 \gamma_1 \Delta_1(\zeta)], \quad 
\end{equation*}
\begin{equation*}
    \varphi_{12}(\zeta)= \frac{\Omega_2}{\Psi^*}[\gamma_1 \Delta_2(\zeta)-\gamma_2 \Delta_4(\zeta)], \quad  \varphi_{21}(\zeta)= \Psi_2 \left[ \Delta_5(\zeta) -\frac{\Omega_1}{\Omega_2} \Delta_6(\zeta)\right],
\end{equation*}
\begin{equation*}
   \Delta_5(\zeta)= \gamma_1 [e^{-\alpha_1 \zeta D} + \Delta_1(\zeta) \cosh (\alpha_1 \zeta D)], \quad \Delta_6(\zeta)= \gamma_2 [e^{-\alpha_2 \zeta D} + \Delta_3(\zeta) \cosh (\alpha_2 \zeta D)].
\end{equation*}
In order to estimate $\Upsilon_{ij}(x_1,s)$, the first order approximation of $\Delta_i(\zeta)$ $(i=1,2,3,...,6)$ for large $d$ has been considered and thus:
\begin{equation}
\begin{aligned}
\Upsilon_{11}(X_3,s) = \Upsilon_{22}(X_3,s) &= -\frac{2\Omega_2 \gamma_1}{\Psi^*} \sum_{n=1}^{\infty} \left[ \frac{s+X_3}{(2n)^2 \alpha_1^2 D^2 + (s+X_3)^2} + \frac{s-X_3}{(2n)^2 \alpha_1^2 D^2 + (s-X_3)^2} \right] \\
&\quad + \frac{2\Omega_1 \gamma_2}{\Psi^*} \sum_{n=1}^{\infty} \left[ \frac{s+X_3}{(2n)^2 \alpha_2^2 D^2 + (s+X_3)^2} + \frac{s-X_3}{(2n)^2 \alpha_2^2 D^2 + (s-X_3)^2} \right]
\end{aligned}
\label{eq38}
\end{equation}

\begin{equation}
\begin{aligned}
\Upsilon_{12}(X_3,s) &= \frac{2 \Omega_2^2 \gamma_1}{\Psi_2 \Psi^{*2}} 
\sum_{n=1}^{\infty} \left[ 
\frac{s+X_3}{(2n-1)^2 \alpha_1^2 D^2 + (s+X_3)^2} + 
\frac{s-X_3}{(2n-1)^2 \alpha_1^2 D^2 + (s-X_3)^2} 
\right] \\
&\quad - \frac{2 \Omega_2 \Omega_1 \gamma_2}{\Psi_2 \Psi^{*2}} 
\sum_{n=1}^{\infty} \left[ 
\frac{s+X_3}{(2n-1)^2 \alpha_2^2 D^2 + (s+X_3)^2} + 
\frac{s-X_3}{(2n-1)^2 \alpha_2^2 D^2 + (s-X_3)^2} 
\right]
\end{aligned}
\label{eq39}
\end{equation}
\begin{equation}
\begin{aligned}
\Upsilon_{21}(X_3,s) &= 2 \Psi_2 \gamma_1 
\sum_{n=1}^{\infty} \left[ 
\frac{s+X_3}{(2n-1)^2 \alpha_1^2 D^2 + (s+X_3)^2} + 
\frac{s-X_3}{(2n-1)^2 \alpha_1^2 D^2 + (s-X_3)^2} 
\right] \\
&\quad - \frac{2 \Psi_2 \Omega_1 \gamma_2}{\Omega_2} 
\sum_{n=1}^{\infty} \left[ 
\frac{s+X_3}{(2n-1)^2 \alpha_2^2 D^2 + (s+X_3)^2} + 
\frac{s-X_3}{(2n-1)^2 \alpha_2^2 D^2 + (s-X_3)^2} 
\right]
\end{aligned}
\label{eq40}
\end{equation}
Expanding $\Upsilon_{ij}$ $(i,j=1,2)$ in powers of $1/D$ (for large D) and using the series expansion: 
\[
\frac{\pi^2}{8} = \sum_{n=1}^\infty \frac{1}{(2n-1)^2}, \quad \frac{\pi^2}{12} = \sum_{n=1}^\infty \frac{(-1)^{n+1}}{n^2}, \quad \text{and} \quad \frac{\pi^2}{6} = \sum_{n=1}^\infty \frac{1}{n^2},
\]
Eqs. (\ref{eq38})-\ref{eq40} reduces to:
\be
\Upsilon_{11}(X_3,s)= \Upsilon_{22}(X_3,s)= -\frac{\Psi_3 \pi^2}{6 D^2}s, \quad \Upsilon_{12}(X_3,s)= \frac{\Omega_2 \Psi_3 \pi^2}{2 \Psi^* \Psi_2 D^2}s, \quad \Upsilon_{21}(X_3,s)= \frac{\Psi_2 \Psi_3 \Psi^* \pi^2}{2 \Omega_2 D^2}s,
\label{eq41}
\ee
where $\Psi_3=\frac{1}{\Psi^*}\left( \frac{\Omega_2 \gamma_1}{\alpha_1^2}-\frac{\Omega_1 \gamma_2}{\alpha_2^2} \right).$
The asymptotic expansion of $ \Phi_i(s^2)$ is considered in the form as:
\be
\Phi_i(s^2)= \Phi_i^{(0)}(s^2)+\frac{1}{D^2} \Phi_i^{(1)}(s^2)+ O\left( \frac{1}{D^4}\right).
\label{eq42}
\ee
With use of Eqs. (\ref{eq41}) and (\ref{eq42}) and comparing the coefficient of constant terms and terms containing $1/d^2$ from Eqs. (\ref{eq35}) and (\ref{eq36}), it leads to the following integral equations:
\begin{align}
\int_a^b \frac{2 s \: \Phi_1^{(0)}(s^2)}{s^2-X_3^2}ds=&-2 \Psi_1 \sigma_1 (X_3),
\label{eq43} \\
\int_a^b \frac{2 s \: \Phi_2^{(0)}(s^2)}{s^2-X_3^2}ds=&\:2 \Psi_2 \sigma_2 (X_3),
\label{eq44} \\
\int_a^b \frac{2 s \: \Phi_1^{(1)}(s^2)}{s^2-X_3^2}ds=&\frac{\pi^2 \Psi_3}{6}\int_a^b s \Phi_1^{(0)}(s^2) ds- \frac{\pi^2 \Omega_2 \Psi_3}{2 \Psi_2 \Psi^* }\int_a^b s \Phi_2^{(0)}(s^2) ds,
\label{eq45} \\
\int_a^b \frac{2 s \: \Phi_2^{(1)}(s^2)}{s^2-X_3^2}ds=&\frac{\pi^2 \Psi_3}{6}\int_a^b s \Phi_2^{(0)}(s^2) ds- \frac{\pi^2 \Psi_2\Psi_3\Psi^*}{2 \Omega_2 }\int_a^b s \Phi_1^{(0)}(s^2) ds.
\label{eq46}
\end{align}
Consider the edge of moving collinear cracks subjected to a force of intensity \( \sigma_1 \), which remains constant within the crack front. Additionally, a normal punched pressure load \( \sigma_2(X_3) \) is assumed to be constant and equal to \( \sigma_2 \). By applying the Hilbert transformation \cite{srivastava1970xx}, the expressions for \( \Phi_1(s^2) \) and \( \Phi_2(s^2) \) are derived as follows:
\begin{align}
\Phi_1(s^2) &= \frac{-2\Psi_1 \sigma_1}{\pi} \sqrt{\frac{s^2-a^2}{b^2-s^2}} + \frac{C_1}{\sqrt{(s^2-a^2)(b^2-s^2)}} 
 + \frac{1}{D^2} \left[\frac{G_1^*}{\pi} \sqrt{\frac{s^2-a^2}{b^2-s^2}} 
+ \frac{C_3}{\sqrt{(s^2-a^2)(b^2-s^2)}} \right],  
\label{Phi1} \\
\Phi_2(s^2) &= \frac{2 \Psi_2 \sigma_2}{\pi} \sqrt{\frac{s^2-a^2}{b^2-s^2}} 
+ \frac{C_2}{\sqrt{(s^2-a^2)(b^2-s^2)}}+ \frac{1}{D^2} \left[\frac{G_2^*}{\pi} \sqrt{\frac{s^2-a^2}{b^2-s^2}} 
+ \frac{C_4}{\sqrt{(s^2-a^2)(b^2-s^2)}} \right],
\label{Phi2}
\end{align}
where,
\begin{equation}
\begin{aligned}
    G_1^*=&\frac{\pi^3 \Psi_3}{12}\left[- \frac{\Psi_1 \sigma_1}{\pi}(b^2-a^2)+C_1\right]-\frac{\pi^3\Omega_2 \Psi_3}{4\Psi_2\Psi^*} \left[ \frac{\Psi_2 \sigma_2}{\pi}(b^2-a^2)+C_2\right],\\
    G_2^*=&\frac{\pi^3 \Psi_3}{12}\left[ \frac{\Psi_2 \sigma_2}{\pi}(b^2-a^2)+C_2\right]-\frac{\pi^3\Psi_3\Psi_2\Psi^*}{4\Psi_2}\left[ \frac{-\Psi_1 \sigma_1}{\pi}(b^2-a^2) +C_1\right],
    \end{aligned}
\end{equation}
 and the constants \( C_i \) are to be determined to satisfy the integral constraint (\ref{eq28}).
\subsection{Stress intensity factor and crack opening displacement}
\label{Stress intensity factor and crack opening displacement}
The expression for the stress intensity factor for the propagation of a Mode-I crack in a crystalline monoclinic strip at the ends of the crack is given by:  
\begin{align}
K_I^{(a)} &= \lim_{X_3 \to a^-} \sqrt{2(a - X_3)} \, \tau_{22}(X_3,0),   \qquad 0 < X_3 < a, \label{SIF1}\\ 
K_I^{(b)} &= \lim_{X_3 \to b^+} \sqrt{2(X_3 - b)} \, \tau_{22}(X_3,0),   \qquad X_3 > b. \label{SIF2}
\end{align}
Substituting the values from Eqs.~(\ref{eq22}), (\ref{relation_M4M3}), (\ref{eq26}), (\ref{eq29}), (\ref{eq30}), (\ref{Phi1}), and (\ref{Phi2}) into Eqs.~(\ref{SIF1}) and (\ref{SIF2}), the final expression for the stress intensity factor at the ends of the crack, i.e., at \( X_3 = a \) and \( X_3 = b \), is given by:
\begin{align}
K_I^{(a)}&=\frac{\pi}{2 \Psi_1}\frac{1}{\sqrt{a}\sqrt{b^2-a^2}}\left[ C_1+\frac{C_3}{D^2}\right], \label{eq53}\\
K_I^{(b)}&= \frac{\pi}{2\Psi_1}\frac{1}{\sqrt{b}\sqrt{b^2-a^2}} \left\{ \frac{-2\Psi_1 \sigma_1}{\pi} (b^2-a^2)+\frac{G_1^* (b^2-a^2)}{\pi D^2}-\left[C_1+\frac{C_3}{D^2}\right]\right\}.
\label{eq54}
\end{align}
The crack opening displacement (COD) for Mode I crack propagation, is given by
\be
\Delta u_2(X_3,0)= u_2(X_3, 0^+)-u_2(X_3,0^-)=\pi \left[ \frac{\xi_1}{\alpha_1}-\frac{\xi_2 \Omega_1}{\alpha_2 \Omega_2}\right] \int_{X_3}^b \Phi_1(s^2)ds.
\label{COD1}
\ee
By incorporating the expression for \(\Phi_1(s^2)\) from Eq. (\ref{Phi1}) into Eq. (\ref{COD1}), the crack opening displacement (COD) for collinear Griffith cracks in a monoclinic crystalline strip can be expressed as:
\be
\begin{aligned}
\Delta u_2(X_3,0)&=\pi \left[ \frac{\xi_1}{\alpha_1}-\frac{\xi_2 \Omega_1}{\alpha_2 \Omega_2}\right] 
\left\{ \left[\frac{-2\psi_1\sigma_1}{\pi} + \frac{G_1^*}{\pi D^2}\right] [b E(\psi,k)-\frac{a^2}{b}F(\psi,k)]
+ \frac{1}{b} \left[C_1+\frac{C_3}{D^2} \right] F(\psi,\,k) \right\}
\label{eq55}
\end{aligned}
\ee
where \( F(\psi, k) \) and $E(\psi,k)$ are the incomplete elliptic integral of the first kind and second kind \cite{byrd2013handbook}, and  
\[
\psi = \arcsin \sqrt{\frac{b^2 - X_3^2}{b^2 - a^2}}, \quad k = \sqrt{\frac{b^2 - a^2}{b^2}}.
\]

Equations~(\ref{eq53}), (\ref{eq54}), and~(\ref{eq55}) present the final expressions for the stress intensity factors at both crack tips and the crack opening displacement for Mode I crack propagation. These expressions form the key outcome of the study, providing crucial insights into the behavior of collinear cracks in a monoclinic strip subjected to various loading conditions. In the next subsection, the obtained results are verified against well-established studies from the literature. Additionally, the case where the strip is subjected to a point load, instead of a constant normal pressure, is considered to investigate the influence of different loading conditions.

\subsection{Special cases}
\label{Special cases}
\subsubsection{When the edge of moving collinear cracks is subjected to point loading}
\label{When the edge of moving collinear cracks is subjected to point loading}
Let us consider the edge of the moving collinear cracks is loaded at $X_3=\pm X_0$ by a force of intensity $\sigma_1$, which is constant load inside the crack front, therefore the case of point loading is assumed as:
\be
\sigma_1(X_3)=\sigma_1^* \: \delta(X_3- X_0)
\label{eq56}
\ee
where \( \delta(X_3 - X_0) \) represents the Dirac delta function. Applying the Hilbert transformation \cite{srivastava1970xx}, the expressions for \( \Phi_1(s^2) \) and \( \Phi_2(s^2) \) are obtained as follows:
\begin{align}
\Phi_1(s^2) &= \frac{4\Psi_1 \sigma_1^*}{\pi^2} \sqrt{\frac{s^2-a^2}{b^2-s^2}} \sqrt{\frac{b^2-X_0^2}{X_0^2-a^2}} \frac{X_0}{X_0^2-s^2} 
+ \frac{C_1^*}{\sqrt{(s^2-a^2)(b^2-s^2)}} \notag \\  
&\quad + \frac{1}{D^2} \left[\frac{G_3^*}{\pi} \sqrt{\frac{s^2-a^2}{b^2-s^2}} 
+ \frac{C_3^*}{\sqrt{(s^2-a^2)(b^2-s^2)}} \right],  
\label{eq57} \\
\Phi_2(s^2) &= \frac{2 \Psi_2 \sigma_2}{\pi} \sqrt{\frac{s^2-a^2}{b^2-s^2}} 
+ \frac{C_2^*}{\sqrt{(s^2-a^2)(b^2-s^2)}} \notag \\  
&\quad + \frac{1}{D^2} \left[\frac{G_4^*}{\pi} \sqrt{\frac{s^2-a^2}{b^2-s^2}} 
+ \frac{C_4^*}{\sqrt{(s^2-a^2)(b^2-s^2)}} \right],
\label{eq58}
\end{align}
where,
\begin{equation}
\begin{aligned}
    G_3^*=&\frac{\pi^3 \Psi_3}{12}\left[- \frac{4 \Psi_1 \sigma_1^*}{\pi^2}\sqrt{\frac{b^2-X_0^2}{X_0^2-a^2} }X_0+C_1^*\right]-\frac{\pi^3\Omega_2 \Psi_3}{4\Psi_2\Psi^*} \left[ \frac{\Psi_2 \sigma_2}{\pi}(b^2-a^2)+C_2^*\right],\\
    G_4^*=&\frac{\pi^3 \Psi_3}{12}\left[ \frac{\Psi_2 \sigma_2}{\pi}(b^2-a^2)+C_2^*\right]-\frac{\pi^3\Psi_3\Psi_2\Psi^*}{4\Psi_2}\left[ \frac{-4\Psi_1 \sigma_1^*}{\pi^2}\sqrt{\frac{b^2-X_0^2}{X_0^2-a^2} }X_0+C_1^*\right].
    \end{aligned}
\end{equation}
The constants \( C_i^* \) are determined from Eq.~(\ref{eq28}). Consequently, the stress intensity factor for point loading in the propagation of a Mode-I crack within a crystalline monoclinic strip, evaluated at the crack tips, is expressed as:
 \begin{align}
K_I^{(a)*} &= \lim_{X_3 \to a^-} \sqrt{2(a - X_3)} \, \tau_{22}(X_3,0),   \qquad 0 < X_3 < a, \label{eq60}\\ 
K_I^{(b)*} &= \lim_{X_3 \to b^+} \sqrt{2(X_3 - b)} \, \tau_{22}(X_3,0),   \qquad X_3 > b. \label{eq61}
\end{align}
By substituting the values from Eqs.~(\ref{eq22}), (\ref{relation_M4M3}), (\ref{eq26}), (\ref{eq29}), (\ref{eq30}), (\ref{eq57}), and (\ref{eq58}) into Eqs.~(\ref{eq60}) and (\ref{eq61}), the explicit expressions for the stress intensity factor at the crack tips are obtained as:
\begin{align}
K_I^{(a)*}&=\frac{\pi}{2 \Psi_1}\frac{1}{\sqrt{a}\sqrt{b^2-a^2}}\left[ C_1^*+\frac{C_3^*}{D^2}\right], \label{eq62}\\
K_I^{(b)*}&= \frac{\pi}{2\Psi_1}\frac{1}{\sqrt{b}\sqrt{b^2-a^2}} \left\{ \frac{4\Psi_1 \sigma_1^*}{\pi^2} \frac{X_0 (b^2-a^2)}{\sqrt{X_0^2-a^2}\sqrt{b^2-X_0^2}}+\frac{G_3^* (b^2-a^2)}{\pi D^2}-\left[C_1^*+\frac{C_3^*}{D^2}\right]\right\}.
\label{eq63}
\end{align}
The COD for Mode-I crack propagation is given by:
\be
\Delta u_2^*(X_3,0)= u_2(X_3, 0^+)-u_2(X_3,0^-)=\pi \left[ \frac{\xi_1}{\alpha_1}-\frac{\xi_2 \Omega_1}{\alpha_2 \Omega_2}\right] \int_{X_3}^b \Phi_1(s^2)ds.
\label{eq64}
\ee
Substituting the expression for \(\Phi_1(s^2)\) from Eq.~(\ref{eq57}) into Eq.~(\ref{eq64}), the COD for collinear Griffith cracks in a monoclinic crystalline strip is obtained as:
\be
\begin{aligned}
\Delta u_2^*(X_3,0)&=\pi \left[ \frac{\xi_1}{\alpha_1}-\frac{\xi_2 \Omega_1}{\alpha_2 \Omega_2}\right] 
\left\{ \frac{4\psi_1\sigma_1^*}{\pi^2} \sqrt{\frac{b^2-X_0^2}{X_0^2-a^2}} X_0 
\int_{X_3}^b \frac{\sqrt{s^2-a^2}}{\sqrt{b^2-s^2} (X_0^2-s^2)} ds \right.\\
&\quad + \left. \frac{G_3^*}{\pi D^2} [b E(\psi,k)-\frac{a^2}{b}F(\psi,k)]
+ \frac{1}{b} \left[C_1^*+\frac{C_3^*}{D^2} \right] F(\psi,\,k) \right\}
\label{eq65}
\end{aligned}
\ee
\subsubsection{When the strip is considered to be an isotropic layer and in the absence of initial stress}
\label{When the strip is considered to be an isotropic layer and in the absence of initial stress}
Considering the given strip to be isotropic and assuming that no initial stress is present along the \(X_2\) and \(X_3\) directions, the material constants are defined as follows:   
\begin{equation}  
\mu_{44} = \mu, \quad \mu_{22} = \mu_{33} = \lambda + 2\mu, \quad \mu_{23} = \lambda, \quad \mu_{24} = \mu_{34} = 0.  
\end{equation}  
Here, \(\mu\) and \(\lambda\) represent the Lamé constants, which characterize the elastic properties of the isotropic medium. The absence of initial stress is expressed as:  
\begin{equation}  
\tau_{22}^0 = \tau_{33}^0 = 0,  
\end{equation}  
indicating that no pre-existing stress is acting within the material.  
Based on these assumptions, the expression for the stress intensity factor as given in Eqs. (\ref{eq53}) and (\ref{eq54}) at the crack tips,  is transformed as:
\begin{align}
K_I^{(a)}&=\frac{\pi}{2 \bar\Psi_1}\frac{1}{\sqrt{a}\sqrt{b^2-a^2}}\left[ C_1+\frac{C_3}{D^2}\right], \label{eq68}\\
K_I^{(b)}&= \frac{\pi}{2\bar\Psi_1}\frac{1}{\sqrt{b}\sqrt{b^2-a^2}} \left\{ \frac{2 \bar\Psi_1 \sigma_1}{\pi} (b^2-a^2)+\frac{\overline G_1^* (b^2-a^2)}{\pi D^2}-\left[C_1+\frac{C_3}{D^2}\right]\right\}.
\label{eq69}
\end{align}
where,
\begin{equation}
\begin{aligned}
\overline G_1^*=&\frac{\pi^3 \bar\Psi_3}{12}\left[- \frac{\bar\Psi_1 \sigma_1}{\pi}(b^2-a^2)+C_1\right]-\frac{\pi^3 \bar\Omega_2 \bar\Psi_3}{4\bar\Psi_2 \bar\Psi^*} \left[ \frac{\bar\Psi_2 \sigma_2}{\pi}(b^2-a^2)+C_2\right], \notag
    \end{aligned}
\end{equation}
with
\be
\begin{aligned}
&\bar\Psi_3=\frac{1}{\bar\Psi^*}\left( \frac{\bar\Omega_2 \bar\gamma_1}{\alpha_1^2}-\frac{\bar\Omega_1 \bar\gamma_2}{\alpha_2^2} \right), \quad
\bar \Psi_1=\frac{\bar \Omega_2}{\bar\Psi^*}, \quad \bar\Psi_2= \frac{1}{\bar\Psi^*}\left[\frac{\bar\xi_1 \bar\Omega_2}{\alpha_1}-\frac{\bar\xi_2 \bar\Omega_1}{\alpha_2} \right], \quad \bar\Psi^*= \bar\Omega_1 \bar\gamma_2-\bar\Omega_2 \bar\gamma_1, \\
&\bar\gamma_1= \lambda + \bar\xi_1(\lambda+2 \mu),
\quad \bar\gamma_2= \lambda + \bar\xi_2(\lambda+2 \mu),\quad  \bar\Omega_1=\mu\left(\alpha_1-\frac{\bar\xi_1}{\alpha_1}\right),
\quad \bar\Omega_2=\mu\left(\alpha_2-\frac{\bar\xi_2}{\alpha_2}\right), \\
& \bar\xi_1 = \frac{\alpha_1^2 - \left( \frac{\lambda +2 \mu}{\mu} - \frac{v^2}{v_c^2} \right)}
{\frac{\lambda+\mu}{\mu}}, \quad \bar\xi_2 = \frac{\alpha_2^2 - \left( \frac{\lambda +2 \mu}{\mu} - \frac{v^2}{v_c^2} \right)}
{\frac{\lambda+\mu}{\mu}}.
\notag
\end{aligned}
\ee
Eqs. (\ref{eq68}) and (\ref{eq69}) represent the closed-form expressions for the stress intensity factor corresponding to the propagation of a mode-I collinear crack in a homogeneous isotropic strip. These expressions match the special case of the well-known prior result given by Das \textit{et al.} \cite{das1992four} when the infinite elastic medium, which is isotropic, consists of two collinear cracks, thereby verifying the present study.

In the next section, numerical simulations and graphical visualizations of the stress intensity factor and crack opening displacement are presented to examine the effects of various material parameters and dynamic loading on these quantities.
\section{Numerical simulation and graphical analysis}
\label{Numerical simulation and graphical analysis}
Numerical simulations are conducted in MATLAB to analyze the variation in the stress intensity factor and crack opening displacement, as obtained in Eqs. (\ref{eq53}), (\ref{eq54}), and (\ref{eq55}), under different material and loading conditions. The study considers two monoclinic crystalline materials—Lithium Niobate (LiNbO$_3$) and Lithium Tantalate (LiTaO$_3$)—along with an isotropic material for comparison. The material constants used in the analysis are listed in Table \ref{table:material_constants}. Figures \ref{Figure 2}–\ref{Figure 4} present the graphical results illustrating the influence of parameters such as crack velocity, position, length, applied pressure, and initial stress (compressive and tensile) on stress intensity factor and crack opening displacement. In these figures, the bold line $( \rule[0.5ex]{1em}{1pt} )$ represents Lithium Niobate, the dotted line $(\ldots)$ corresponds to Lithium Tantalate, and the dashed line $(---)$ denotes the isotropic material. Figures \ref{Figure 7} and \ref{Figure 8} depict a special case of the study where the punch pressure is modeled as a point load instead of a uniform normal pressure. The outcomes are compared accordingly, with the marker '$-->$' representing the uniform normal pressure and '$--\circ$' indicating the point loading.
\begin{table}[h]
    \centering
    \caption{Material constants for the considered monoclinic crystalline and isotropic materials \cite{kumar2021analytical}.}
    \renewcommand{\arraystretch}{1.2}
    \begin{tabular}{|c|c|c|c|c|}
        \hline
        \textbf{Matrial constant} & \textbf{Unit} & \textbf{Lithium Niobate} & \textbf{Lithium Tantalate} & \textbf{Isotropic Material} \\
        \hline
        $\mu_{22}$  & N/m$^2$  & $20.3 \times 10^{10}$  & $23.3 \times 10^{10}$  & -  \\
        $\mu_{24}$  & N/m$^2$  & $-0.9 \times 10^{10}$  & $1.1 \times 10^{10}$   & -  \\
        $\mu_{23}$  & N/m$^2$  & $7.5 \times 10^{10}$   & $8.0 \times 10^{10}$   & -  \\
        $\mu_{44}$  & N/m$^2$  & $6.0 \times 10^{10}$   & $9.4 \times 10^{10}$   & -  \\
        $\mu_{33}$  & N/m$^2$  & $24.5 \times 10^{10}$  & $27.5 \times 10^{10}$  & -  \\
        $\mu$       & N/m$^2$  & -                      & -                      & $19.87 \times 10^9$ \\
        $\lambda$   & N/m$^2$  & -                      & -                      & $25.1 \times 10^9$  \\
        $\rho$      & kg/m$^3$ & 4700                   & 7450                   & 4705  \\
        \hline
    \end{tabular}
    \label{table:material_constants}
\end{table}
\subsection{Effect of crack velocity on stress intensity factor for various material parameters}
Figures~\ref{Figure 2} and \ref{Figure 3} illustrate the variation of the normalized stress intensity factor (SIF), $K_I^{(a)}/\sigma_1 \sqrt{b}$ and $K_I^{(b)}/\sigma_1 \sqrt{b}$, with respect to the normalized crack velocity $v/v_c$ at the left and right crack tips, respectively. It is evident that the SIF increases with crack velocity, which can be attributed to the enhanced dynamic effects during rapid crack propagation. These effects limit the redistribution of stress, thereby increasing the stress concentration at the crack tips.
The influence of crack tip ratio, represented by the ratio $a/b$, is shown in Figs.~\ref{Fig_1} and \ref{Fig_7}. As $a/b$ increases, the crack becomes more symmetric about the center, leading to a more uniform stress distribution along the crack faces in an infinite elastic medium. This symmetry reduces the stress concentration near the tips and, in turn, lowers the SIF, indicating improved resistance to crack propagation.
Figures \ref{Fig_2} and \ref{Fig_8} highlight the effect of applied normal pressure ($\sigma_2/\sigma_1$). An increase in this pressure intensifies the normal loading on the crack surfaces, leading to higher stress accumulation at the crack tips and a corresponding rise in SIF at both ends. This suggests that elevated normal stress facilitates crack growth.
The impact of vertical initial stress ($\tau_{22}$) is presented in Figs.~\ref{Fig_3}, \ref{Fig_4}, \ref{Fig_9}, and \ref{Fig_10}. Compressive vertical stress (Figs.~\ref{Fig_3} and \ref{Fig_9}) tends to close the crack faces, which enhances local stress near the crack tips and increases the SIF. In contrast, tensile vertical stress (Figs.~\ref{Fig_4} and \ref{Fig_10}) promotes crack opening, resulting in more evenly distributed stress and a reduction in SIF at both tips.
The effect of horizontal initial stress ($\tau_{33}$) is examined in Figs.~\ref{Fig_5}, \ref{Fig_6}, \ref{Fig_11}, and \ref{Fig__12}. Under compressive horizontal stress (Figs.~\ref{Fig_5} and \ref{Fig_11}), the lateral constraint on the material inhibits crack opening and lowers the stress intensity near the tips. On the other hand, tensile horizontal stress (Figs.~\ref{Fig_6} and \ref{Fig__12}) increases the tendency of the crack to open, which amplifies the localized stress at the crack tips and results in a higher SIF.
Among the materials considered, the isotropic material exhibits the most significant variation in the SIF with increasing crack velocity across all figures, indicating a heightened sensitivity to both external and internal loading conditions. This pronounced sensitivity implies a greater susceptibility to damage. In contrast, Lithium Niobate and Lithium Tantalate show relatively moderate changes in SIF, reflecting more stable crack-tip behavior and improved durability under varying mechanical conditions.

\subsection{Effect of crack position on stress intensity factor for various material parameters}
Figures~\ref{Figure 5} and \ref{Figure 6} illustrate the variation of the normalized stress intensity factor (SIF), $K_I^{(a)}/\sigma_1 \sqrt{b}$ and $K_I^{(b)}/\sigma_1 \sqrt{b}$, with respect to the crack tip ratio $a/b$ at the left and right tips, respectively, while accounting for the influence of varying internal and external material parameters. It is observed that as the ratio $a/b$ increases, the SIF at both tips decreases, indicating a more stable crack configuration.
Figures~\ref{Fig_18} and \ref{Fig_24} depict the variation in SIF due to increasing crack velocity in increments of 0.1. The results show a significant rise in SIF with increasing crack velocity across all materials, with the effect being most pronounced when the velocity reaches $0.8$. Furthermore, Figs. \ref{Fig_19} and \ref{Fig_25} present the influence of increasing applied normal pressure $\sigma_2/\sigma_1$ at the boundary. A positive correlation is observed, where higher applied pressure enhances the SIF, promoting crack propagation.
The effect of vertical initial stress $\tau_{22}$ is shown in Figures \ref{Fig_20}, \ref{Fig_21}, \ref{Fig_26}, and \ref{Fig_27}. Compressive vertical initial stress (Figs.~\ref{Fig_20} and \ref{Fig_26}) contributes to an increase in SIF at both crack tips, due to the induced stress concentration. In contrast, tensile vertical initial stress (Figs.~\ref{Fig_21} and \ref{Fig_27}) facilitates crack opening and leads to a decrease in the SIF at both tips.
Figures~\ref{Fig_22}, \ref{Fig_23}, \ref{Fig_28}, and \ref{Fig_29} represent the effect of horizontal initial stress $\tau_{33}$. An increase in compressive horizontal stress (Figs.~\ref{Fig_22} and \ref{Fig_28}) reduces the SIF due to lateral confinement, while an increase in tensile horizontal stress (Figs.~\ref{Fig_23} and \ref{Fig_29}) enhances crack opening and results in a higher SIF at both crack tips.

\subsection{Effect of material parameters on crack opening displacement}
Figure~\ref{Figure 4} illustrates the variation in crack opening displacement (COD) under the influence of different internal and external parameters for the considered materials. Figure~\ref{Fig_12} shows the COD response to variations in the crack tip ratio $a/b$ at the left crack tip. It is observed that increasing the ratio $a/b$ leads to a decrease in COD across all materials. This occurs because a higher $a/b$ ratio results in a more centrally located crack, promoting a more symmetric stress distribution which reduces the intensity of crack opening.
Figure~\ref{Fig_13} depicts the influence of the applied normal pressure $\sigma_2/\sigma_1$ at the boundary (i.e., punch pressure). As the pressure increases, the COD correspondingly rises. This is due to the fact that higher applied pressure generates greater normal stress along the crack faces, enhancing crack face separation and thus increasing the opening displacement.
The effect of crack velocity is shown in Figure~\ref{Fig_14}, where an increase in crack propagation speed results in a more pronounced crack opening. This behavior is attributed to dynamic effects: higher crack velocities reduce the time for stress redistribution, concentrating stress at the crack tip and promoting greater displacement.
Figure~\ref{Fig_15} investigates the impact of vertical initial compressive stress $\tau_{22}$ on COD. It is observed that increasing compressive vertical stress leads to an increase in COD. This can be explained by the fact that compressive stress perpendicular to the crack plane enhances the stress intensity near the crack tips due to constrained deformation around the crack region. This constraint results in outward deformation near the crack faces, which in turn increases the relative displacement between the crack surfaces.
 As a result, the crack opening becomes more prominent.
Figures~\ref{Fig_16} and \ref{Fig_17} explore the effect of horizontal initial stress $\tau_{33}$ on COD. When a compressive horizontal stress is applied, it acts parallel to the crack faces, constraining lateral expansion and reducing the opening displacement. In contrast, increasing tensile horizontal stress facilitates lateral stretching of the material, thereby promoting crack face separation and increasing COD.
Among the considered materials, Lithium Niobate exhibits the least variation in COD, indicating relatively stable crack-opening behavior under different loading conditions. This can be attributed to its strong piezoelectric coupling and mechanical stiffness. Conversely, Lithium Tantalate and the isotropic material demonstrate greater sensitivity to parameter variations.

To visualize these variations, a surface plot is provided in Fig. \ref{Figure 44}, illustrating the change in COD with respect to key parameters for Lithium Niobate (LiNbO$_3$). Specifically, Fig. \ref{Fig_122} examines the effect of crack position ($a/b$), Fig. \ref{Fig_133} investigates the influence of crack velocity ($v/v_c$), Fig. \ref{Fig_144} highlights the impact of applied punch pressure ($\sigma_2/\sigma_1$), and Fig.~\ref{Fig_155} demonstrates how the layer height ($D/b$) affects COD. Together, these figures provide a comprehensive 3D view of how COD varies with these parameters.
\bfg[htbp]
\centering
\begin{subfigure}[b] {0.49\textwidth}
\includegraphics[width=\textwidth ]{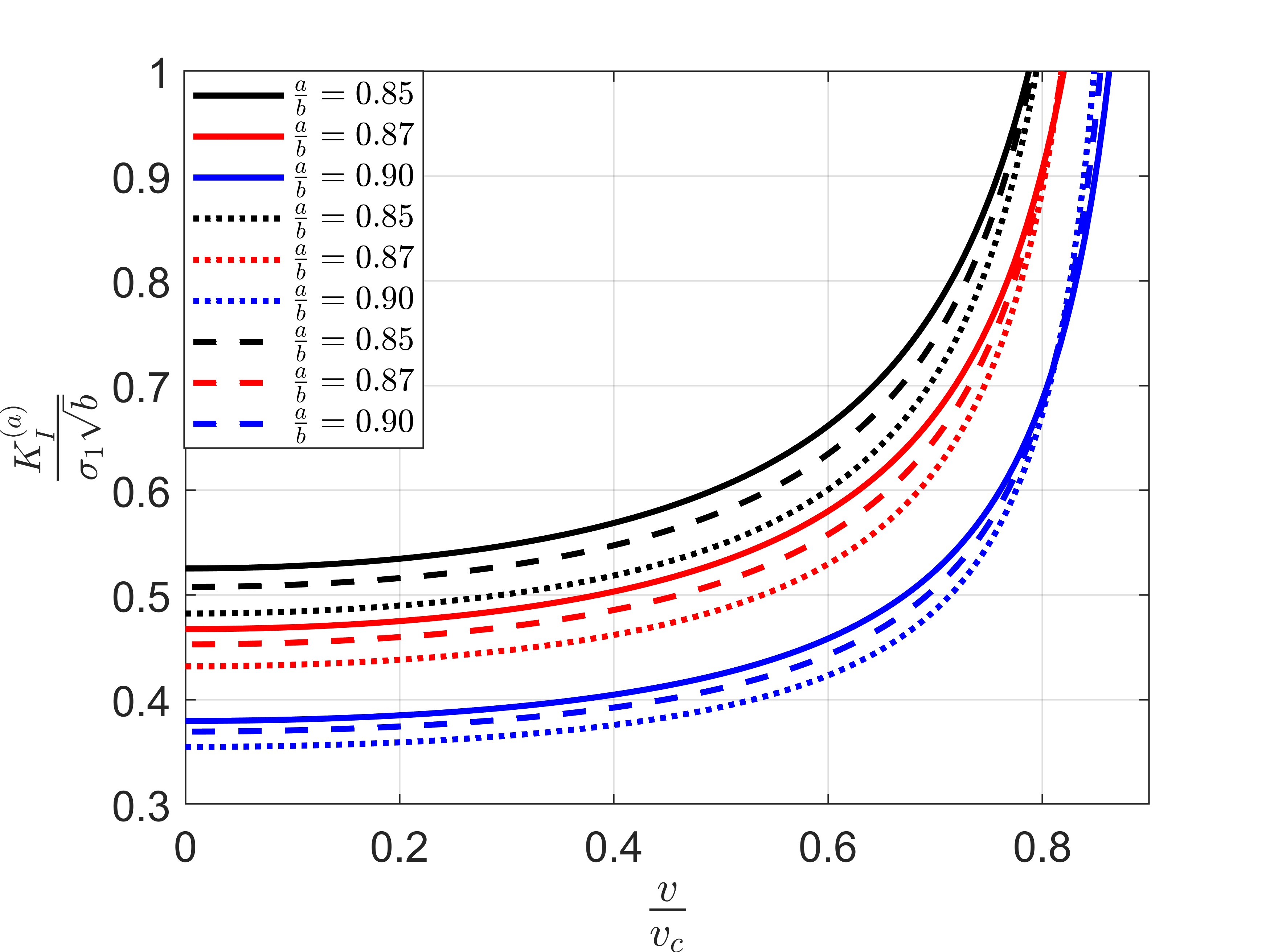}
\caption{}
\label{Fig_1}
\end{subfigure}
~
\begin{subfigure}[b] {0.49\textwidth}
\includegraphics[width=\textwidth ]{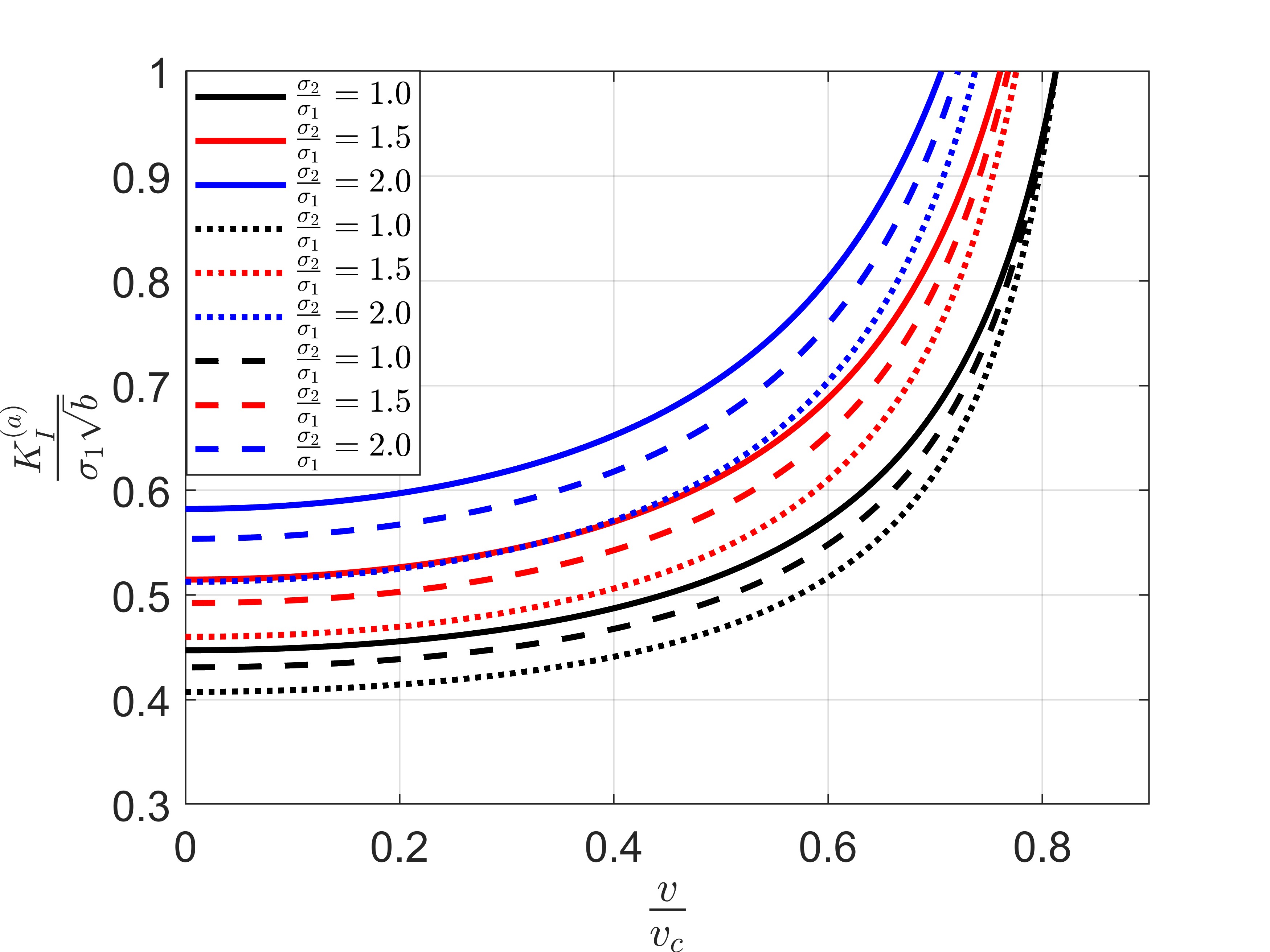}
\caption{}
\label{Fig_2}
\end{subfigure}
~
\begin{subfigure}[b] {0.49\textwidth}
\includegraphics[width=\textwidth ]{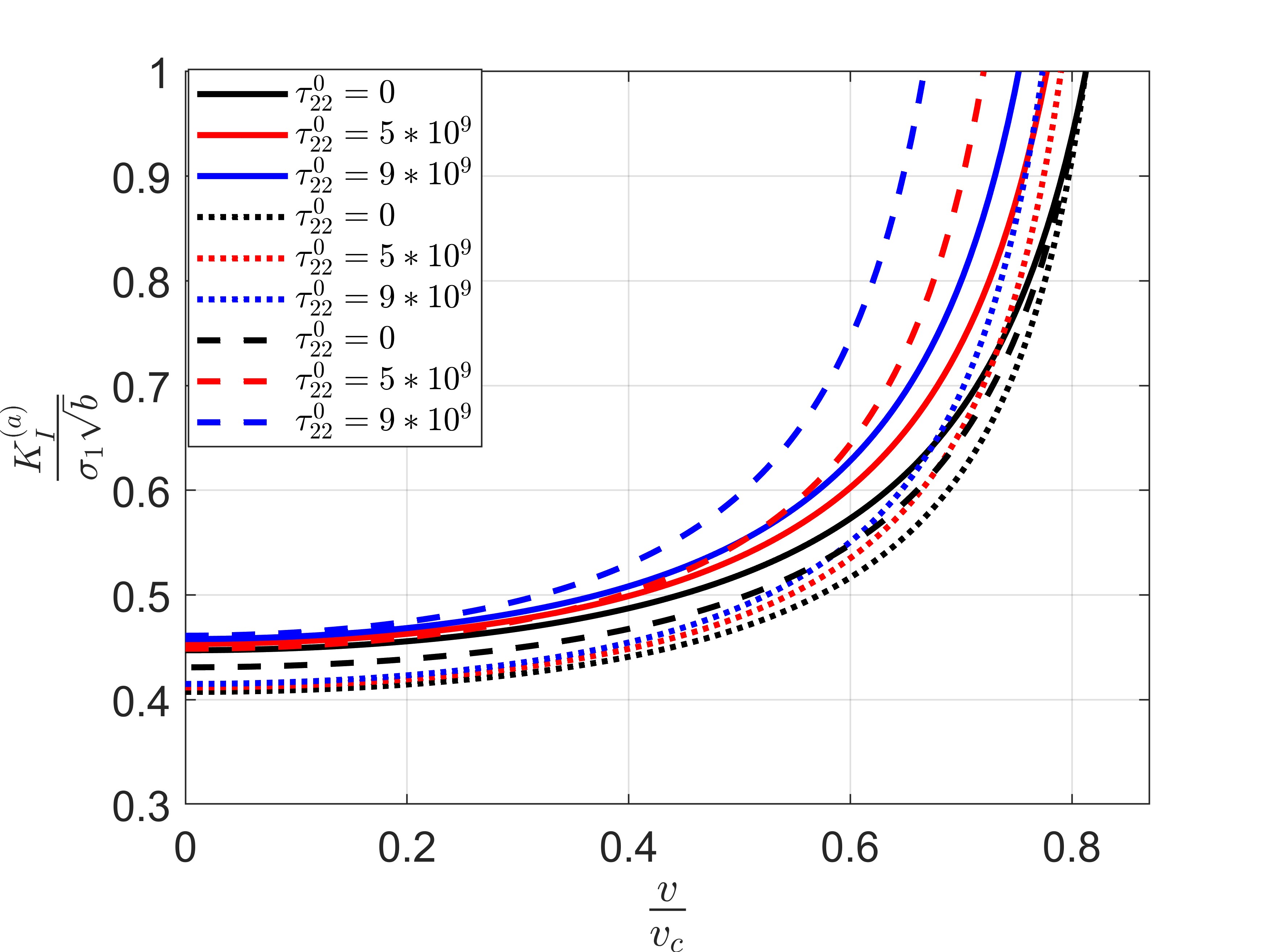}
\caption{}
\label{Fig_3}
\end{subfigure}
~
\begin{subfigure}[b] {0.49\textwidth}
\includegraphics[width=\textwidth ]{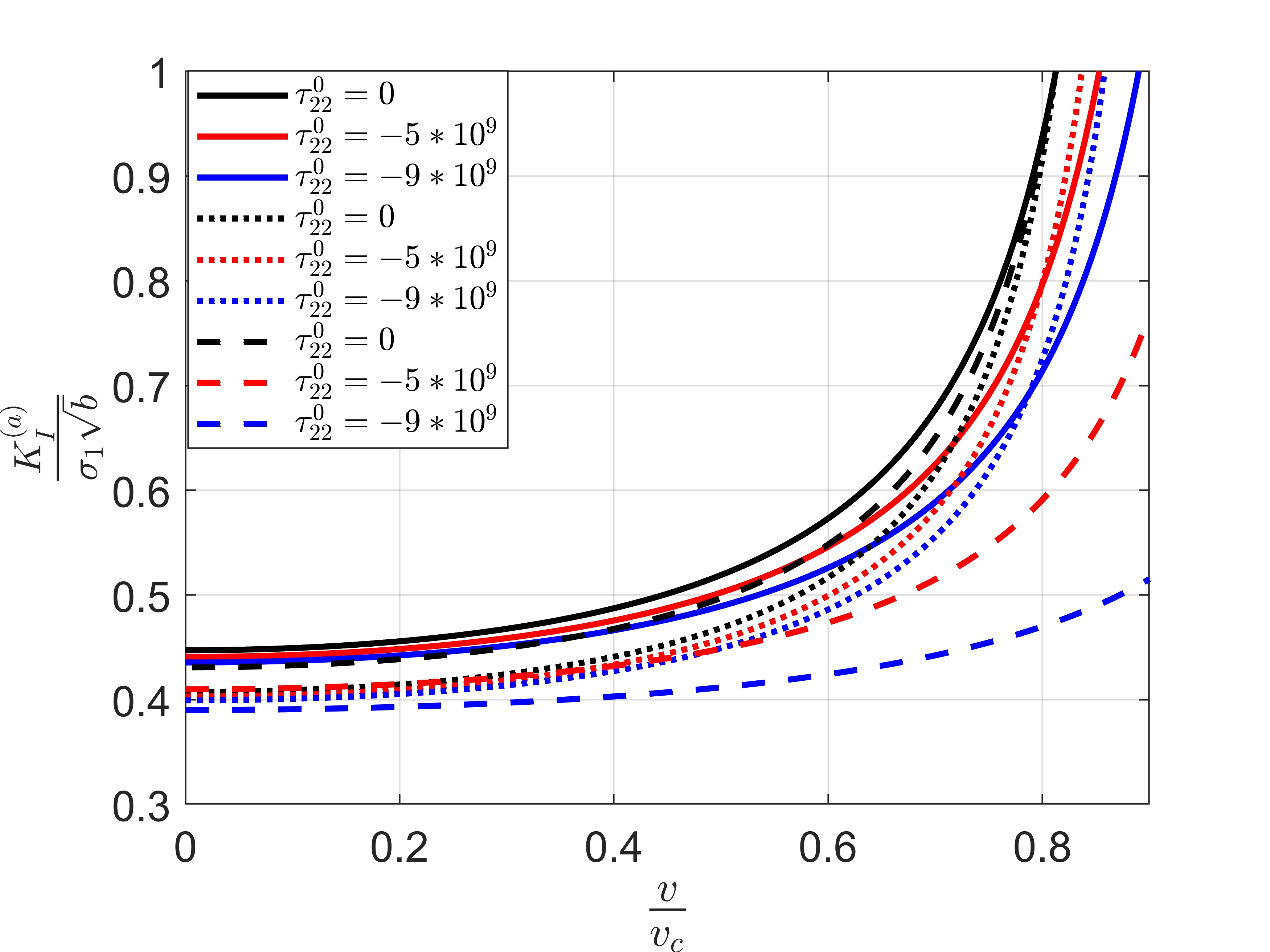}
\caption{}
\label{Fig_4}
\end{subfigure}
~
\begin{subfigure}[b] {0.49\textwidth}
\includegraphics[width=\textwidth ]{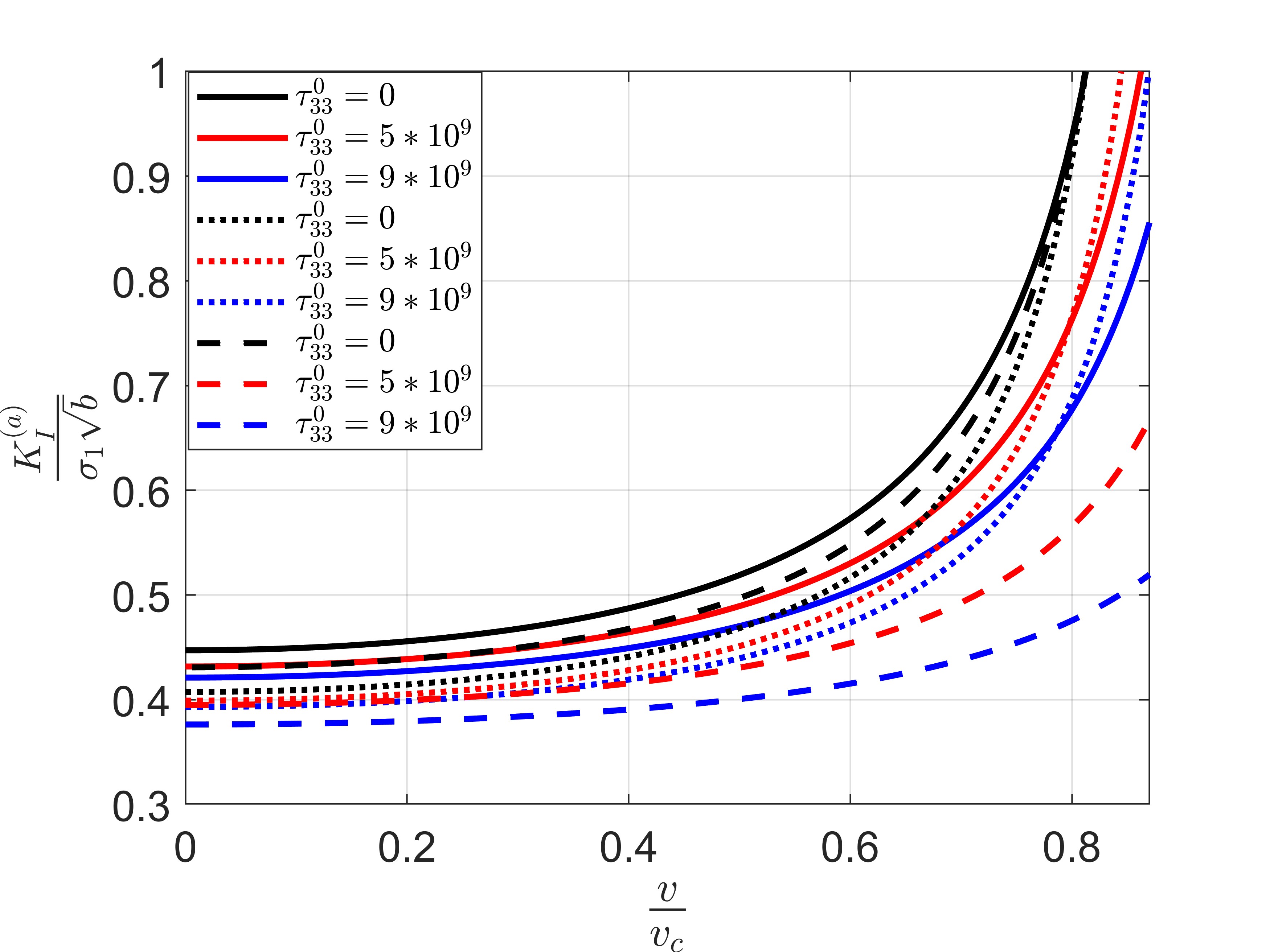}
\caption{}
\label{Fig_5}
\end{subfigure}
~
\begin{subfigure}[b] {0.49\textwidth}
\includegraphics[width=\textwidth ]{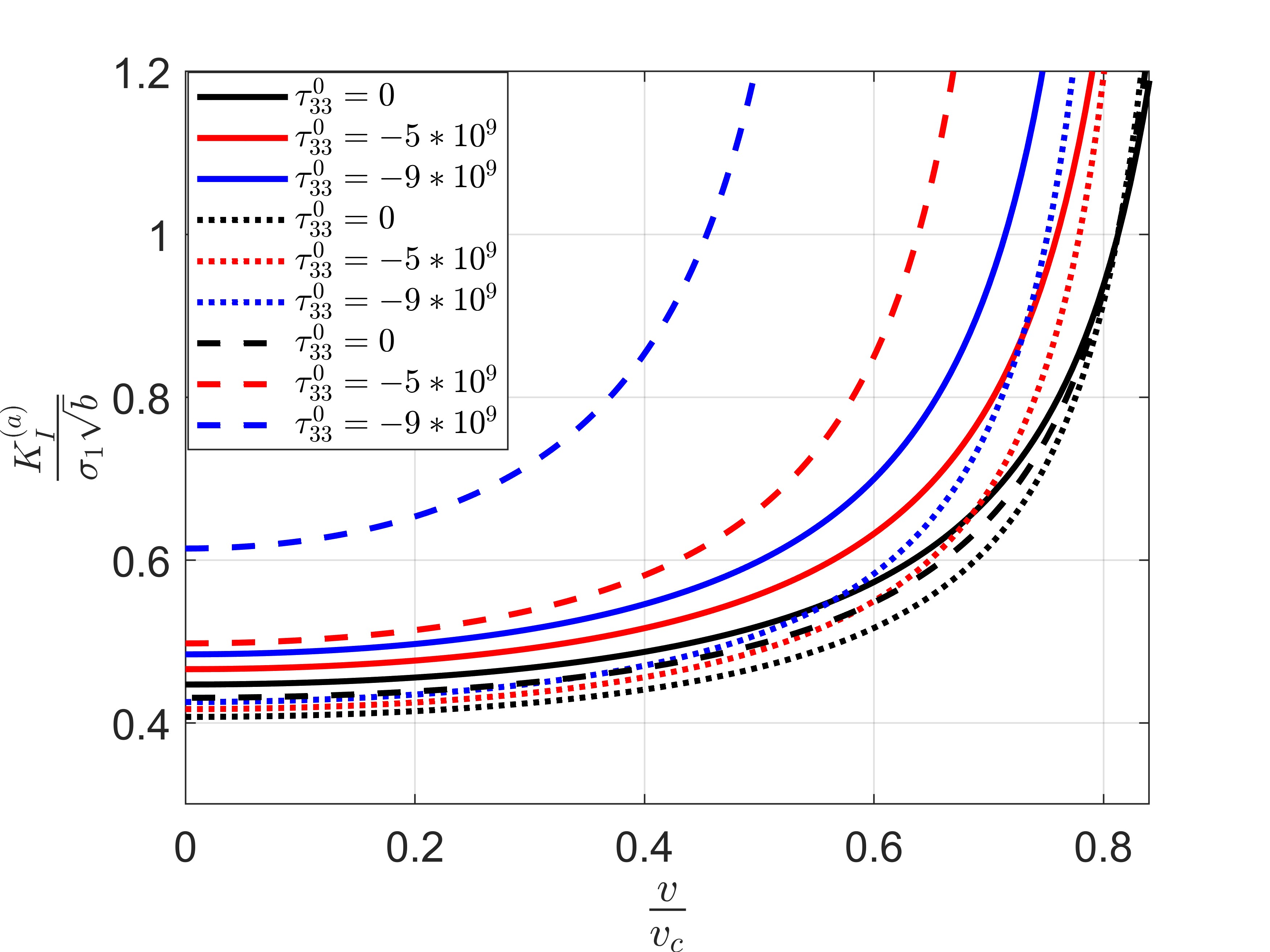}
\caption{}
\label{Fig_6}
\end{subfigure}
\caption{Variation of the dimensionless stress intensity factor $ (K^{(a)}_I/\sigma_1 \sqrt{b} )$ at the left crack tip $\left( X_3 = a \right)$ with respect to the dimensionless crack velocity $\left( v / v_c \right)$, considering the effects of (a) crack position $\left( a / b \right)$, (b) normal pressure $\left( \sigma_2 / \sigma_1 \right)$, (c) initial compressive stress $\tau_{22}$, (d) initial tensile stress $\tau_{22}$, (e) initial compressive stress $\tau_{33}$, and (f) initial tensile stress $\tau_{33}$.}

\label{Figure 2}
\efg

\bfg[htbp]
\centering
\begin{subfigure}[b] {0.49\textwidth}
\includegraphics[width=\textwidth ]{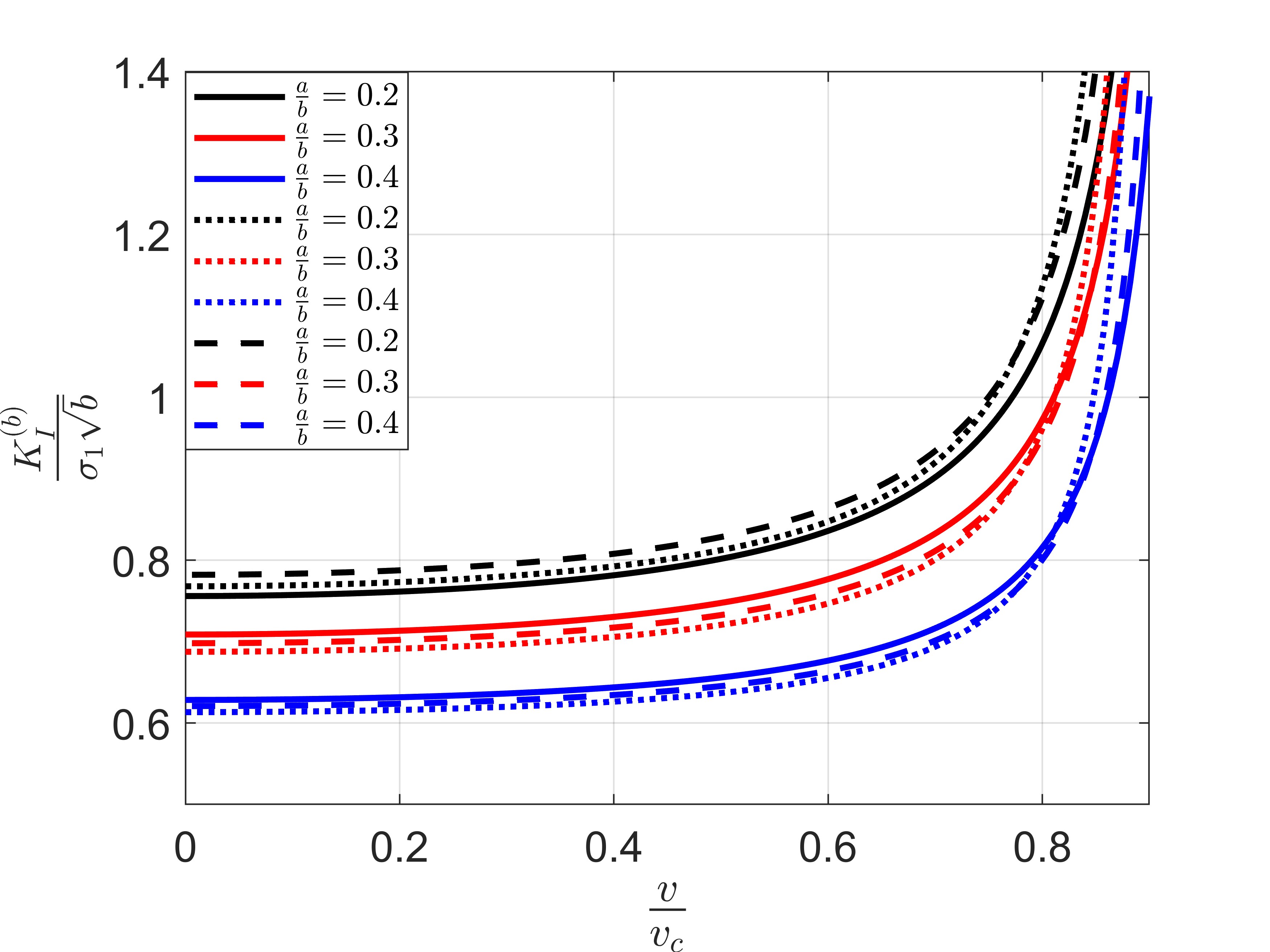}
\caption{}
\label{Fig_7}
\end{subfigure}
~
\begin{subfigure}[b] {0.49\textwidth}
\includegraphics[width=\textwidth ]{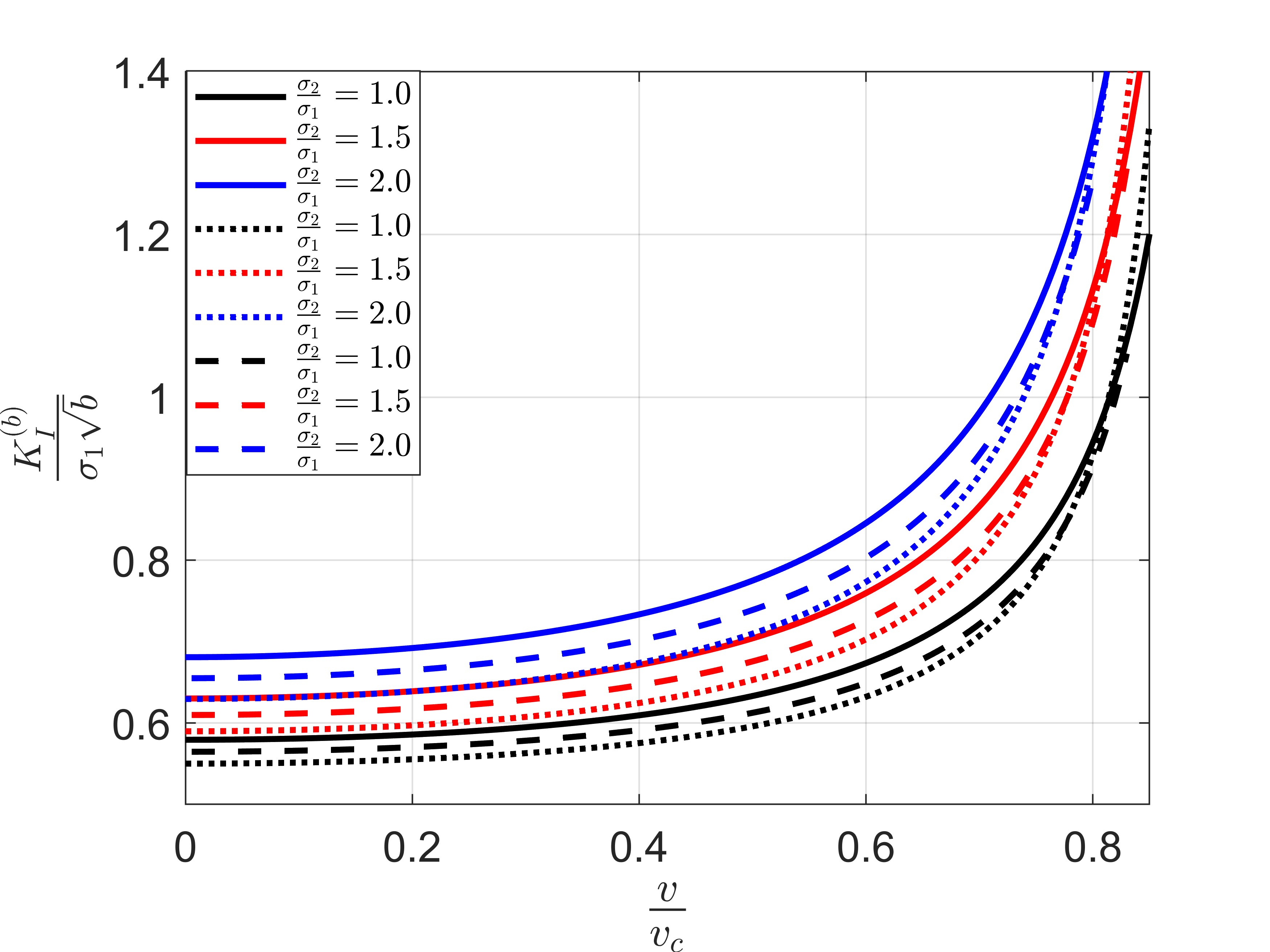}
\caption{}
\label{Fig_8}
\end{subfigure}
~
\begin{subfigure}[b] {0.49\textwidth}
\includegraphics[width=\textwidth ]{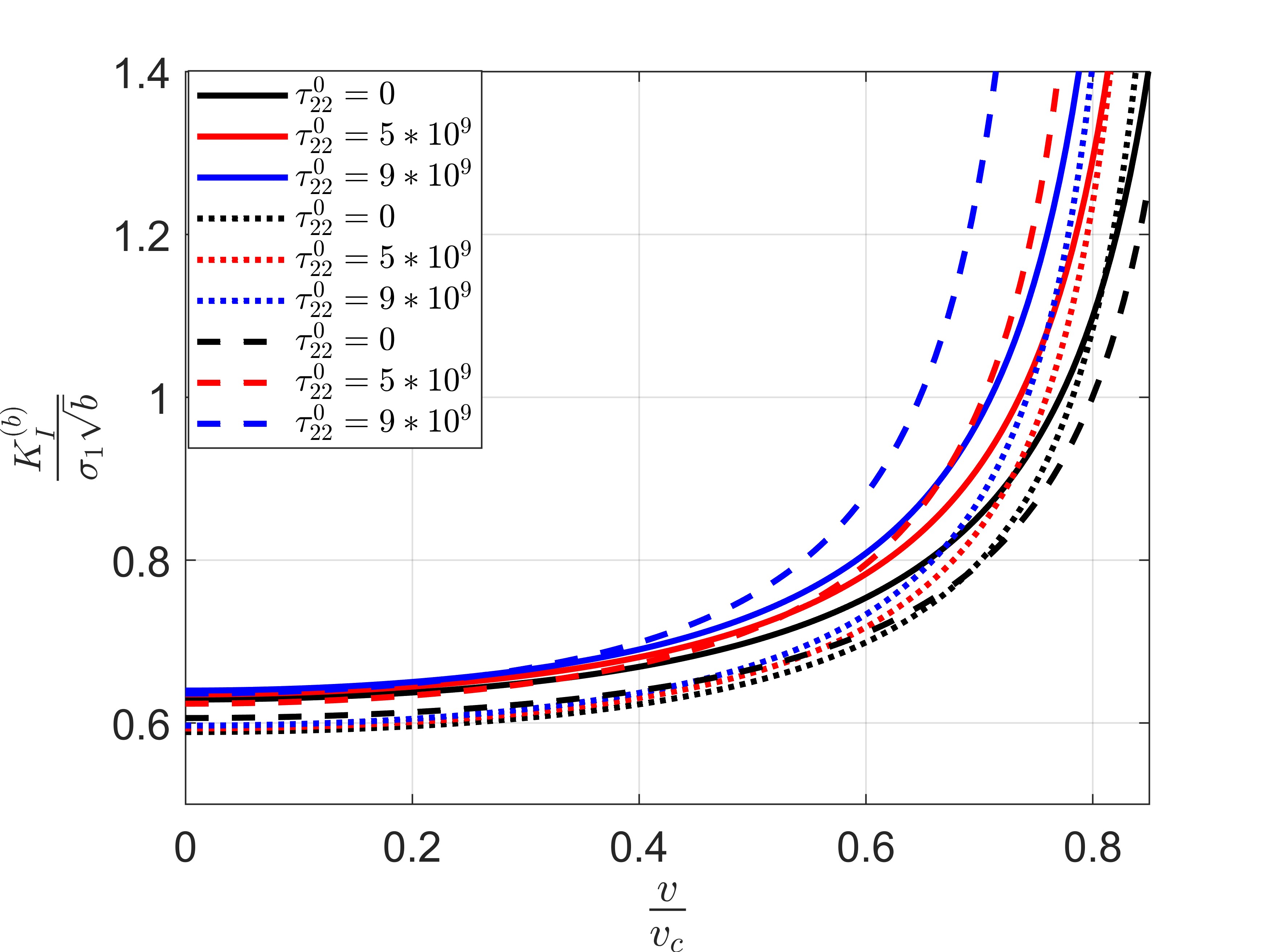}
\caption{}
\label{Fig_9}
\end{subfigure}
~
\begin{subfigure}[b] {0.49\textwidth}
\includegraphics[width=\textwidth ]{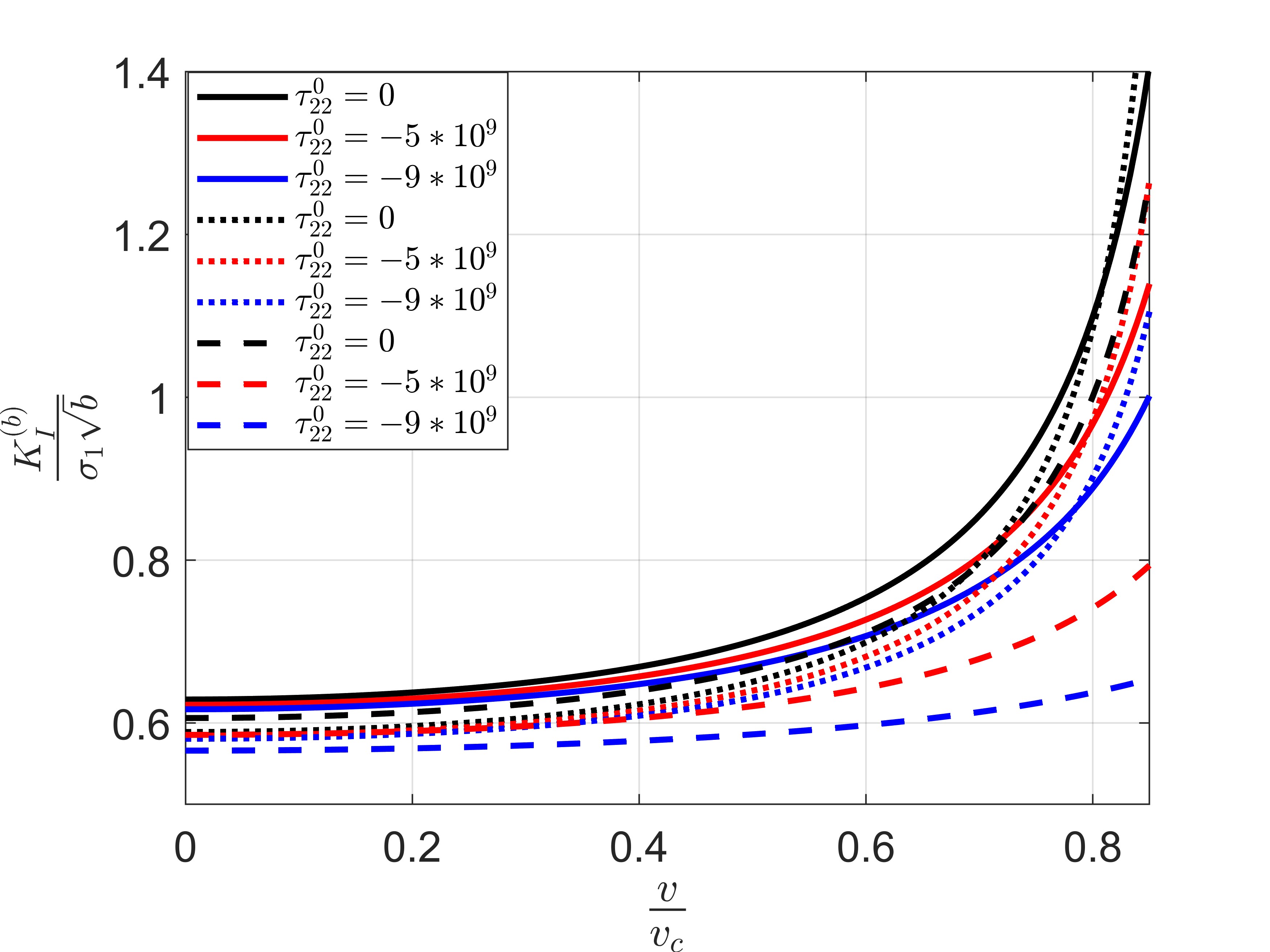}
\caption{}
\label{Fig_10}
\end{subfigure}
~
\begin{subfigure}[b] {0.49\textwidth}
\includegraphics[width=\textwidth ]{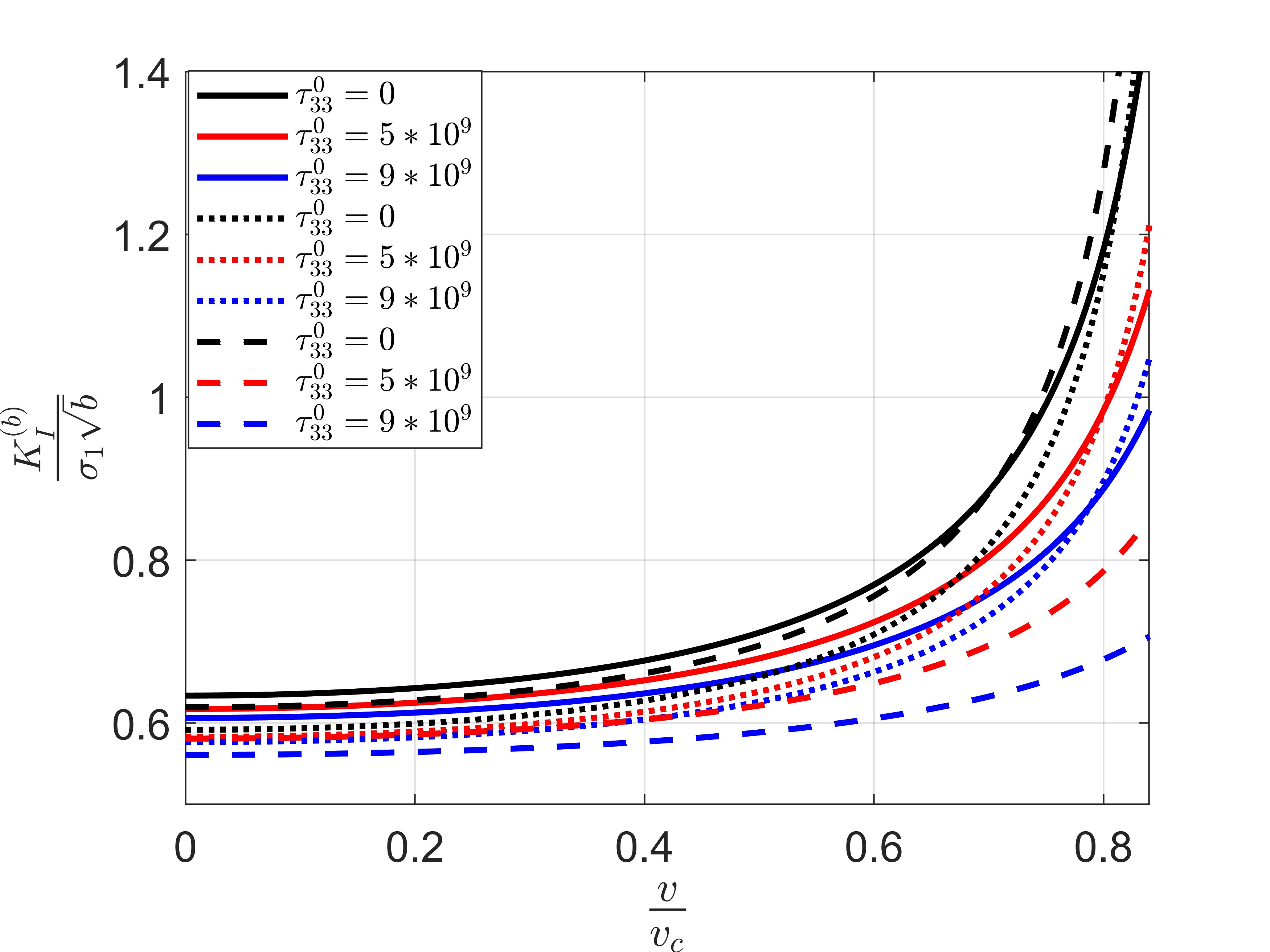}
\caption{}
\label{Fig_11}
\end{subfigure}
~
\begin{subfigure}[b] {0.49\textwidth}
\includegraphics[width=\textwidth ]{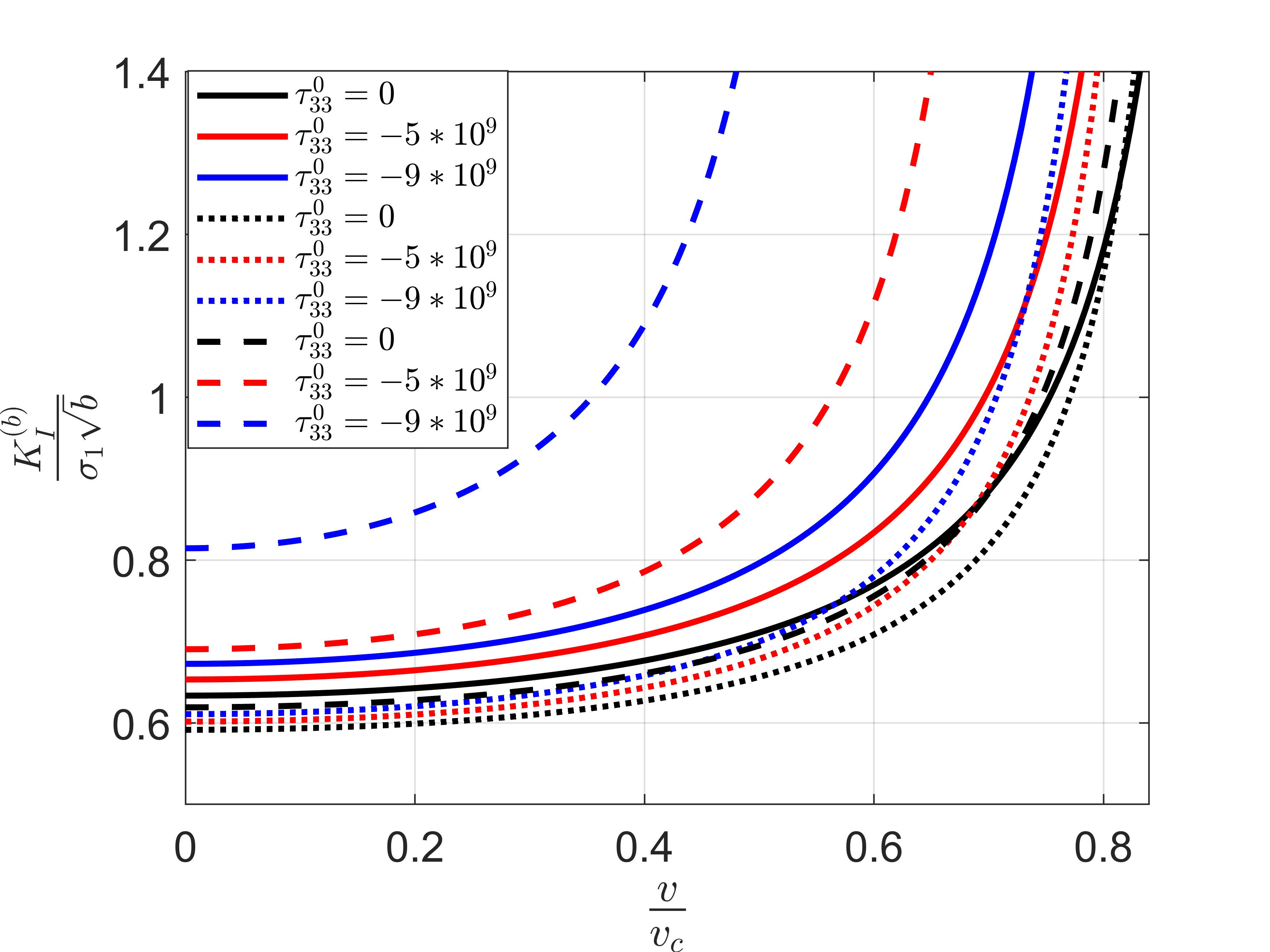}
\caption{}
\label{Fig__12}
\end{subfigure}
\caption{Variation of the dimensionless stress intensity factor $ (K^{(b)}_I/\sigma_1 \sqrt{b} )$ at the right crack tip $\left( X_3 = b \right)$ with respect to the dimensionless crack velocity $\left( v / v_c \right)$, considering the effects of (a) crack position $\left( a / b \right)$, (b) normal pressure $\left( \sigma_2 / \sigma_1 \right)$, (c) initial compressive stress $\tau_{22}$, (d) initial tensile stress $\tau_{22}$, (e) initial compressive stress $\tau_{33}$, and (f) initial tensile stress $\tau_{33}$.}

\label{Figure 3}
\efg

\bfg[htbp]
\centering
\begin{subfigure}[b] {0.49\textwidth}
\includegraphics[width=\textwidth ]{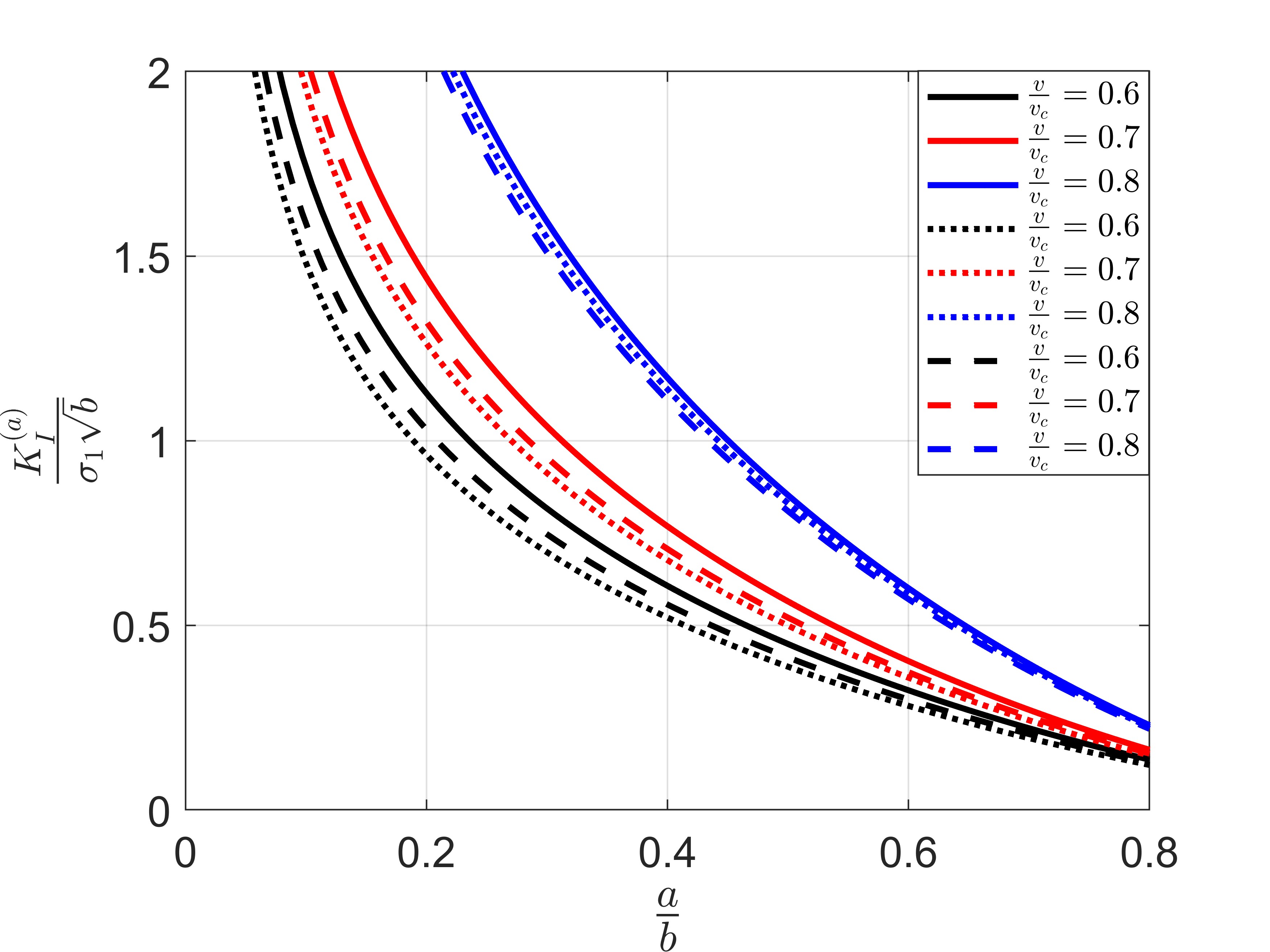}
\caption{}
\label{Fig_18}
\end{subfigure}
~
\begin{subfigure}[b] {0.49\textwidth}
\includegraphics[width=\textwidth ]{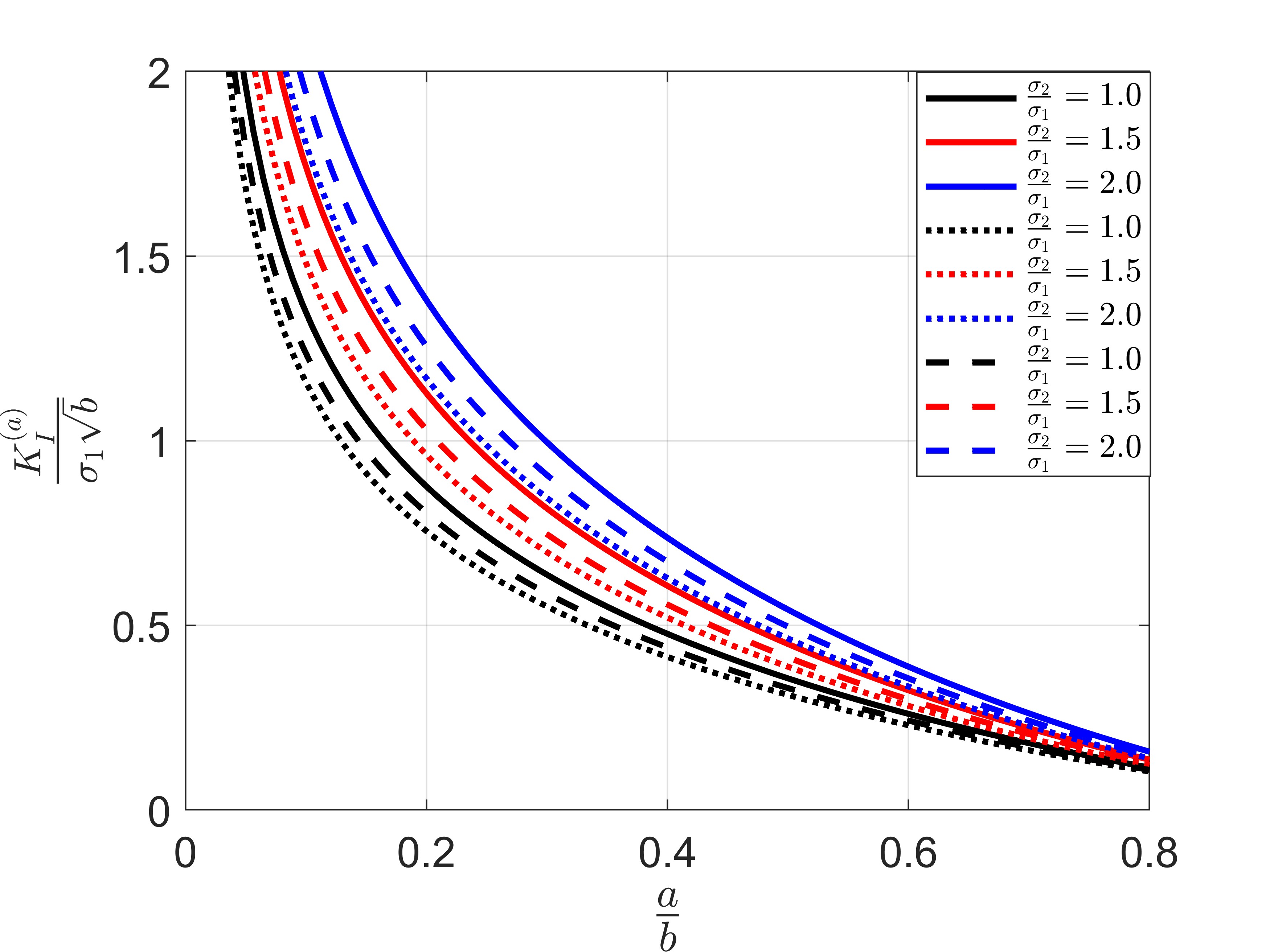}
\caption{}
\label{Fig_19}
\end{subfigure}
~
\begin{subfigure}[b] {0.49\textwidth}
\includegraphics[width=\textwidth ]{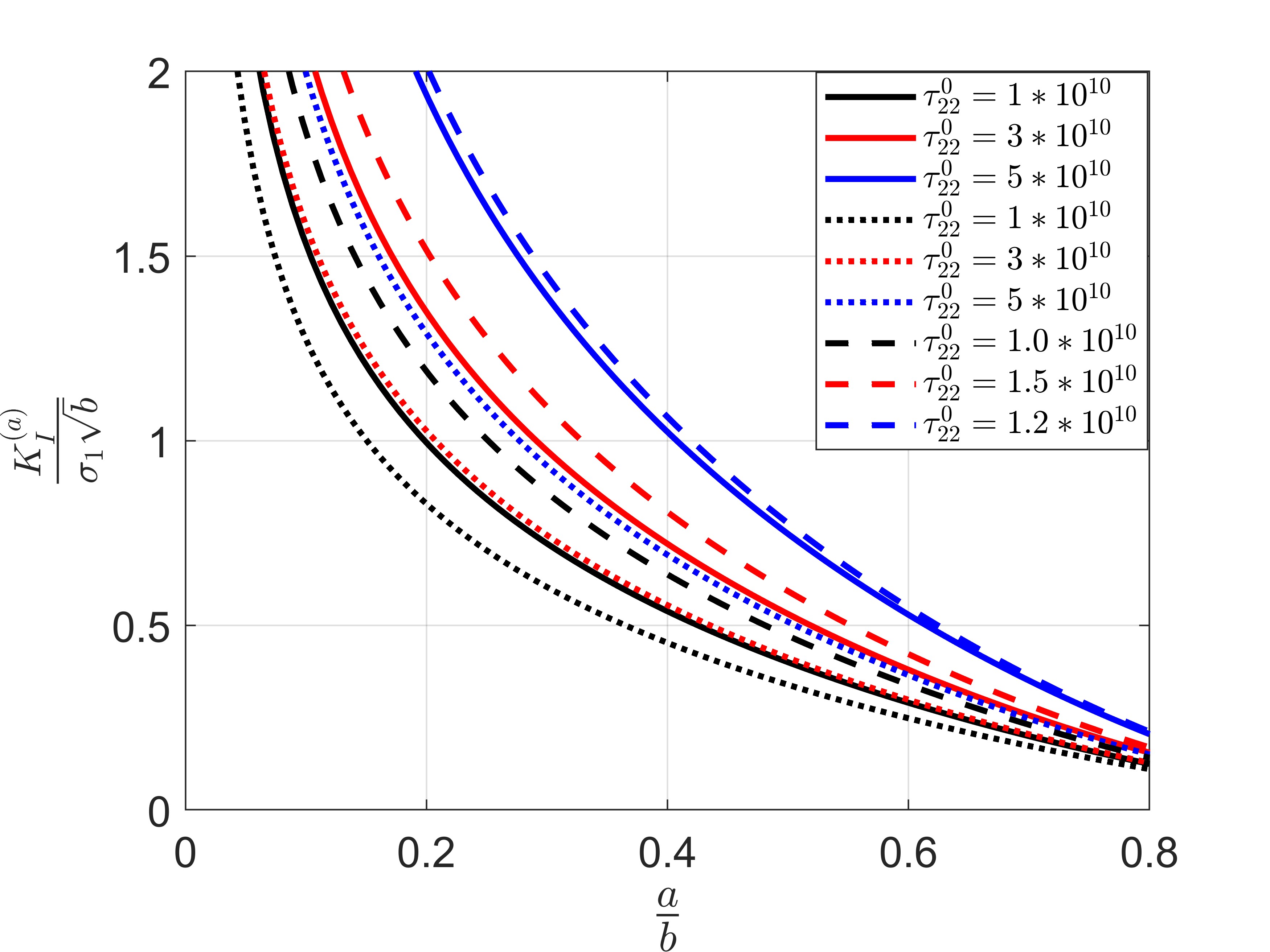}
\caption{}
\label{Fig_20}
\end{subfigure}
~
\begin{subfigure}[b] {0.49\textwidth}
\includegraphics[width=\textwidth ]{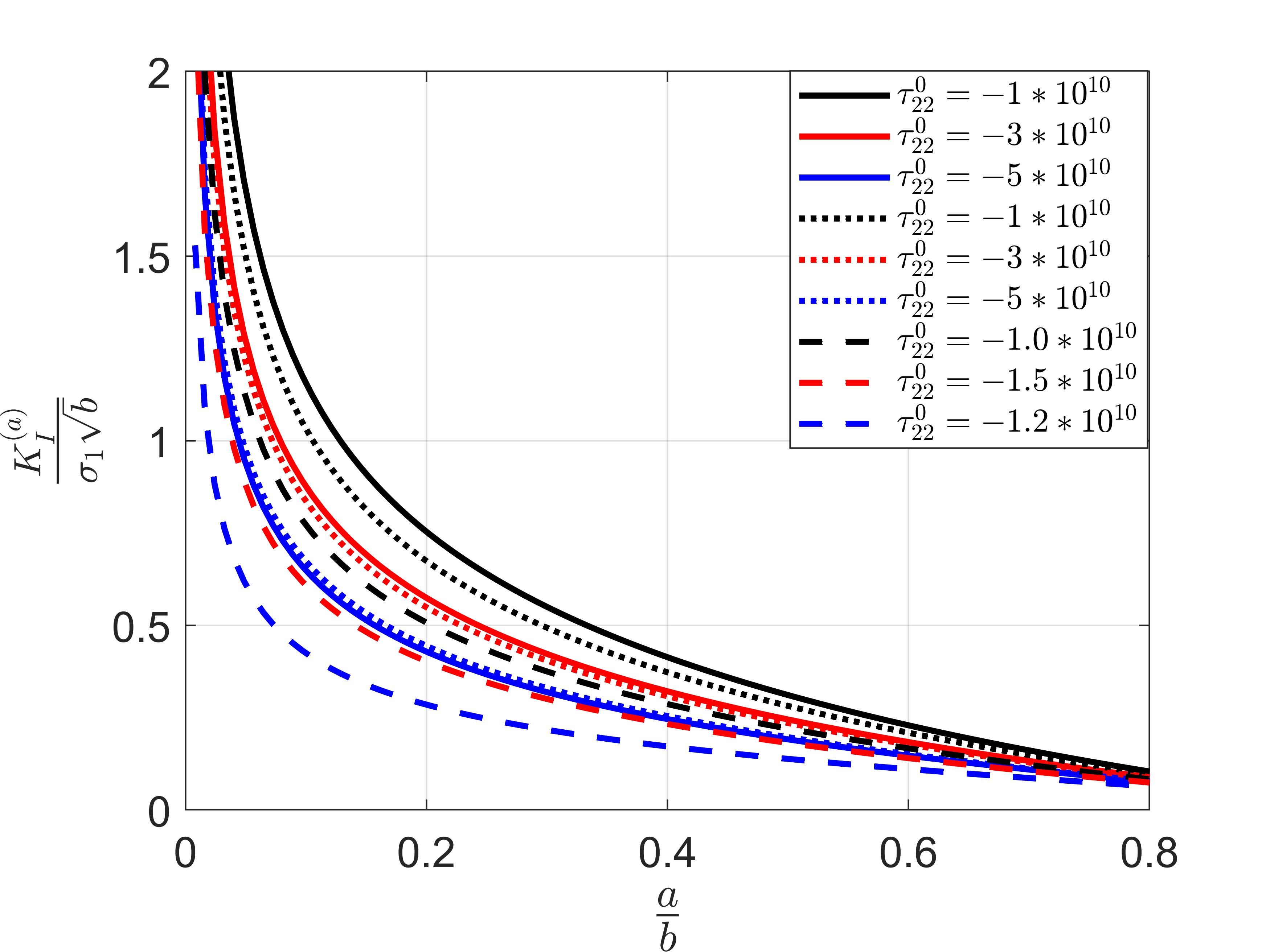}
\caption{}
\label{Fig_21}
\end{subfigure}
~
\begin{subfigure}[b] {0.49\textwidth}
\includegraphics[width=\textwidth ]{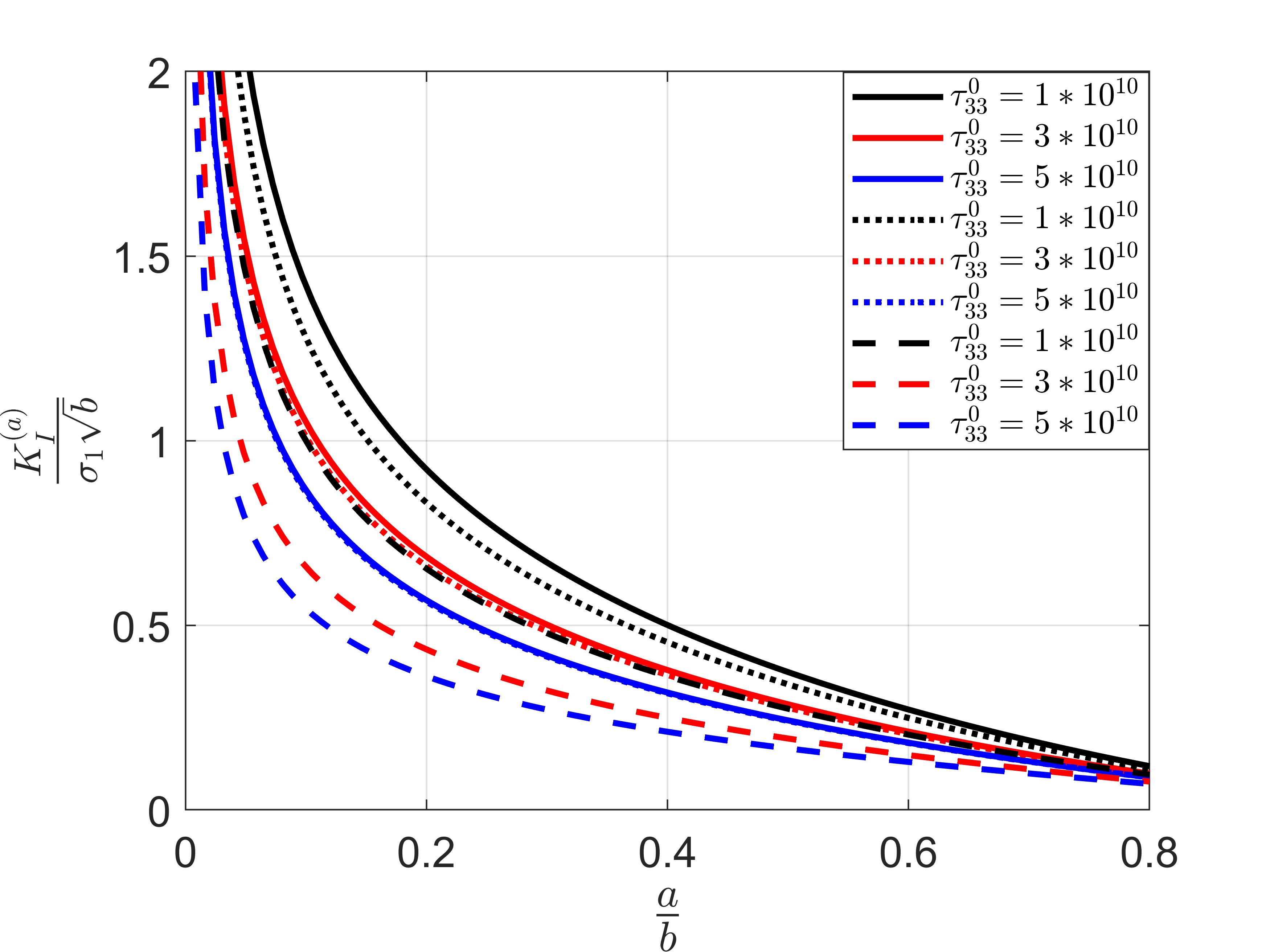}
\caption{}
\label{Fig_22}
\end{subfigure}
~
\begin{subfigure}[b] {0.49\textwidth}
\includegraphics[width=\textwidth ]{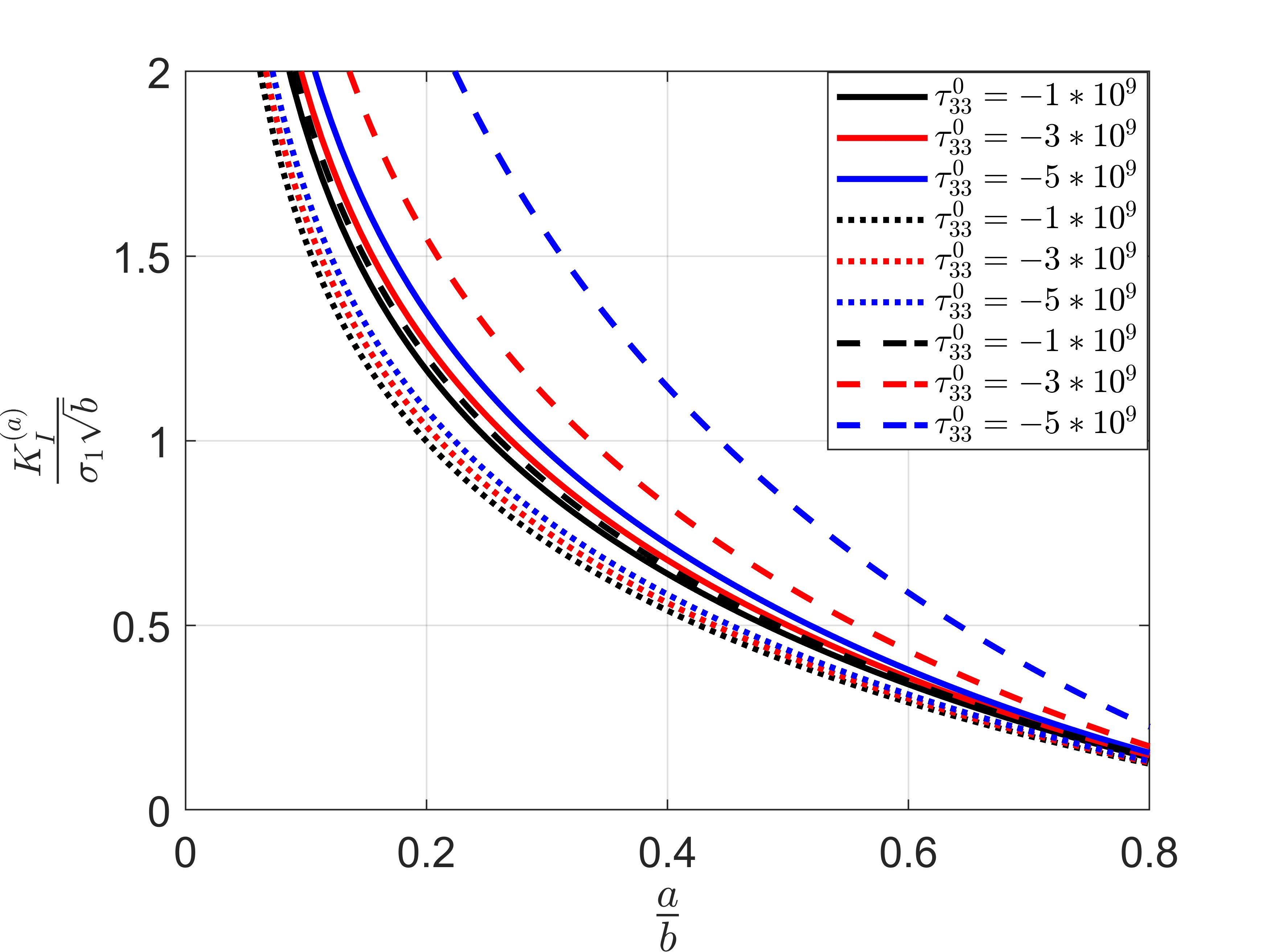}
\caption{}
\label{Fig_23}
\end{subfigure}
\caption{Variation of the dimensionless stress intensity factor $ (K^{(a)}_I/\sigma_1 \sqrt{b} )$ at the left crack tip $\left( X_3 = a \right)$ with respect to the dimensionless crack position $\left( a / b \right)$, considering the effects of (a) crack velocity $\left( v / v_c \right)$, (b) normal pressure $\left( \sigma_2 / \sigma_1 \right)$, (c) initial compressive stress $\tau_{22}$, (d) initial tensile stress $\tau_{22}$, (e) initial compressive stress $\tau_{33}$, and (f) initial tensile stress $\tau_{33}$.}

\label{Figure 5}
\efg

\bfg[htbp]
\centering
\begin{subfigure}[b] {0.49\textwidth}
\includegraphics[width=\textwidth ]{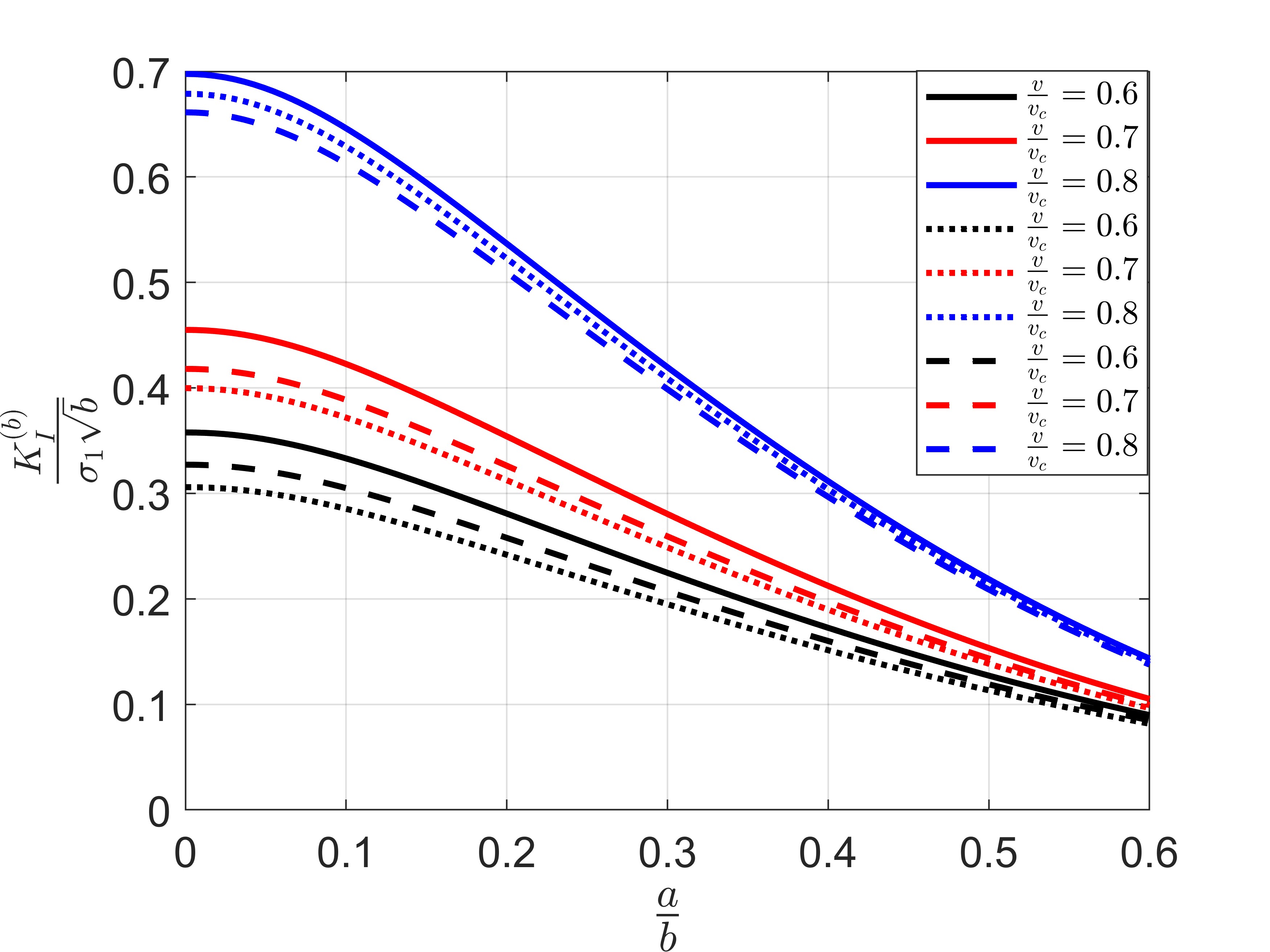}
\caption{}
\label{Fig_24}
\end{subfigure}
~
\begin{subfigure}[b] {0.49\textwidth}
\includegraphics[width=\textwidth ]{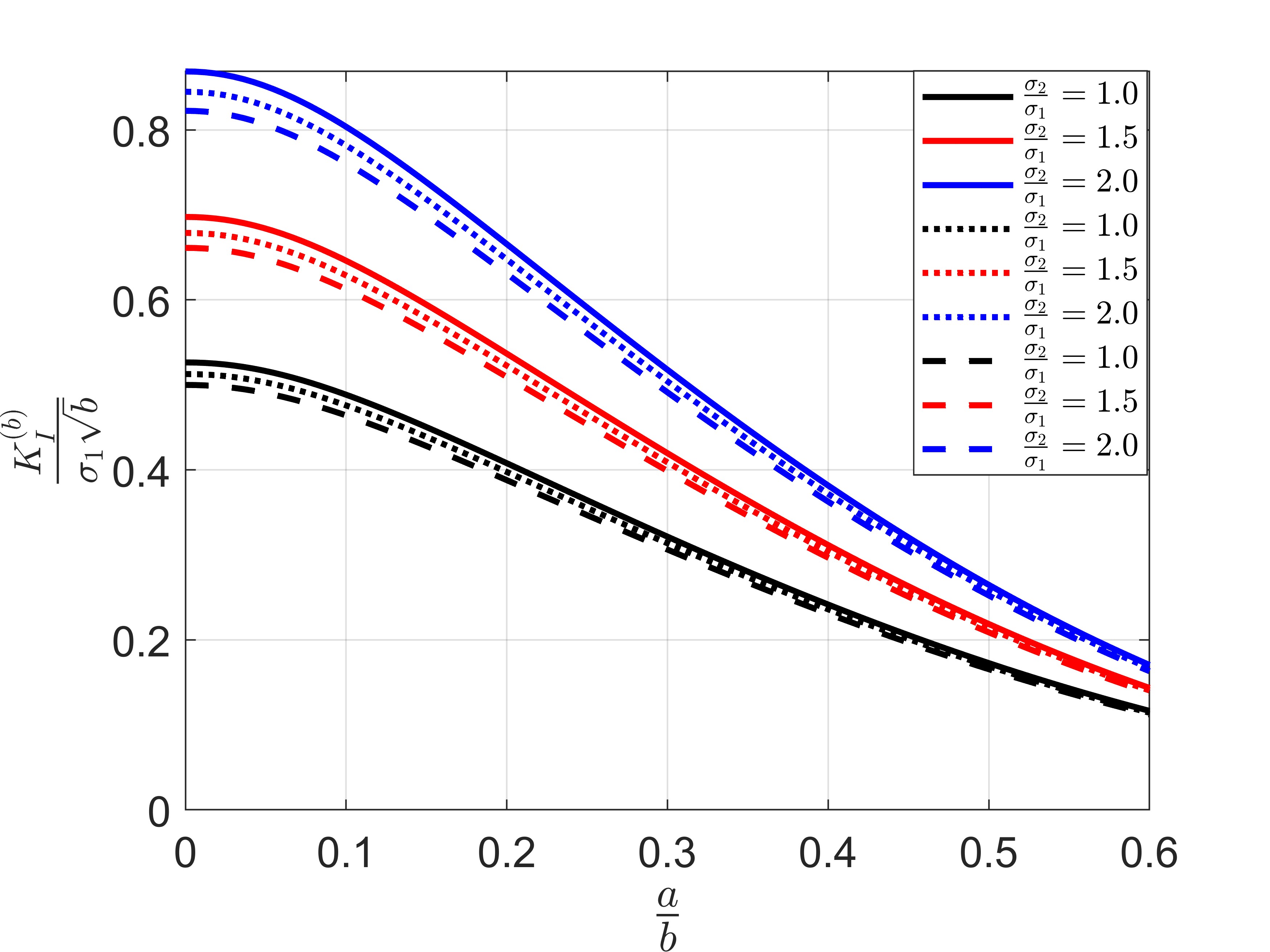}
\caption{}
\label{Fig_25}
\end{subfigure}
~
\begin{subfigure}[b] {0.49\textwidth}
\includegraphics[width=\textwidth ]{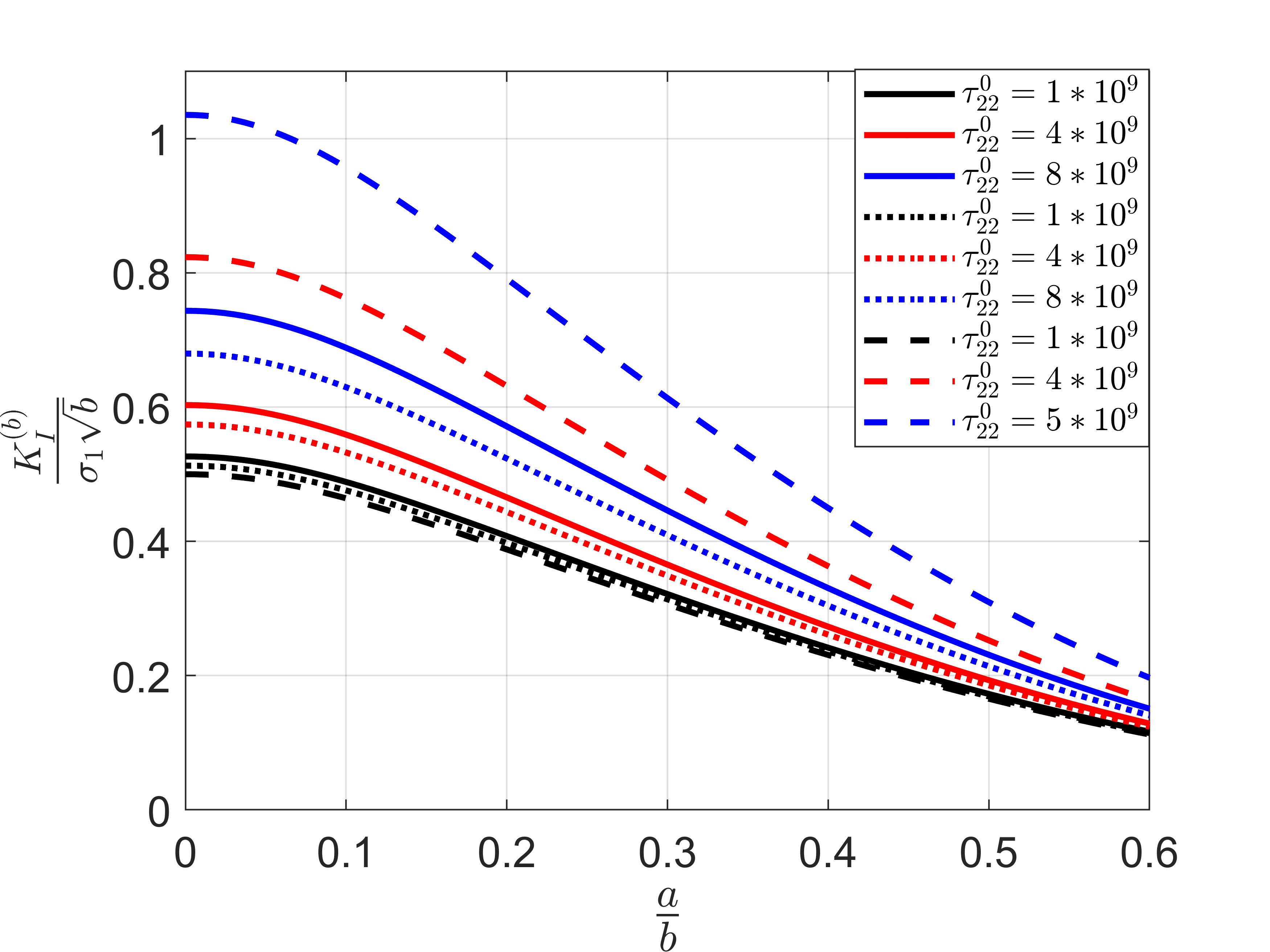}
\caption{}
\label{Fig_26}
\end{subfigure}
~
\begin{subfigure}[b] {0.49\textwidth}
\includegraphics[width=\textwidth ]{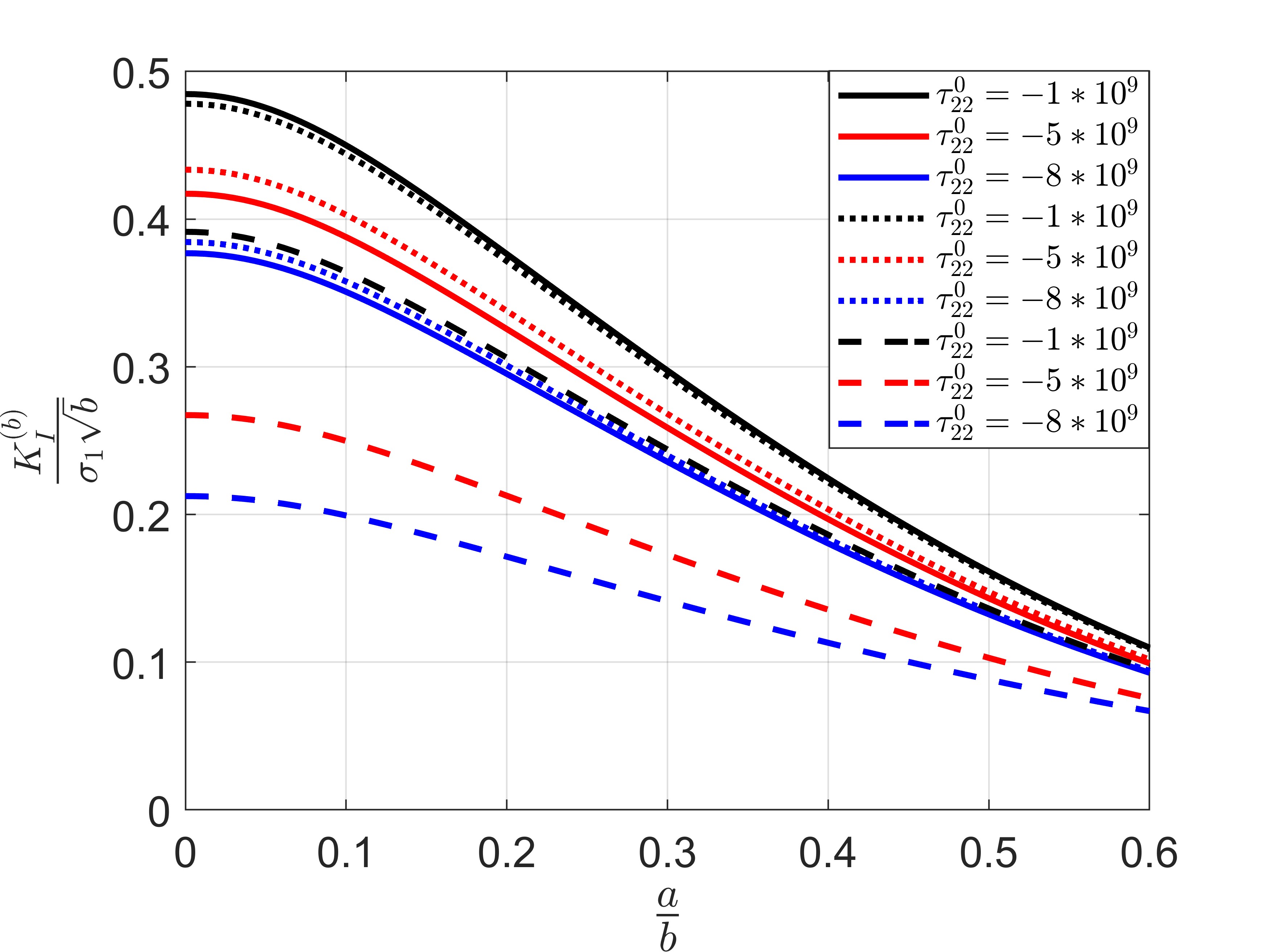}
\caption{}
\label{Fig_27}
\end{subfigure}
~
\begin{subfigure}[b] {0.49\textwidth}
\includegraphics[width=\textwidth ]{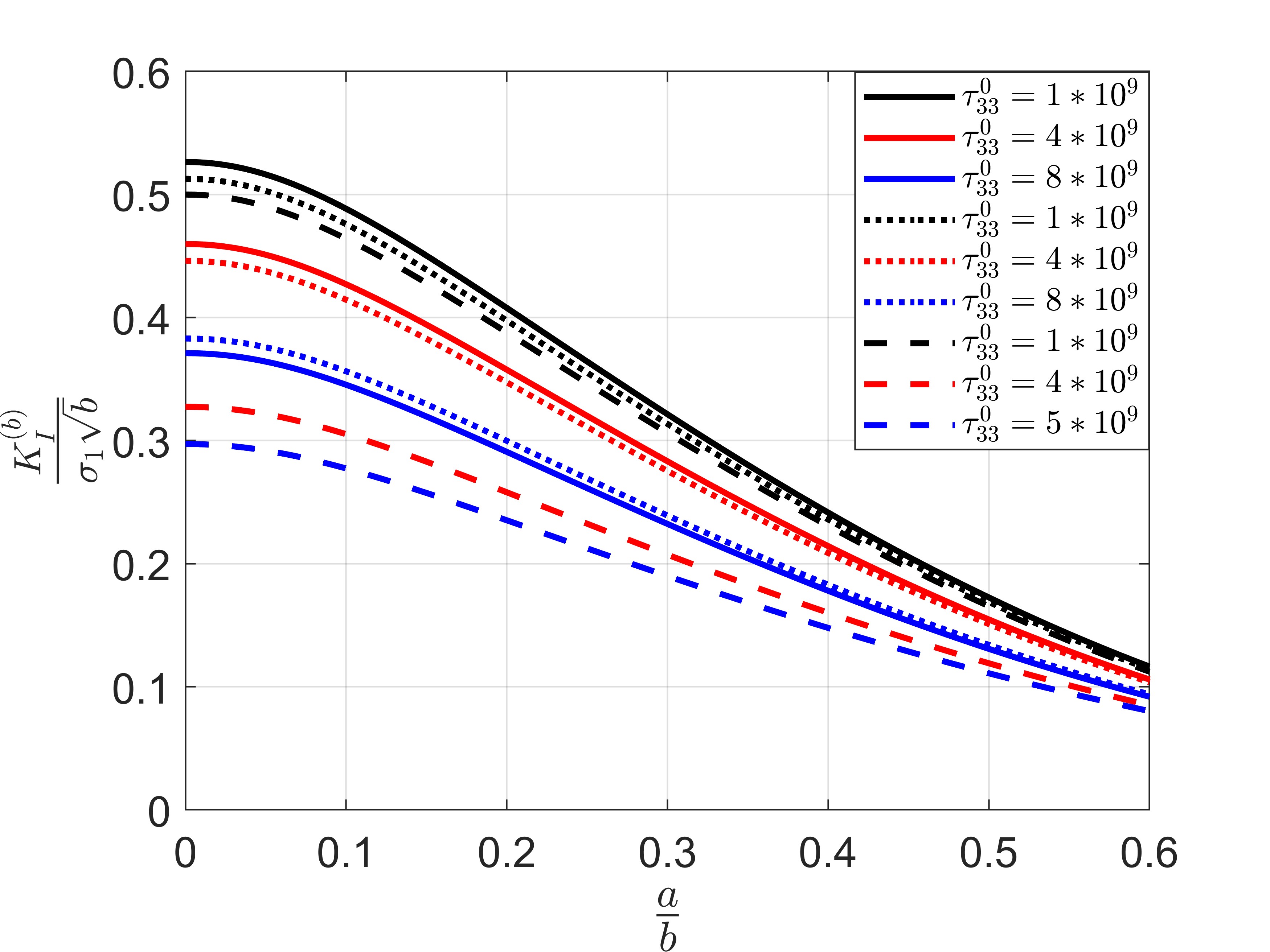}
\caption{}
\label{Fig_28}
\end{subfigure}
~
\begin{subfigure}[b] {0.49\textwidth}
\includegraphics[width=\textwidth ]{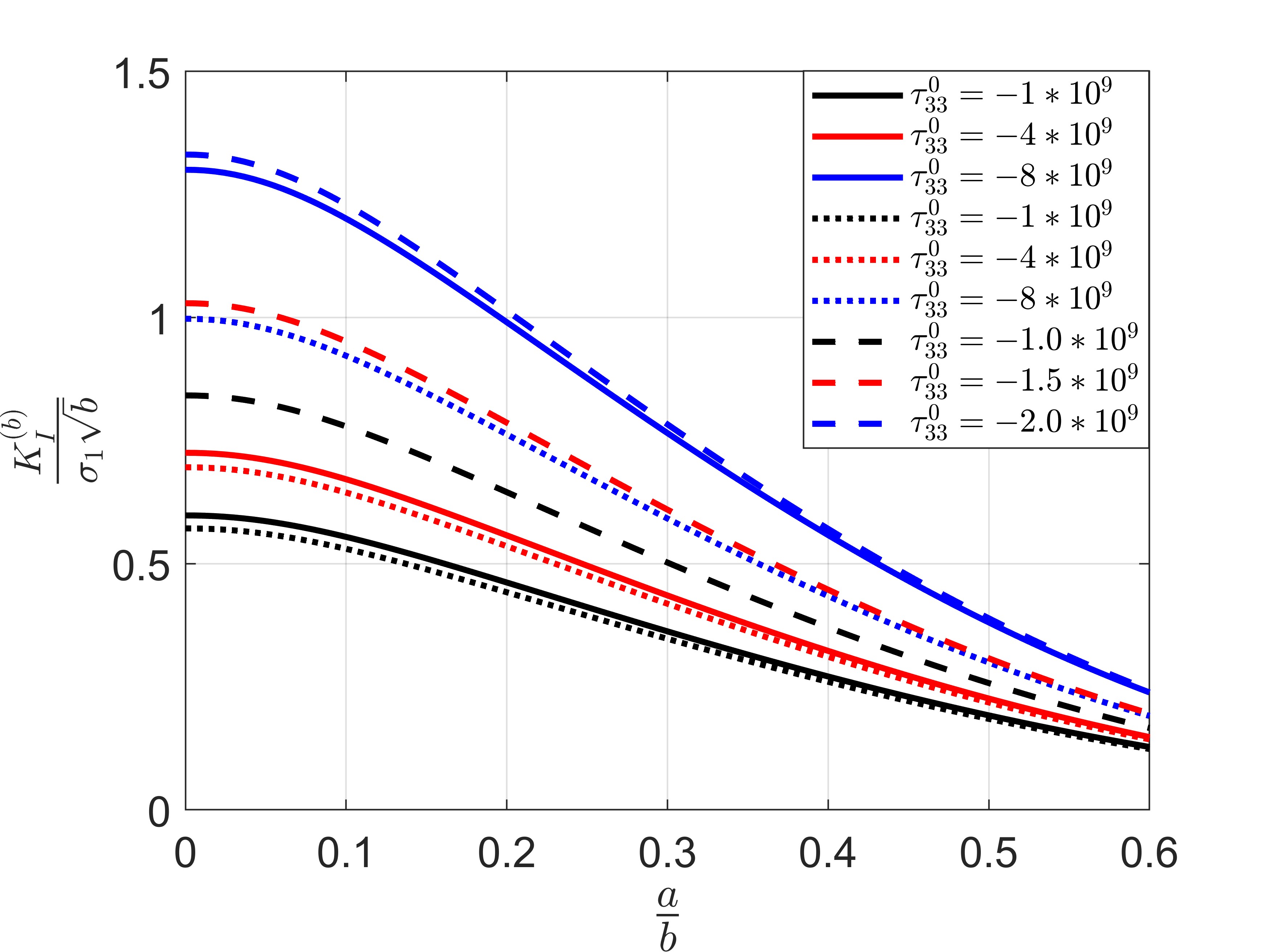}
\caption{}
\label{Fig_29}
\end{subfigure}
\caption{Variation of the dimensionless stress intensity factor $ (K^{(b)}_I/\sigma_1 \sqrt{b} )$ at the right crack tip $\left( X_3 = b \right)$ with respect to the dimensionless crack position $\left( a / b \right)$, considering the effects of (a) crack velocity $\left( v / v_c \right)$, (b) normal pressure $\left( \sigma_2 / \sigma_1 \right)$, (c) initial compressive stress $\tau_{22}$, (d) initial tensile stress $\tau_{22}$, (e) initial compressive stress $\tau_{33}$, and (f) initial tensile stress $\tau_{33}$.}

\label{Figure 6}
\efg

\bfg[htbp]
\centering
\begin{subfigure}[b] {0.49\textwidth}
\includegraphics[width=\textwidth ]{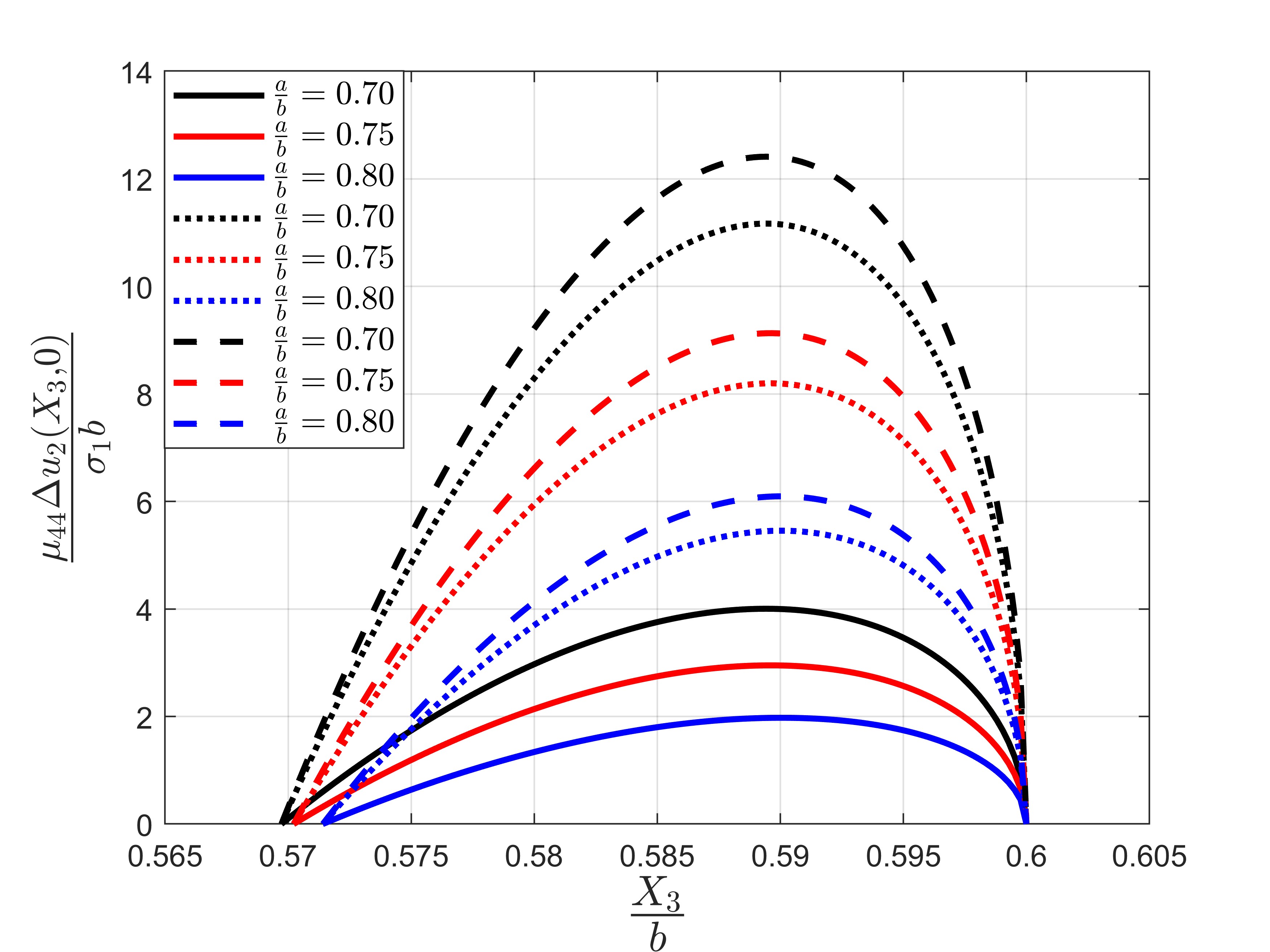}
\caption{}
\label{Fig_12}
\end{subfigure}
~
\begin{subfigure}[b] {0.49\textwidth}
\includegraphics[width=\textwidth ]{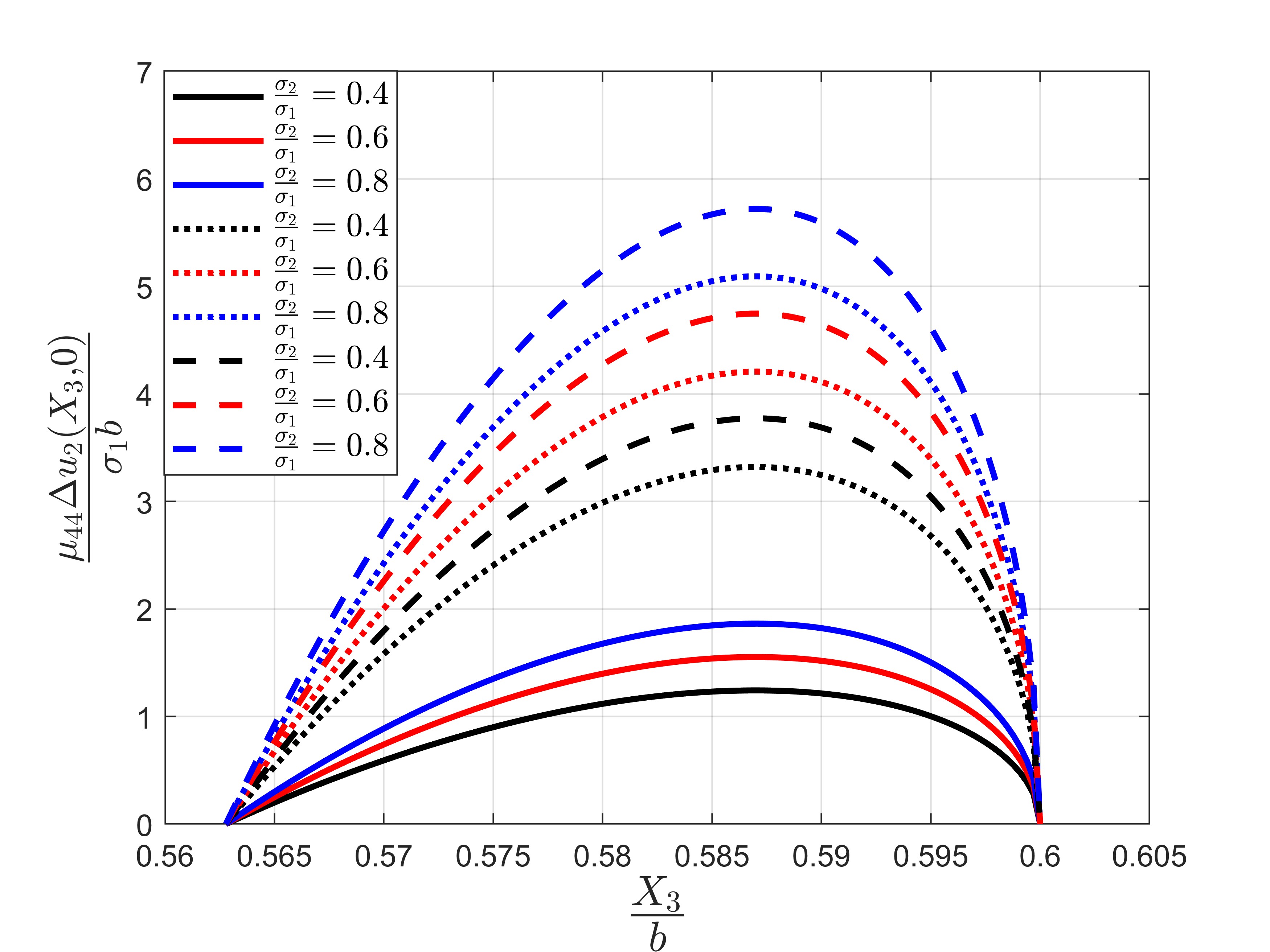}
\caption{}
\label{Fig_13}
\end{subfigure}
~
\begin{subfigure}[b] {0.49\textwidth}
\includegraphics[width=\textwidth ]{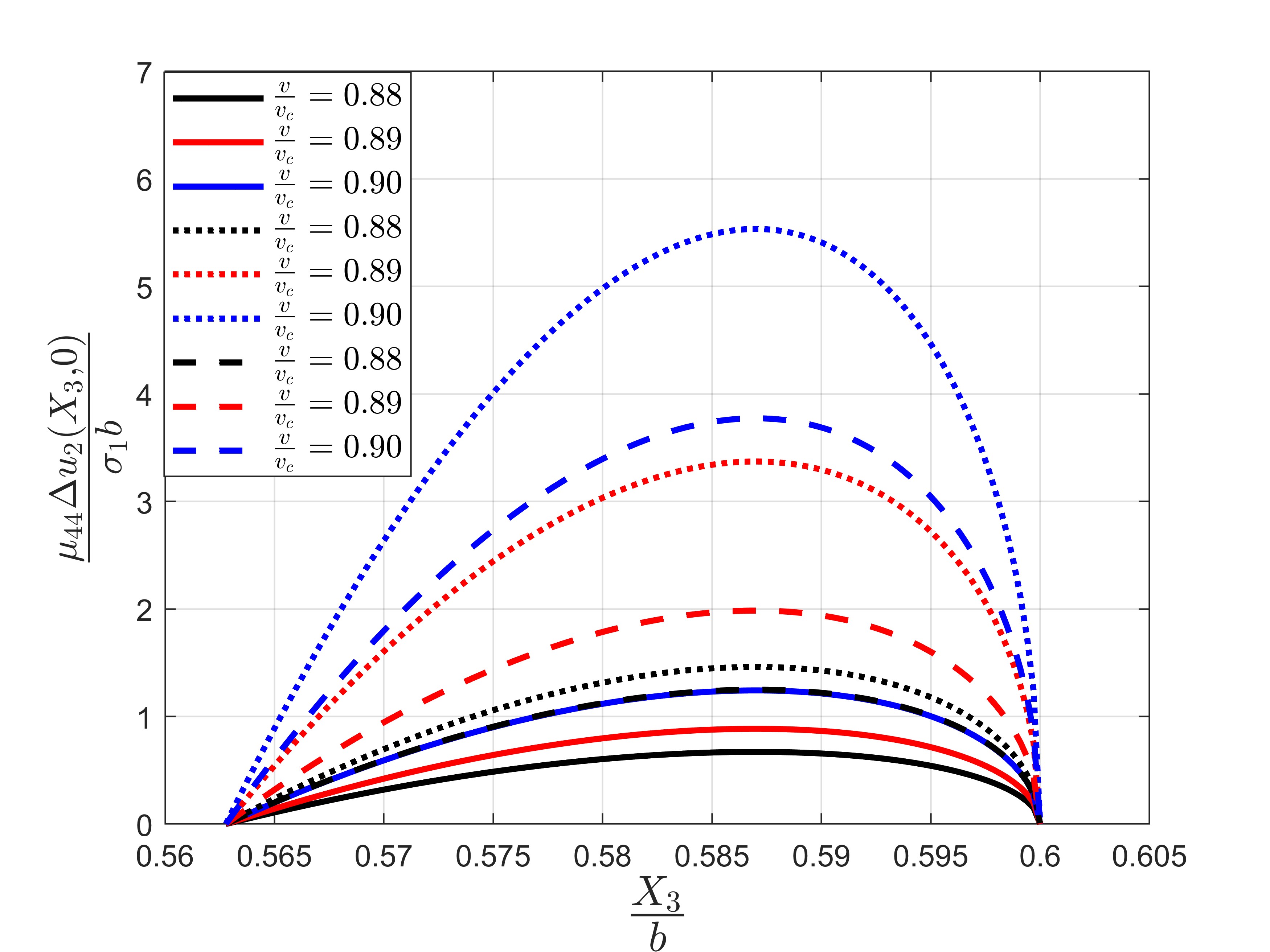}
\caption{}
\label{Fig_14}
\end{subfigure}
~
\begin{subfigure}[b] {0.49\textwidth}
\includegraphics[width=\textwidth ]{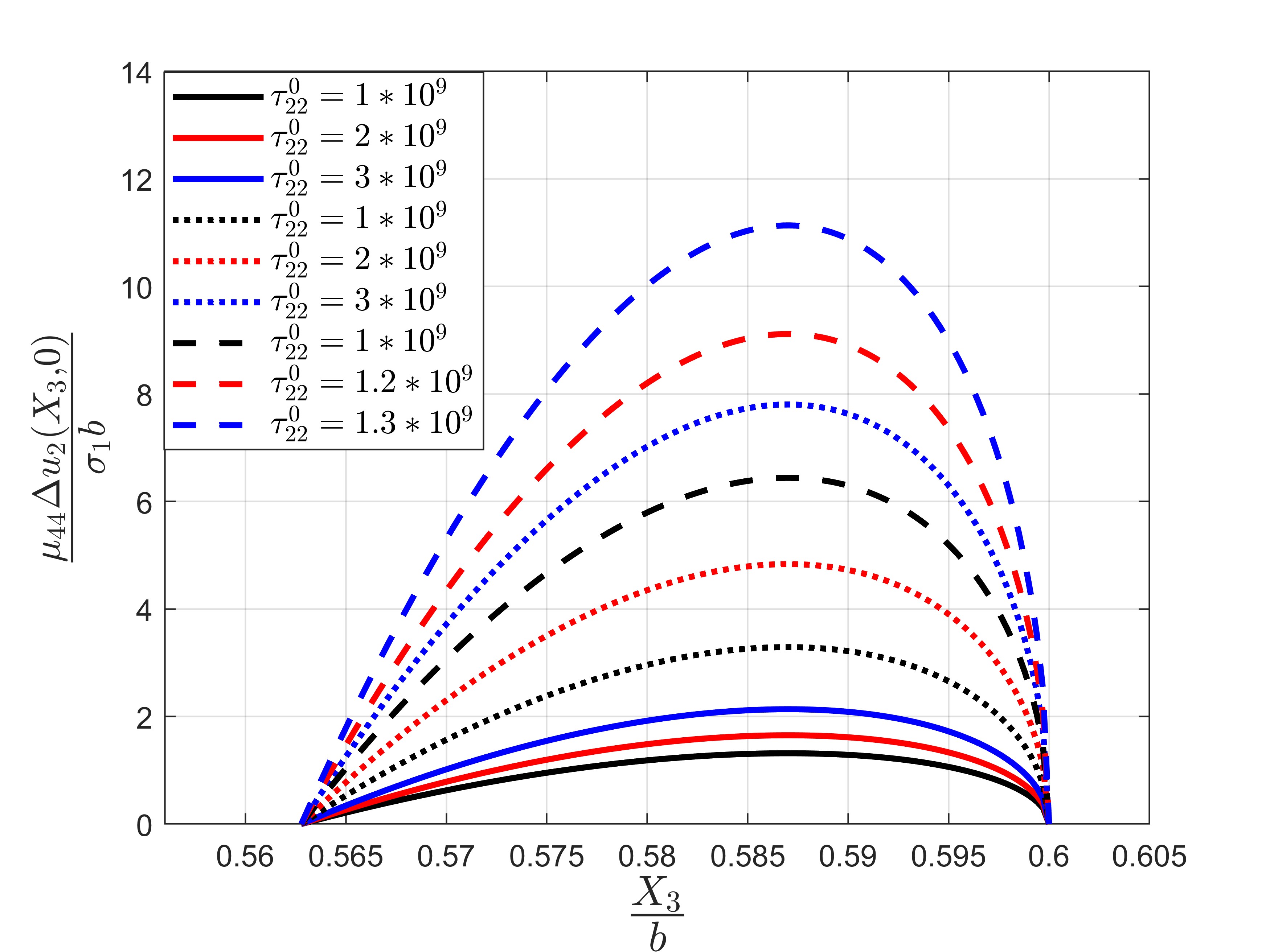}
\caption{}
\label{Fig_15}
\end{subfigure}
~
\begin{subfigure}[b] {0.49\textwidth}
\includegraphics[width=\textwidth ]{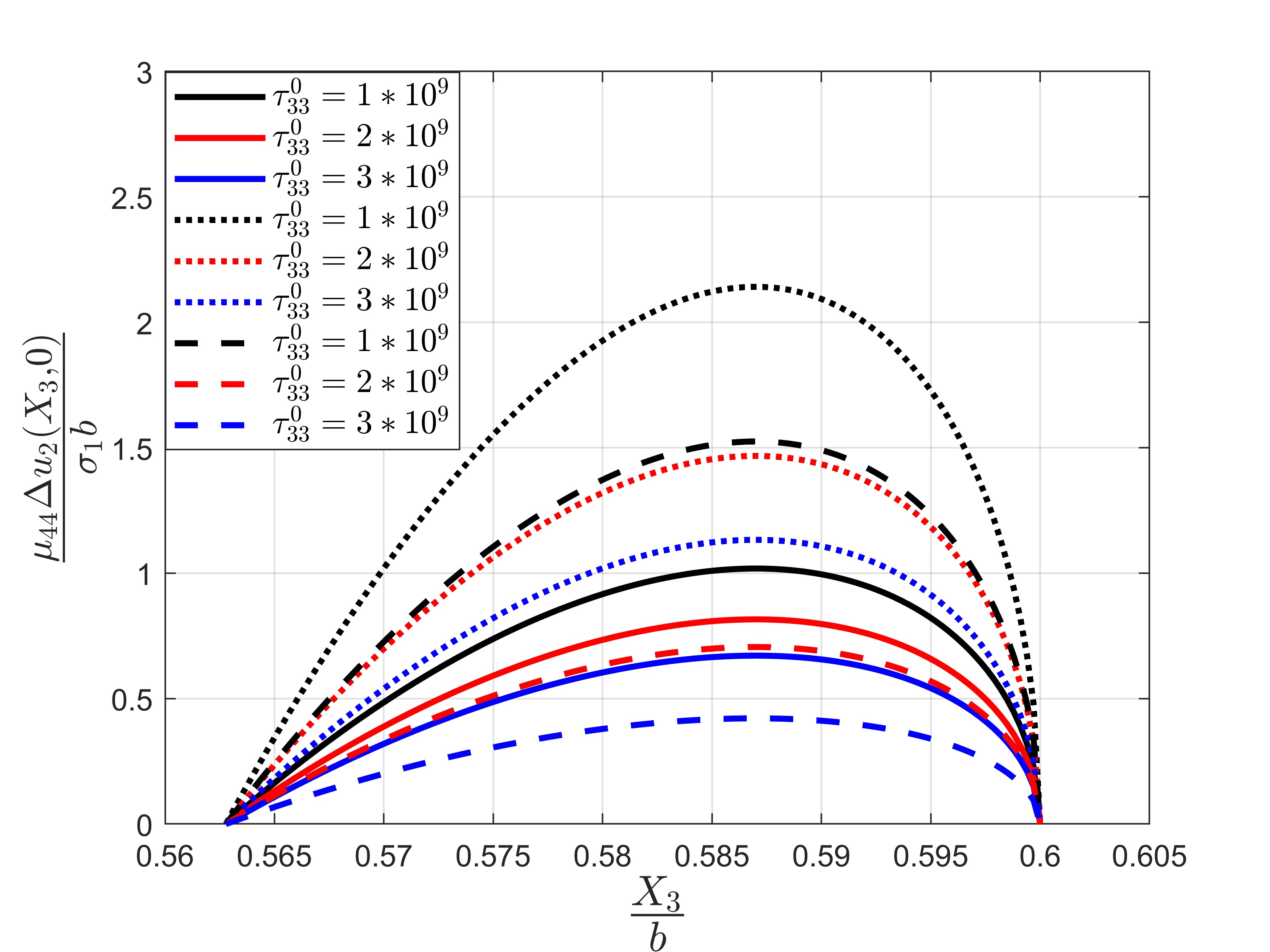}
\caption{}
\label{Fig_16}
\end{subfigure}
~
\begin{subfigure}[b] {0.49\textwidth}
\includegraphics[width=\textwidth ]{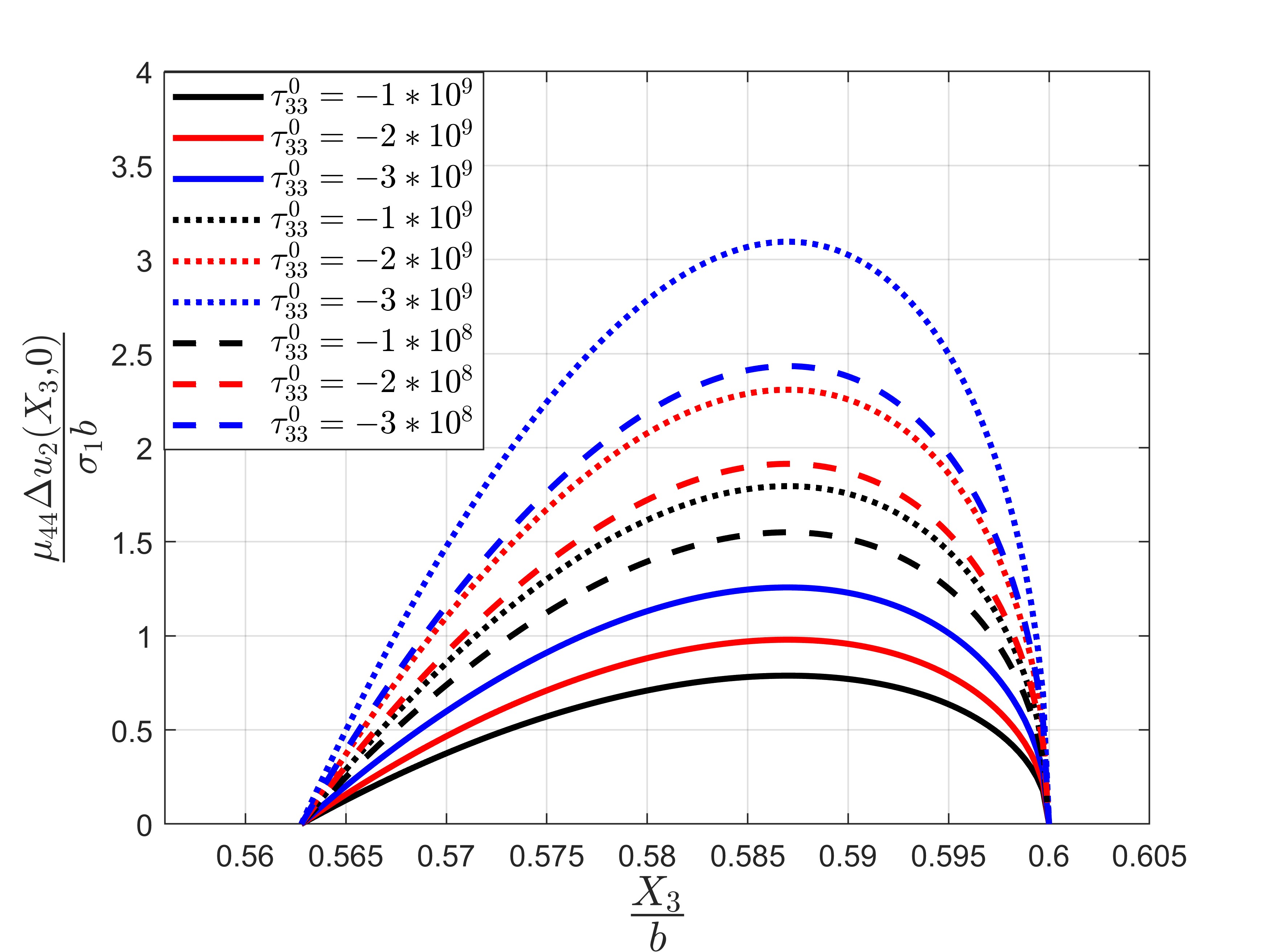}
\caption{}
\label{Fig_17}
\end{subfigure}
\caption{Variation of the dimensionless crack opening displacement $ (\mu_{44}\Delta u_2(X_3,0)/\sigma_1 b )$, considering the effects of (a) crack position $\left( a / b \right)$, (b) normal pressure $\left( \sigma_2 / \sigma_1 \right)$, (c) crack velocity ($v/v_c$), (d) initial compressive stress $\tau_{22}$, (e) initial compressive stress $\tau_{33}$, and (f) initial tensile stress $\tau_{33}$.}

\label{Figure 4}
\efg

\bfg[htbp]
\centering
\begin{subfigure}[b] {0.49\textwidth}
\includegraphics[width=\textwidth ]{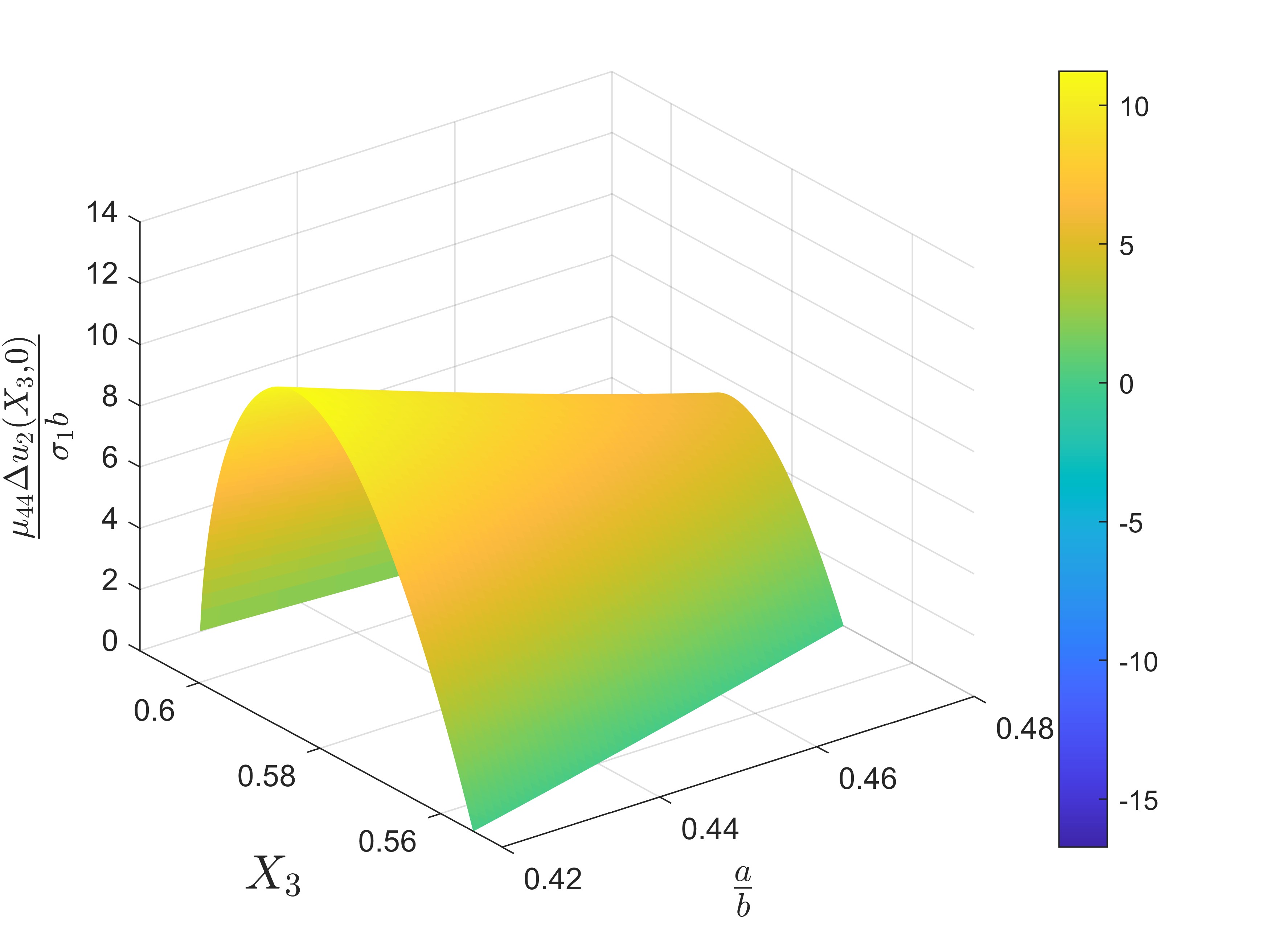}
\caption{}
\label{Fig_122}
\end{subfigure}
~
\begin{subfigure}[b] {0.49\textwidth}
\includegraphics[width=\textwidth ]{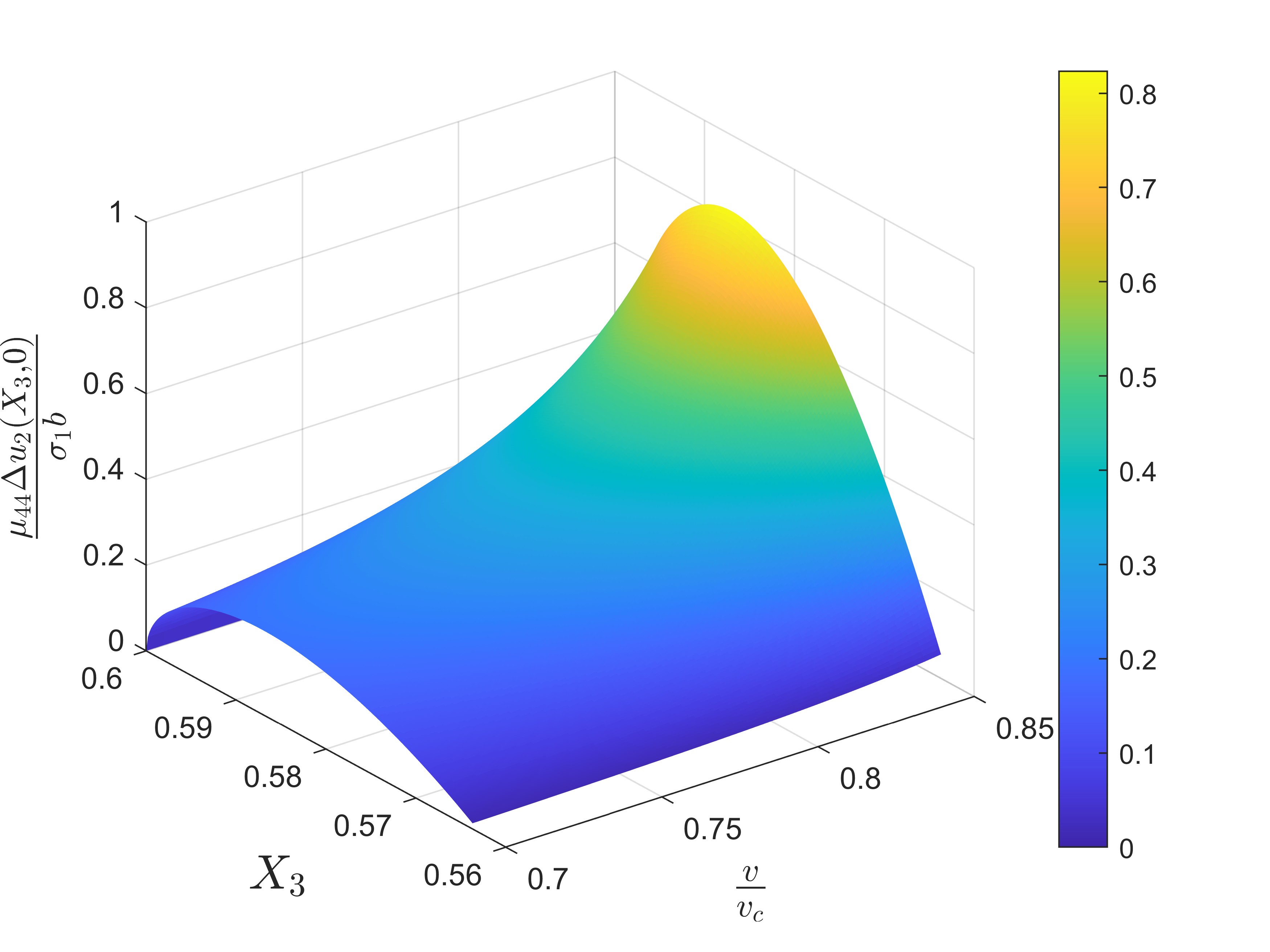}
\caption{}
\label{Fig_133}
\end{subfigure}
~
\begin{subfigure}[b] {0.49\textwidth}
\includegraphics[width=\textwidth ]{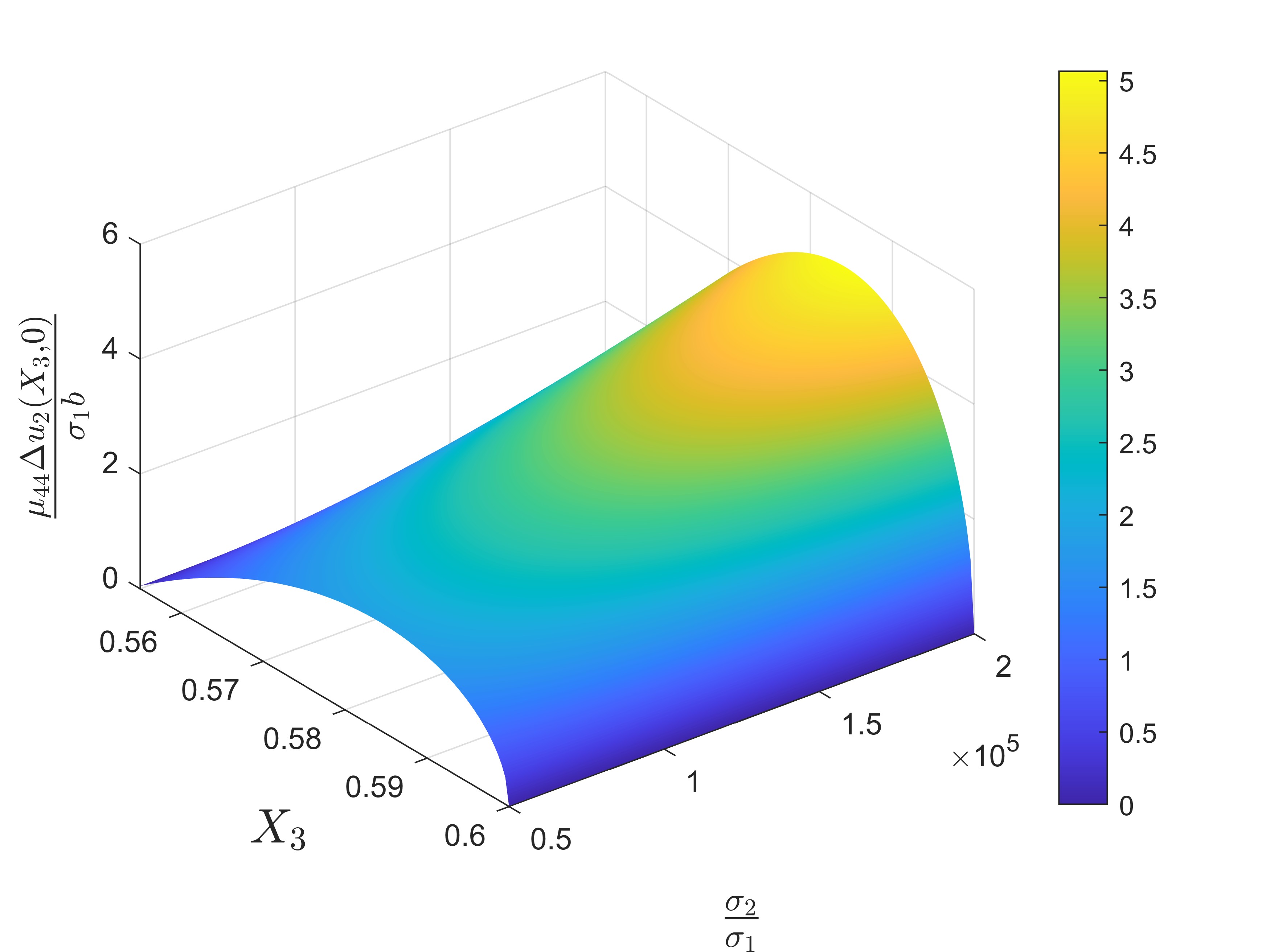}
\caption{}
\label{Fig_144}
\end{subfigure}
~
\begin{subfigure}[b] {0.49\textwidth}
\includegraphics[width=\textwidth ]{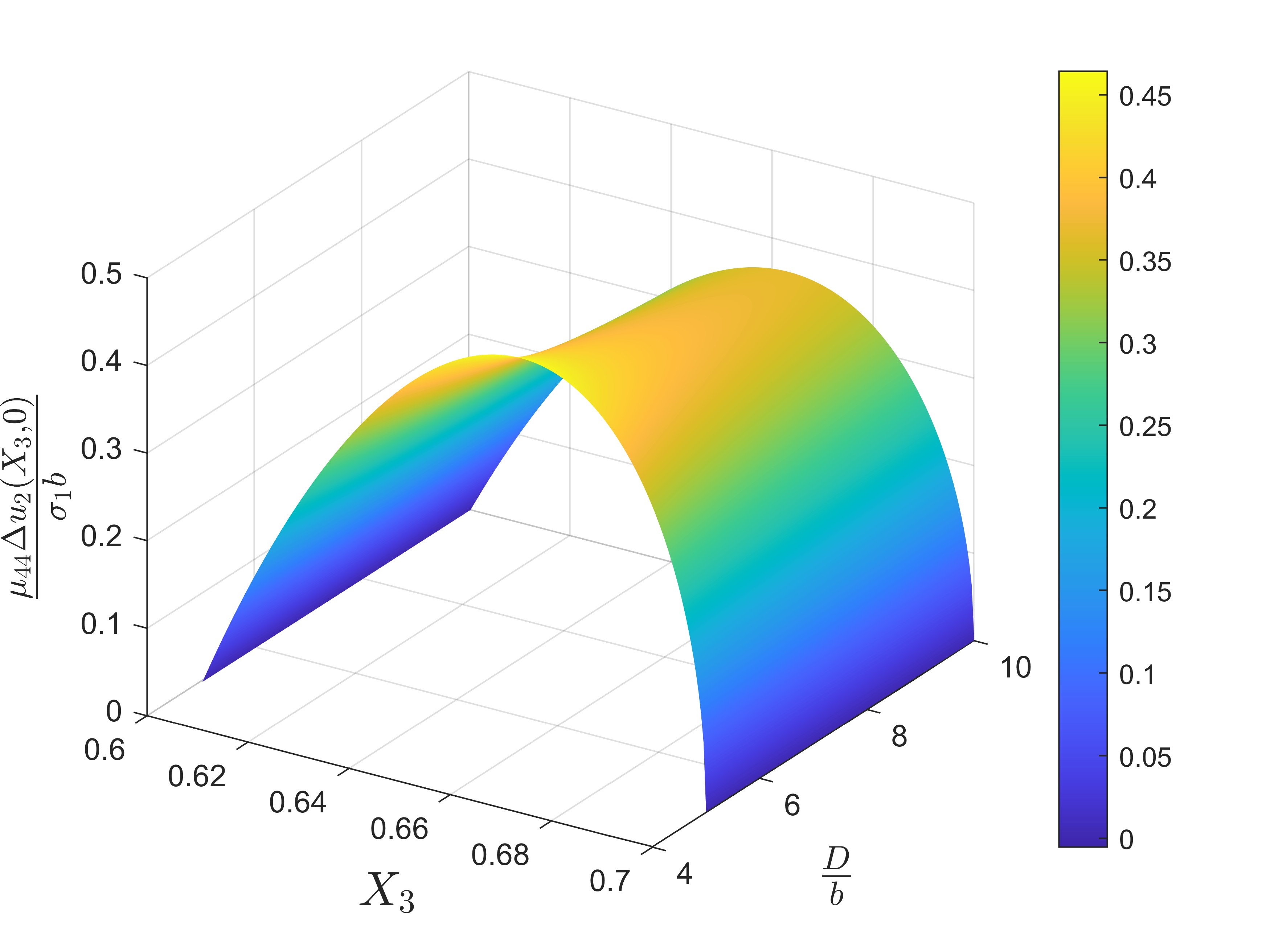}
\caption{}
\label{Fig_155}
\end{subfigure}
\caption{Surface plot illustrating the variation in crack opening displacement $ (\mu_{44}\Delta u_2(X_3,0)/\sigma_1 b )$ with respect to: (a) crack position ($a/b$), (b) crack velocity ($v/v_c$), (c) applied punch pressure ($\sigma_2/\sigma_1$), and (d) height of the layer ($D/b$) for Lithium Niobate (LiNbO$_3$).}

\label{Figure 44}
\efg

\subsection{Special cases: comparative analysis under point loading and uniform normal pressure}
To better understand the influence of localized loading, Figures \ref{Figure 7} and \ref{Figure 8} illustrate a special case where the strip is subjected to point loading at $X_0/b=0.3$. This case is compared against the general scenario involving a uniform normal constant pressure applied along the strip. For this purpose, Eqs.~(\ref{eq53}), (\ref{eq54}), (\ref{eq62}), and~(\ref{eq63}) have been utilized to compute the stress intensity factor (SIF) at both crack tips, specifically at $X_3 = a$ and $X_3 = b$, under three conditions: compressive initial stress, tensile initial stress, and the absence of initial stress.
Based on this comparative study, the following key observations have been made:
\begin{itemize}
    \item[(a)] Figure~\ref{Figure 7} shows that the SIF at the left crack tip ($X_3 = a$) is significantly higher when the strip is subjected to point loading compared to the case of uniform pressure. This indicates a greater stress concentration under localized loading. In contrast, Figure~\ref{Figure 8} reveals that the SIF at the right crack tip ($X_3 = b$) is lower under point loading than under uniform pressure, highlighting the sensitivity of SIF distribution to the type of external loading.

    \item[(b)] It is also observed that when the strip is subjected to compressive initial stress, i.e., $\tau_{22} = \tau_{33} = 10^{10}$, the SIF is lower than that in the case without initial stress. On the other hand, the SIF increases when the strip is under tensile initial stress, i.e., $\tau_{22} = \tau_{33} = -10^{10}$. This behavior underscores the significant effect of the initial stress state on fracture characteristics.

\end{itemize}
\bfg[htbp]
\centering
\begin{subfigure}[b] {0.49\textwidth}
\includegraphics[width=\textwidth ]{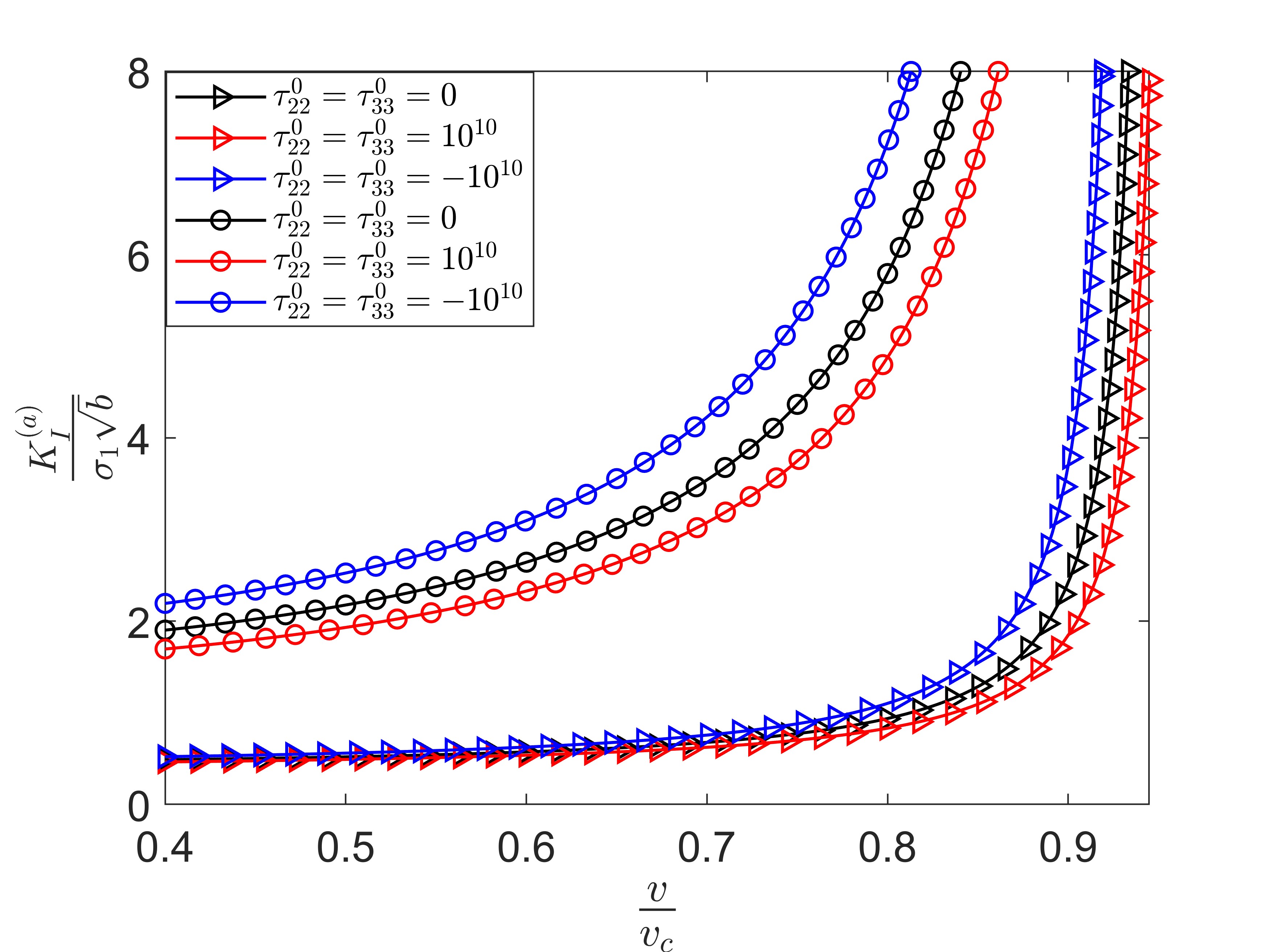}
\caption{}
\label{Fig_30}
\end{subfigure}
~
\begin{subfigure}[b] {0.49\textwidth}
\includegraphics[width=\textwidth ]{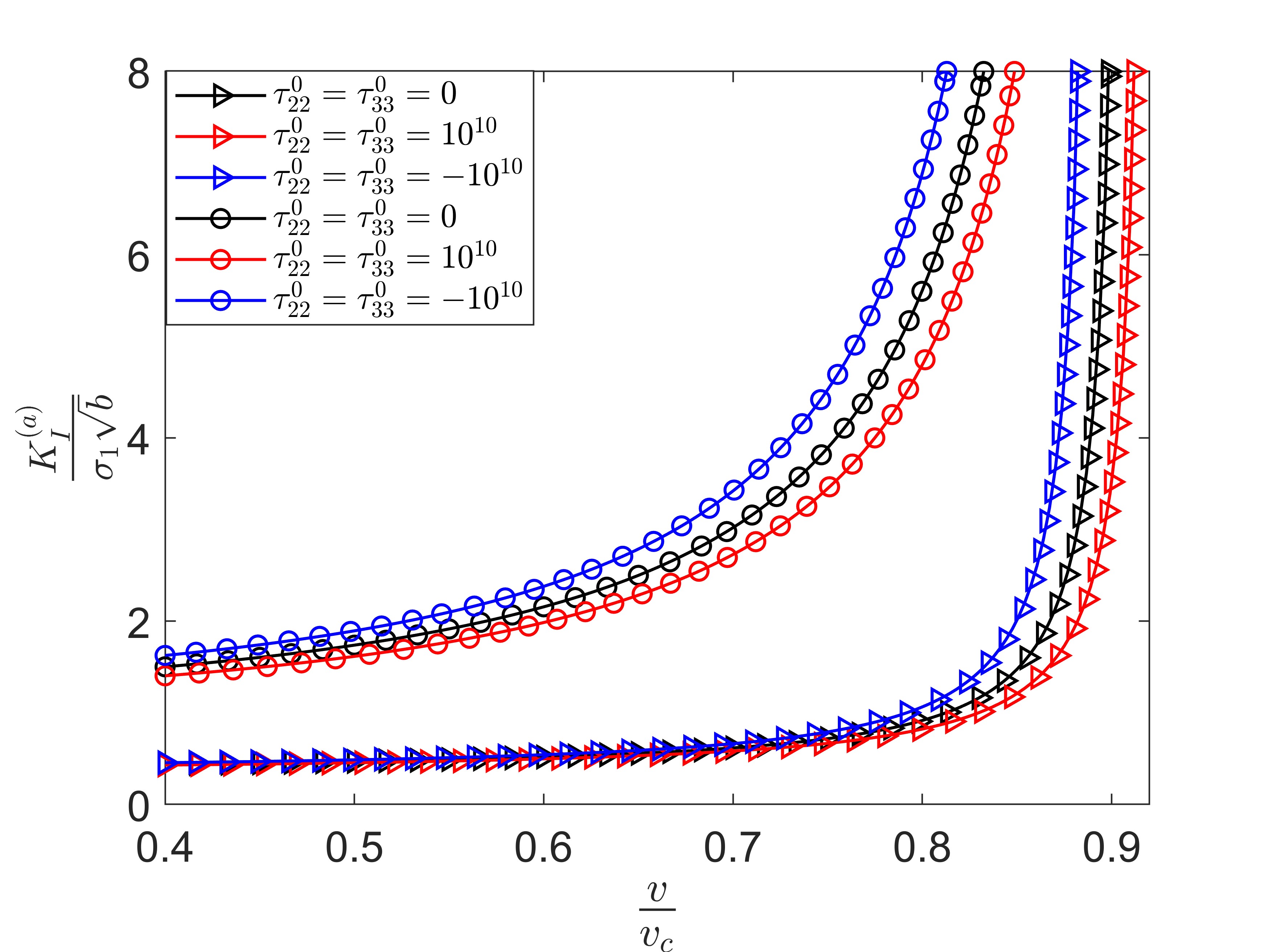}
\caption{}
\label{Fig_31}
\end{subfigure}
~
\begin{subfigure}[b] {0.49\textwidth}
\includegraphics[width=\textwidth ]{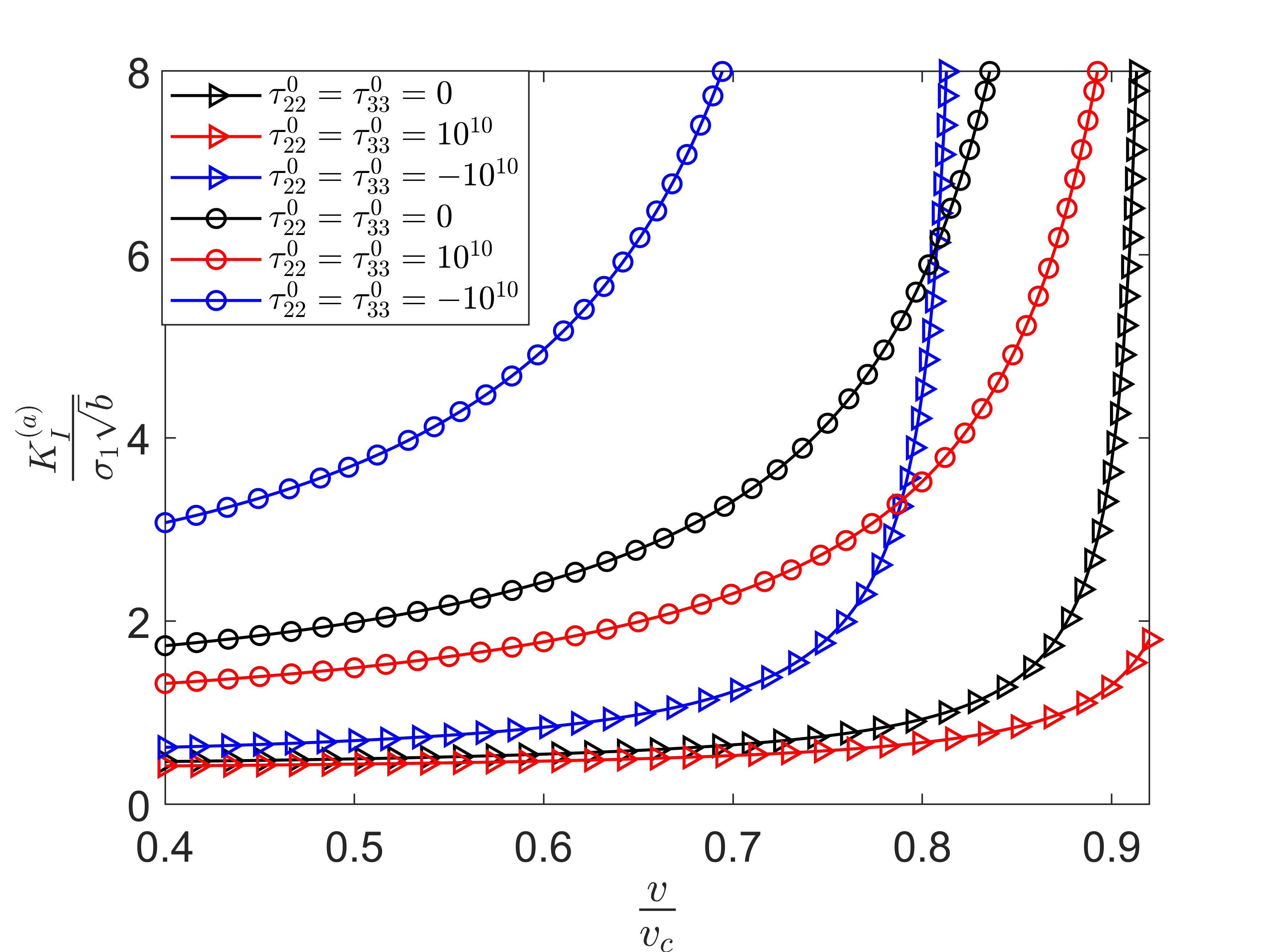}
\caption{}
\label{Fig_32}
\end{subfigure}
\caption{Variation of the dimensionless stress intensity factor $(K^{(a)}_I/\sigma_1\sqrt{b})$ at the left crack tip ($X_3 = a$) versus the dimensionless crack velocity $\left(v/v_c\right)$ for a strip subjected to uniform normal constant pressure (marker “$-->$”) or point loading (marker “$--0$”) under initial compressive, tensile, and pressure-free conditions for (a) Lithium Niobate, (b) Lithium Tantalate, and (c) an isotropic material.}
\label{Figure 7}
\efg
\bfg[htbp]
\centering
\begin{subfigure}[b] {0.49\textwidth}
\includegraphics[width=\textwidth ]{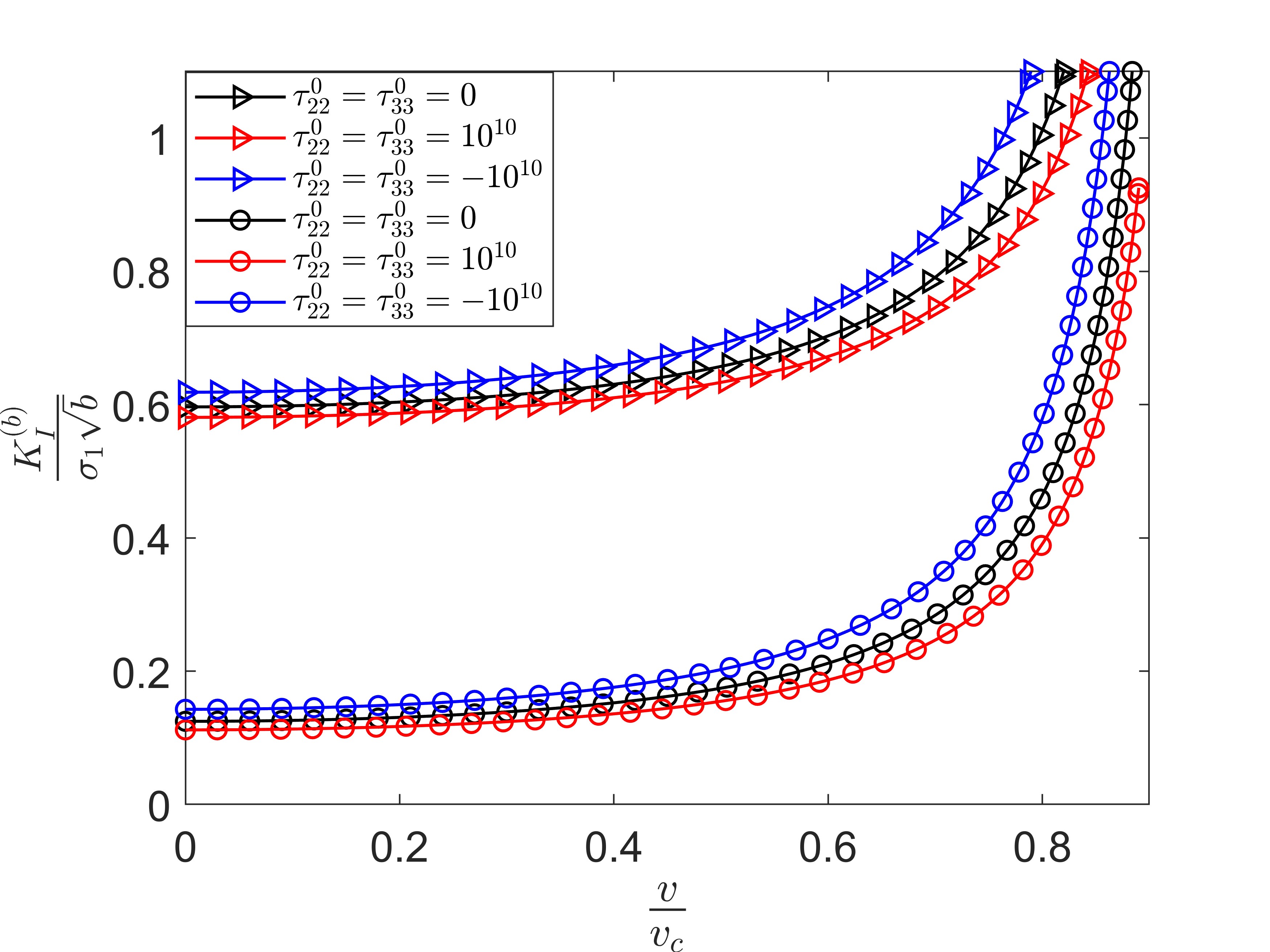}
\caption{}
\label{Fig_33}
\end{subfigure}
~
\begin{subfigure}[b] {0.49\textwidth}
\includegraphics[width=\textwidth ]{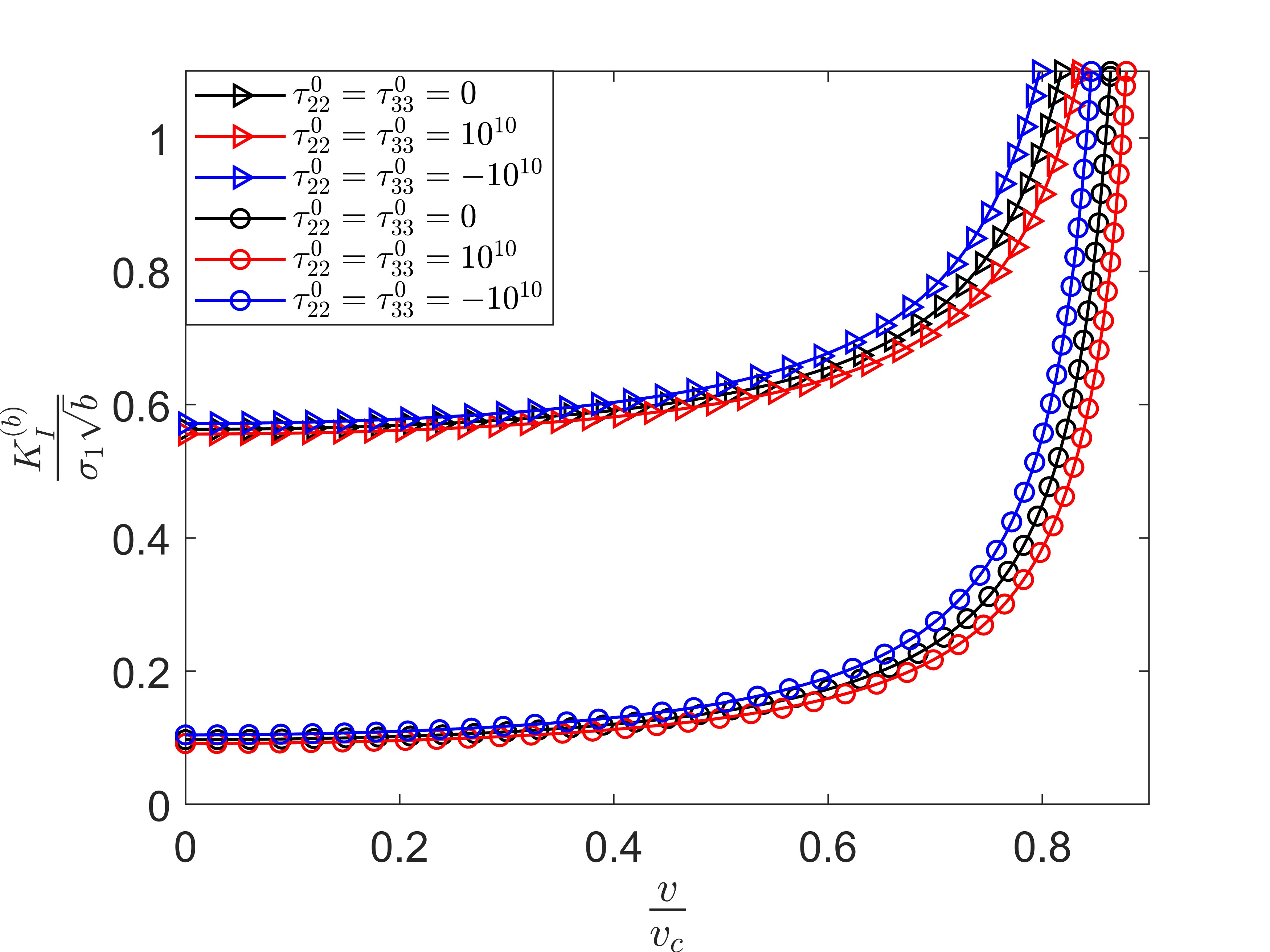}
\caption{}
\label{Fig_34}
\end{subfigure}
~
\begin{subfigure}[b] {0.49\textwidth}
\includegraphics[width=\textwidth ]{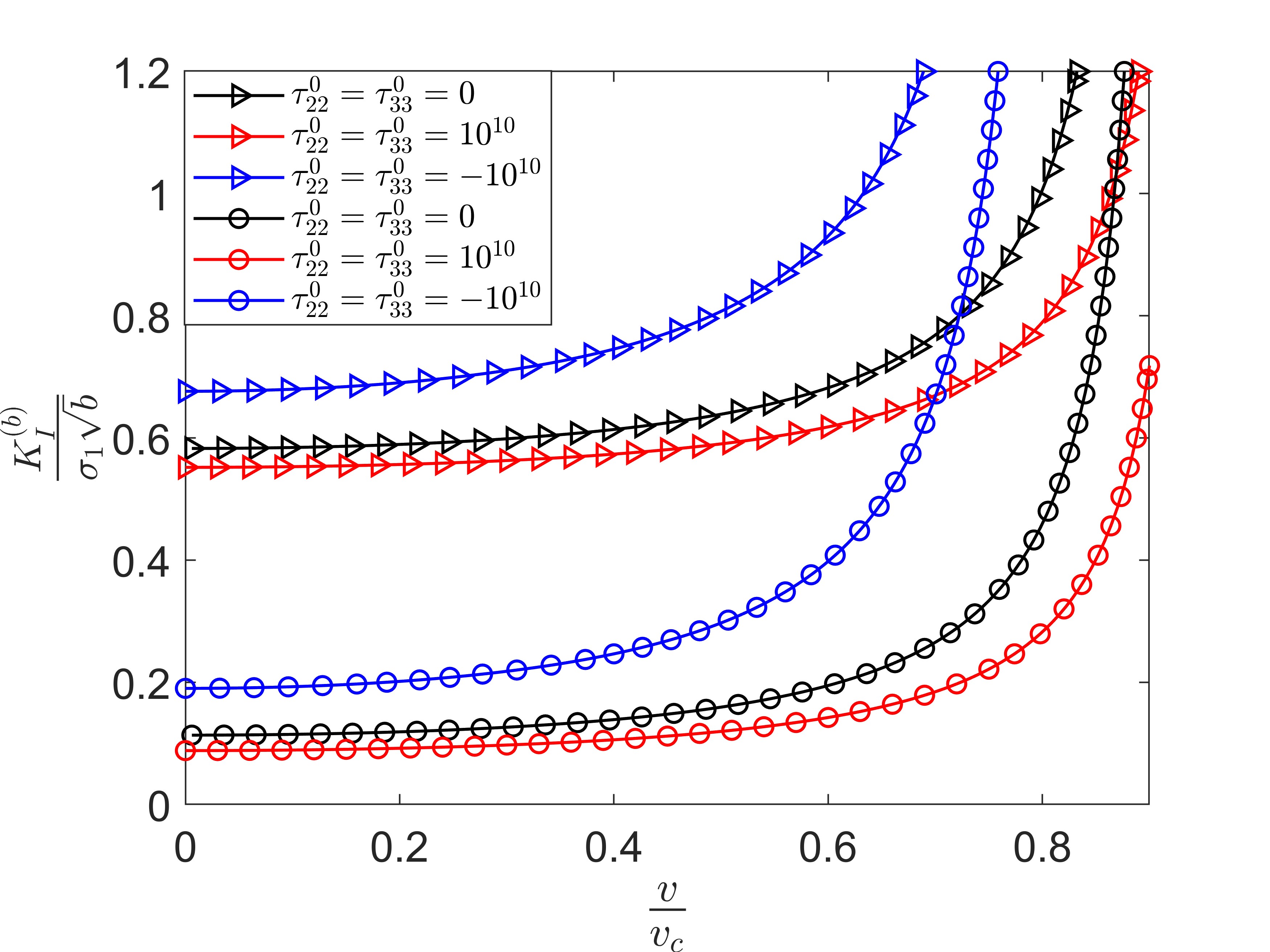}
\caption{}
\label{Fig_35}
\end{subfigure}
\caption{Variation of the dimensionless stress intensity factor $(K^{(b)}_I/\sigma_1\sqrt{b})$ at the right crack tip ($X_3 = b$) versus the dimensionless crack velocity $\left(v/v_c\right)$ for a strip subjected to uniform normal constant pressure (marker “$-->$”) or point loading (marker “$--0$”) under initial compressive, tensile, and pressure-free conditions for (a) Lithium Niobate, (b) Lithium Tantalate, and (c) an isotropic material.}

\label{Figure 8}
\efg
\newpage
\section{Conclusion}
\label{Conclusion}
The present study investigates the propagation of moving collinear Griffith cracks under Mode I in a pre-stressed monoclinic crystalline material layer subjected to punch loading at the crack location. Closed-form expressions for the stress intensity factor (SIF) and crack opening displacement (COD) have been derived using an analytical formulation that incorporates the Hilbert transform within a Galilean-transformed moving reference frame. To explore the material behavior under various loading conditions, the study considers two crystalline monoclinic materials, Lithium Niobate (LiNbO$_3$) and Lithium Tantalate (LiTaO$_3$), and compares their responses with those of an isotropic material. The analysis examines the effects of material anisotropy, initial stress fields, punch loading, and crack velocity on the dynamic fracture parameters. The key findings are as follows:
\begin{itemize}
    \item An increase in crack velocity results in elevated SIF values at both crack tips, demonstrating a direct correlation between crack propagation speed and stress concentration. In contrast, an increase in the crack tip ratio (i.e., a decrease in crack length) reduces the SIF at both crack tips, underscoring the sensitivity of fracture behavior to geometric asymmetry in crack configuration.
\item It is observed that compressive initial stress applied vertically to the material strip leads to an increase in the intensity of the SIF at both crack tips, whereas an increase in the magnitude of vertically applied tensile initial stress results in a decrease in SIF intensity. This signifies that compressive stress tends to promote crack propagation by enhancing stress concentration, while tensile stress offers a stabilizing effect, potentially delaying fracture under dynamic loading conditions.
\item It has been found that compressive initial stress applied horizontally leads to a decrease in SIF at both crack tips, whereas the application of tensile horizontal initial stress results in an increase in SIF intensity. This indicates that horizontal compressive stress helps to suppress crack growth by reducing stress concentration, while tensile stress in the same direction facilitates crack propagation under dynamic conditions.
 \item It has been found that Lithium Niobate exhibits minimal sensitivity to crack opening displacement, whereas Lithium Tantalate and isotropic materials show the highest sensitivity. This suggests that Lithium Niobate is more resistant to crack propagation, while materials like Lithium Tantalate and isotropic materials are more prone to significant displacement under similar loading conditions, making them more vulnerable to fracture under dynamic stresses.
\item It has been observed that the SIF is higher when a constant normal pressure is applied, compared to the case of point loading at the right crack tip. Conversely, at the left crack tip, the SIF is higher under point loading than when constant normal pressure is applied. This indicates that the nature of the loading (constant pressure vs. point loading) affects the distribution and intensity of stresses differently at the two crack tips.
\end{itemize}
This analysis will be valuable in predicting the lifespan of materials under various loading conditions, aiding in the prediction of fracture and failure in materials used in various sensors and devices, such as SAW sensors, MEMS, and biosensors.

\noindent {\bf Acknowledgements}\\
The authors sincerely acknowledge the National Institute of Technology Hamirpur for providing the necessary facilities to Ms. Diksha during her PhD, and the University Grants Commission (UGC) for supporting this research through a fellowship.\\
\noindent {\bf Conflicts of Interest}\\
The authors declare that there are no conflicts of interest related to this work.\\
\noindent {\bf Code Availability}\\
The code used in this study was entirely developed by the authors in MATLAB, without the use of any third-party software or external code libraries.
\bibliographystyle{unsrt}
\bibliography{refrences}
\end{document}